\documentclass{aastex631}

\submitjournal{AJ on 2024-09-20}

\shortauthors{Couperus et al.}

\begin{document}

\title{The Solar Neighborhood LII: M Dwarf Twin Binaries --- Presumed Identical Twins Appear Fraternal in Variability, Rotation, H$\alpha$, and X-rays}

\correspondingauthor{Andrew A. Couperus}
\email{andcoup1@gmail.com}

\author[0000-0001-9834-5792]{Andrew A. Couperus}
\affiliation{Department of Physics and Astronomy, Georgia State University, Atlanta, GA 30302, USA}
\affiliation{RECONS Institute, Chambersburg, PA 17201, USA}

\author[0000-0002-9061-2865]{Todd J. Henry}
\affiliation{RECONS Institute, Chambersburg, PA 17201, USA}

\author[0000-0001-5643-8421]{Rachel A. Osten}
\affiliation{Space Telescope Science Institute, Baltimore, MD 21218, USA}
\affiliation{Center for Astrophysical Sciences, Department of Physics and Astronomy, Johns Hopkins University, Baltimore, MD 21218, USA}

\author[0000-0003-0193-2187]{Wei-Chun Jao}
\affiliation{Department of Physics and Astronomy, Georgia State University, Atlanta, GA 30302, USA}

\author[0000-0002-1864-6120]{Eliot Halley Vrijmoet}
\affiliation{Five College Astronomy Department, Smith College, Northampton, MA 01063, USA}
\affiliation{RECONS Institute, Chambersburg, PA 17201, USA}

\author[0000-0002-9811-5521]{Aman Kar}
\affiliation{Department of Physics and Astronomy, Georgia State University, Atlanta, GA 30302, USA}
\affiliation{RECONS Institute, Chambersburg, PA 17201, USA}

\author[0000-0003-2159-1463]{Elliott Horch}
\affiliation{Department of Physics, Southern Connecticut State University, New Haven, CT 06515, USA}




\begin{abstract}
\noindent
We present an investigation into the rotation and stellar activity of four fully convective M dwarf `twin' wide binaries. Components in each pair have (1) astrometry confirming they are common-proper-motion binaries, (2) {\it Gaia} $BP$, $RP$, and 2MASS $J$, $H$, and $K_s$ magnitudes matching within 0.10~mag, and (3) presumably the same age and composition. We report long-term photometry, rotation periods, multi-epoch H$\alpha$ equivalent widths, X-ray luminosities, time series radial velocities, and speckle observations for all components. Although it might be expected for the twin components to have matching magnetic attributes, this is not the case. Decade-long photometry of GJ~1183~AB indicates consistently higher spot activity on A than B, a trend matched by A appearing 58$\pm$9\% stronger in $L_X$ and 26$\pm$9\% stronger in H$\alpha$ on average --- this is despite similar rotation periods of A=0.86d and B=0.68d, thereby informing the range in activity for otherwise identical and similarly-rotating M dwarfs. The young $\beta$ Pic Moving Group member 2MA~0201+0117~AB displays a consistently more active B component that is 3.6$\pm$0.5 times stronger in $L_X$ and 52$\pm$19\% stronger in H$\alpha$ on average, with distinct rotation at A=6.01d and B=3.30d. Finally, NLTT~44989~AB displays remarkable differences with implications for spindown evolution --- B has sustained H$\alpha$ emission while A shows absorption, and B is $\geq$39$\pm$4 times stronger in $L_X$, presumably stemming from the surprisingly different rotation periods of A=38d and B=6.55d. The last system, KX~Com, has an unresolved radial velocity companion, and is therefore not a twin system.
\end{abstract}




\section{Introduction} \label{sec:intro}

M dwarfs comprise about three-quarters of all stars in the solar neighborhood \citep{2006AJ....132.2360H, Henry_2018}, and presumably all over the local Universe. They span roughly a factor of eight in mass across their spectral range \citep{Benedict_2016, 2019ApJ...871...63M}, with stars above $\sim$0.35\(\textup{M}_\odot\) hosting partially convective (PC) structures like our Sun while cases below $\sim$0.35\(\textup{M}_\odot\) are fully convective (FC) without a radiative core \citep{1997A&A...327.1039C}. These low-mass stars also host strong magnetic fields that cause starspots, faculae, flares, chromospheric activity, X-ray coronal enhancement, radio emission, and more \citep[and references therein]{2021isma.book.....B}. Among the consequences are flux variations spanning timescales of minutes to decades that manifest across a range of wavelengths.

Investigating this magnetic behavior is critical for understanding the potential habitability of any exoplanets orbiting these stars. M dwarfs are likely candidates for hosting detectable Earth-like exoplanets \citep{2023AJ....165..265M}, upon which the incident stellar flux directly influences atmospheric and surface conditions. The high activity levels of M dwarfs may have dangerous consequences for habitability \citep{2007AsBio...7...30T, Shields_2016}, or may be advantageous in the drive for prebiotic chemistry and evolution of any lifeforms on planetary surfaces \citep{2017ApJ...843..110R}. Regardless, there is interest in studying the poorly understood dynamos of FC M dwarfs \citep{Shulyak_2015}, as they allow an exploration of dynamo theory in a new regime for comparison to the solar magnetic dynamo. The final motivation we posit is the most fundamental, that of understanding the astrophysical behavior of the most abundant stars in the Universe.

Together, these factors generate the desire for reliable predictions of the evolving stellar activity throughout the lives of M dwarfs. Fully assessing their activity is intrinsically tied to understanding the spindown process for these stars as well, because rotation drives activity through the stellar dynamo. However, work by \cite{Newton_2016,Newton_2018} uncovered a strong bimodality in the observed rotation periods of FC M dwarfs, with \cite{Jao_2023} (hereafter J23) finding significantly higher historical magnetic braking strengths in the most massive FC M dwarfs compared to PC stars and lower mass FC stars. The bimodality is likely caused at least in part by a poorly-understood temporary phase of very rapid spindown in these FC stars' lifetimes, with the mass-dependent age of onset being variable across stars and likely linked to the initial rotation period and evolution of the magnetic field morphology \citep[and references therein; Pass et al.~2024 is hereafter referred to as P24]{Garraffo2018, Pass2022, Pass_2023, 2023MNRAS.526..870S, Pass_2024_ApJ}.

Rotational evolution presumably drives activity evolution but, somewhat surprisingly, significant activity level differences have been observed for otherwise nearly identical M dwarfs. For example, the two components in the BL+UV Ceti binary system (GJ 65 AB) have virtually identical masses \citep[A=0.120$\pm$0.003\(\textup{M}_\odot\), B=0.117$\pm$0.003\(\textup{M}_\odot\);][]{Benedict_2016} and rotation periods \citep[A=0.24d, B=0.23d;][]{Barnes_2017}, and are presumably of the same age.  However, they display incongruous star spot distributions \citep{Barnes_2017}, markedly different magnetic field strengths and topologies \citep{Kochukhov_2017}, mismatched X-ray variability \citep{Audard_2003}, and different levels of radio emission \citep[and references therein]{Audard_2003, 2024arXiv240617280P}. The difference in X-ray variability was recently observed to have possibly normalized to roughly similar activity levels compared to 18 years earlier \citep{Wolk-UVCet-CoolStars21}, while the radio differences have persisted over several decades. Radio emission differences have also been found between the similar M dwarf components in Ross 867-8 \citep{2020A&A...633A.130Q}. In addition, \cite{Gunning_2014} found marked differences in chromospheric H$\alpha$ activity between near-equal-mass components in several M dwarf wide binaries, although rotation periods were unavailable, thereby limiting a fully contextualized interpretation of their results.

A dynamo bistability has been proposed to explain some of the magnetic field mismatches in late-type M dwarfs, where two distinct dynamo states could emerge from similar initial fundamental stellar parameters \citep{Gastine_2013}. Other efforts have instead implicated long-term stellar cycles with dynamically changing magnetic structures to explain some mismatches in various mass and rotation regimes \citep{Kitchatinov_2014, Farrish_2021}. Altogether, the cases of observed activity differences in otherwise similar M dwarfs remain inadequately understood, indicating that work remains to be done to improve any predictions about their magnetic attributes and consequent effects on orbiting exoplanets.

In an effort to improve our understanding of M dwarf magnetism and rotation, here we report first results from our investigation of a sample of 36 `twin' M dwarf wide binaries, under the aegis of the REsearch Consortium On Nearby Stars (RECONS; \href{www.recons.org}{www.recons.org}).  For each of the pairs, we seek to determine if the twin components show the same or meaningfully different magnetic properties and rotation. Any observed mismatches in the rotation periods between twin stars could imply stochasticity and set constraints on the spindown process, while differences in activity for otherwise similarly rotating twin stars sets constraints on the potential intrinsic scatter in magnetic activity for even equal mass/age/composition/rotation stars. Of particular note is the understudied long-term years-to-decades variability, where out-of-phase stellar magnetic cycles may be the cause of some activity differences at a given epoch of observation.

Results from this twin study will be split into a two-paper series. This first effort outlines our overall methodology while focusing on a subset of four intriguing sets of twins for which a variety of observations have revealed activity differences. A second forthcoming paper will then discuss the remaining 32 twin systems and overall results for our cumulative sample (Couperus et al.~in prep).

This first paper is split into seven further sections: \S\ref{sec:Sample} outlines our sample, followed by details of each observing campaign in \S\ref{sec:ObsDatRed}. In \S\ref{sec:Results} and \S\ref{sec:rot-activity-results} we present our results, and we give additional notes on the systems in \S\ref{sec:SysNotes}. We then discuss the results in \S\ref{sec:discussion} and summarize the key insights in \S\ref{sec:conclusions}. Additional materials are provided in Appendix \ref{sec:appendix}.

\section{Sample} \label{sec:Sample}

The RECONS Twins sample was constructed by searching \textit{Gaia} DR2 \citep{2018A&A...616A...1G} for common proper motion wide binaries with nearly identical components. First, we extracted M stars within 50~pc by selecting for \texttt{parallax > 20~mas} and \texttt{$BP-RP$ > 2.0}. We then selected source pairs with angular separations of 4$\arcsec$--300$\arcsec$, which allows sources to be resolved in many observing programs while still close enough to fit within typical detector fields of view. Pairs with components having $BP$, $RP$, or 2MASS \textit{J}, \textit{H}, or $K_s$ differing by $>$ 0.10~mag were then removed in order to select only `twin' stars with nearly identical magnitudes across the optical and near-IR wavelengths, where M dwarfs emit most of their light. Finally, we removed pairs whose component parallax distances differed by more than 1~pc. No additional criteria such as parallax error cutoffs or proper motion matches were needed (pairs all have proper motion components matching within a few mas/yr). This yielded an all-sky sample of 36 M dwarf twin binaries that are still astrometrically associated and pass these same cuts in updated \textit{Gaia} DR3 data \citep{GaiaDR3}. To further confirm the binary nature of our stars we crossmatched with the SUPERWIDE catalog of \cite{2020ApJS..247...66H}, finding all four systems considered here to be real wide binaries at $>$99.99\% probability.

The four systems we highlight in this current publication are GJ~1183~AB, KX~Com~A-BC, 2MA~0201+0117~AB, and NLTT~44989~AB, chosen for their standout activity behaviors amongst our sample and their inclusion in the \textit{Chandra} X-ray study detailed here. ``KX~Com" nominally refers to what we call our A component, and we added the name association to what we call our B component and its subsequently discovered C companion discussed later in \S\ref{subsubsec:rv-results} and \S\ref{subsec:KXCOMAB}. The designation NLTT~44989 specifically refers to A, while B is NLTT~44988, but we refer to them as NLTT~44989~A and B for clarity throughout this paper; this system also goes by the name LP 920-61 AB. A and B labels were decided following some existing catalog component names (GJ~1183~A and B), but were otherwise chosen by DR2 $BP$ brightness. The A or B distinction is somewhat arbitrary for these twin stars because different measurements or catalogs will often flip-flop which is the brighter A star, so precise coordinates may prove more useful for the interested investigator.

\begin{figure}[t]
\centering
\includegraphics[scale=0.22]{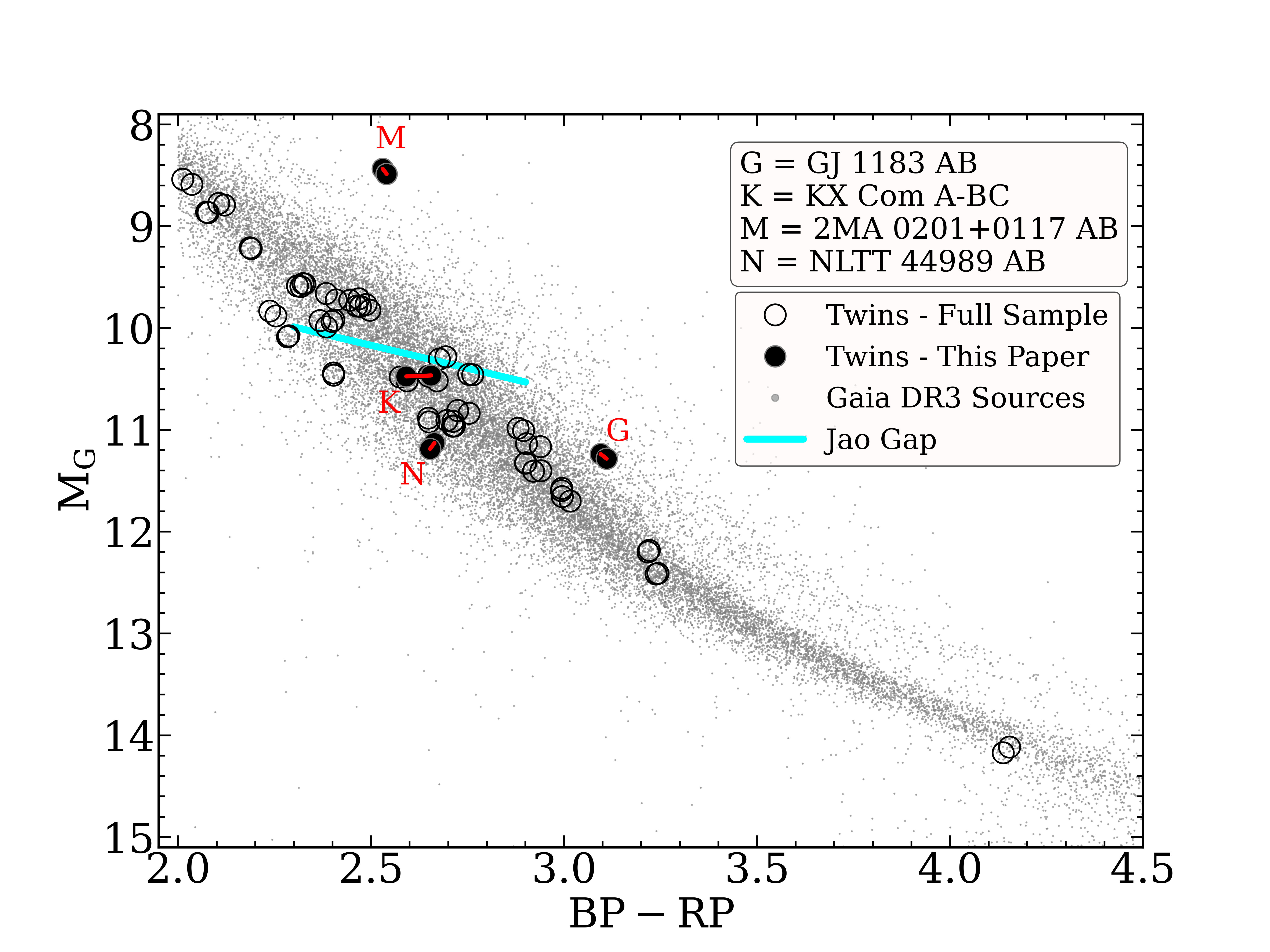}
\figcaption{An observational Hertzsprung–Russell Diagram using \textit{Gaia} DR3 magnitudes and parallaxes. Grey points show a sample of \textit{Gaia} sources within 50 parsecs to illustrate the main sequence. Large black circles indicate stars in our sample of 36 M dwarf twin binary pairs. The four systems examined in this paper are filled in and labeled, with red lines connecting the two components in those pairs. A diagonal cyan line represents the gap marking the transition between partially and fully convective M dwarfs near $\sim$0.35\(\textup{M}_\odot\) \citep{2018ApJ...861L..11J, Jao_2023}, offset downward by 0.05~mag to approximately match the middle of the gap instead of the upper edge. \label{fig:HRD}}
\end{figure}

The twin binaries are shown on an observational Hertzsprung–Russell Diagram in Figure \ref{fig:HRD}, where the four systems of interest for this paper are highlighted with solid points connected by short red lines. As expected, this Figure confirms that no binary giants were mistakenly captured in our search. The significantly elevated pair is 2MA~0201+0117~AB, a member of the young 25-Myr $\beta$ Pictoris association \citep{2015A&A...583A..85A, 2017A&A...607A...3M}; the other slightly elevated system is GJ~1183~AB. All components in the four systems targeted here are fully convective stars  --- three systems are below the partially/fully convective transition gap of \cite{2018ApJ...861L..11J} indicated with the diagonal cyan line, whereas 2MA~0201+0117~AB has been functionally fully convective throughout its brief life because it is a pre-main-sequence (PMS) system. We also note that none of the four systems --- except KX~Com~A-BC, a non-twin triple --- land within the activity dip sub-gap region identified in J23 between $M_G$ = 10.3--10.8.
%
\begin{deluxetable}{lccrrrrrrrc}[t!]
\tablewidth{0pt}
\tablecaption{Four Twin Systems Explored In This Paper --- Astrometry\label{tab:SampleTable-astr}}
\tablehead{
\colhead{Name} & \colhead{RA} & \colhead{Dec} & \colhead{$\pi$} & \colhead{$\mu_\alpha$} & \colhead{$\mu_\delta$} & \colhead{Ang. Sep.} & \colhead{2d Sep.} & \colhead{RUWE} & \colhead{RV$_{Gaia}$} & \colhead{IPDfmp} \\
\colhead{} & \colhead{[ICRS-2016]} & \colhead{[ICRS-2016]} & \colhead{[mas]} & \colhead{[mas/yr]} & \colhead{[mas/yr]} & \colhead{[arcsec]} & \colhead{[AU]} & \colhead{} & \colhead{[km/s]} & \colhead{[\%]}
}
\startdata
GJ~1183~A       &  14 27 55.69  &  $-$00 22 30.5  &  57.01  &  $-$361.15  &      41.70  &  13.07  &  229.3  &  1.510  &  $-$11.72$\pm$3.27  & 0 \\
GJ 1183 B       &  14 27 56.01  &  $-$00 22 18.3  &  57.03  &  $-$363.17  &      52.54  &         &         &  1.524  &       \nodata       & 0 \\
KX Com A        &  12 56 52.80  &  $+$23 29 50.7  &  36.61  &      70.27  &       4.10  &  7.72   &  211.2  &  1.234  &   $-$7.89$\pm$0.84  & 0 \\
KX Com BC       &  12 56 52.24  &  $+$23 29 50.2  &  36.52  &      74.89  &       8.56  &         &         &  1.436  &  $-$10.81$\pm$4.90  & 0 \\
2MA 0201+0117 A &  02 01 47.00  &  $+$01 17 05.1  &  20.30  &      74.77  &   $-$49.21  &  10.45  &  514.7  &  1.404  &      5.18$\pm$0.95  & 0 \\
2MA 0201+0117 B &  02 01 46.85  &  $+$01 17 15.3  &  20.31  &      75.86  &   $-$46.73  &         &         &  1.501  &      5.99$\pm$2.84  & 1 \\
NLTT 44989 A    &  17 33 05.98  &  $-$30 35 10.1  &  54.68  &  $-$113.37  &  $-$123.01  &  4.75   &  86.9   &  1.001  &     40.09$\pm$0.68  & 0 \\
NLTT 44989 B    &  17 33 05.62  &  $-$30 35 11.3  &  54.61  &  $-$121.23  &  $-$122.86  &         &         &  1.167  &     44.42$\pm$1.16  & 39 \\
\enddata
\tablecomments{All astrometric information is from \textit{Gaia} DR3. Physical separations are from 2d projections on the sky assuming an average of the two component distances. Separations for KX~Com refer to A-BC, not B-C. GJ~1183~B has no RV available in DR3. See \S\ref{subsec:sample-companion-checks} for a discussion of RUWE, RV$_{Gaia}$, and IPDfmp.}
\end{deluxetable}
%
\begin{deluxetable}{lcccccccc}[t!]
\tablewidth{0pt}
\tablecaption{Four Twin Systems Explored In This Paper --- Photometry and Mass Estimates\label{tab:SampleTable-phot}}
\tablehead{
\colhead{Name} & \colhead{\textit{G}} & \colhead{$BP$} & \colhead{$RP$} & \colhead{\textit{J}} & \colhead{\textit{H}} & \colhead{$K_s$} & \colhead{$M_G$} & \colhead{\textit{Mass}} \\
\colhead{} & \colhead{[mag]} & \colhead{[mag]} & \colhead{[mag]} & \colhead{[mag]} & \colhead{[mag]} & \colhead{[mag]} & \colhead{[mag]} & \colhead{[\(\textup{M}_\odot\)]}
}
\startdata
GJ 1183 A       &  12.46  &  14.27  &  11.17  &  9.31  &  8.70  &  8.40  &  11.24  &  0.21  \\
GJ 1183 B       &  12.50  &  14.33  &  11.22  &  9.35  &  8.76  &  8.46  &  11.28  &  0.21  \\
KX Com A        &  12.66  &  14.07  &  11.48  &  9.86  &  9.33  &  9.09  &  10.47  &  0.32  \\
KX Com BC       &  12.65  &  14.12  &  11.46  &  9.83  &  9.29  &  9.04  &  10.47  & (0.32) \\
2MA 0201+0117 A &  11.90  &  13.26  &  10.73  &  9.10  &  8.46  &  8.26  &   8.44  & (0.54) \\
2MA 0201+0117 B &  11.95  &  13.33  &  10.79  &  9.15  &  8.53  &  8.27  &   8.49  & (0.53) \\
NLTT 44989 A    &  12.44  &  13.91  &  11.25  &  9.61  &  9.06  &  8.80  &  11.13  &  0.25  \\
NLTT 44989 B    &  12.50  &  13.93  &  11.28  &  9.61  &  9.03  &  8.78  &  11.19  &  0.25  \\
\enddata
\tablecomments{Mass estimates are derived from the \cite{Benedict_2016} MLR for main sequence M dwarfs; values in parentheses are less reliable or unreliable estimates, as discussed in \S\ref{sec:Sample}. \textit{Gaia} $G$, $M_G$, $BP$, and $RP$ magnitudes reported here are from DR3, though note our stars were originally selected using DR2 information. $JHK_s$ magnitudes are from 2MASS \citep{2006AJ....131.1163S}.}
\end{deluxetable}

Astrometric and photometric parameters for the four systems targeted in this paper are given in Tables \ref{tab:SampleTable-astr} and \ref{tab:SampleTable-phot}, respectively. All data in Table \ref{tab:SampleTable-astr} are from \textit{Gaia} DR3, as well as the derived $M_G$ values, $G$, $BP$, and $RP$ in Table \ref{tab:SampleTable-phot}. $JHK_s$ values are from 2MASS. The key values have average errors as follows: $\pi$ $\pm$ 0.03 mas, $G$ $\pm$ 0.003 mag, $BP$ $\pm$ 0.007 mag, $RP$ $\pm$ 0.005 mag, $J$ $\pm$ 0.028 mag, $H$ $\pm$ 0.035 mag, and $K_s$ $\pm$ 0.028 mag. Tables are ordered alphabetically by star name, where `2MA' counts as M. The projected separations are greater than 80 AU for all four wide systems, implying orbital periods $\gtrsim$1000 years at their low masses, significantly longer than the $\sim$100d limit on tidal interaction and locking predicted by \cite{2019ApJ...881...88F} --- we thus conclude that present-day stellar tidal interactions between these wide pairs are negligible.

For context, estimated masses are given in Table \ref{tab:SampleTable-phot}, derived from the $V$-band mass-luminosity relation (MLR) for M dwarfs in \cite{Benedict_2016} via a prescription similar to that described in \cite{Vrijmoet_thesis_2023}. Briefly, several hundred M dwarfs on the RECONS long-term 0.9m program \citep{Henry_2018} with measured $M_V$ were used with the $V$-band MLR to estimate their masses. We then correlated these masses with the stars' \textit{Gaia} DR2 $M_{BP}$ values, and fit that relation with a high-order polynomial, which was then used to estimate masses for our twin stars via their \textit{Gaia} DR2 $M_{BP}$ values. Masses are shown in parentheses for the unresolved KX~Com~BC component, as well as 2MA~0201+0117~A and B, for the latter pair because they are PMS stars and therefore provide upper limit mass estimates at best. Regardless of the exact mass estimation method, our magnitude criteria are ultimately the fundamental observables that select our pairs to be twins in mass; we presume they host functionally identical ages, compositions, and environments as well under the assumption the binary components formed together and are co-eval.

We also employed the BANYAN $\Sigma$ tool of \cite{banyan_2018}, which uses a Bayesian analysis to probabilistically determine a target's candidate membership in nearby young stellar associations based on inputs optionally combining astrometry, radial velocities, or photometric distance constraints. We utilized \textit{Gaia} DR3 astrometry and ran the analysis both with and without our weighted mean CHIRON radial velocities (\S\ref{subsec:chiron-results}) given the binary nature of our targets. The results indicate GJ~1183~A has a 13\% chance of membership in the young Carina-Near association, but only for the A component and only when excluding radial velocities --- this low probability, combined with the low number of 13 stars used to define the group in BANYAN $\Sigma$, leads us to disregard the possible membership. The results do correctly support 2MA~0201+0117~AB belonging to the $\beta$ Pictoris association, and otherwise find no membership probabilities $>$2\% for the other stars considered here.

\subsection{Higher Order Multiplicity Checks} \label{subsec:sample-companion-checks}

Our intended comparisons between binary components require the stars be true twins, so it is crucial to search for any higher order companions --- especially unresolved ones --- that could disrupt the components' twin natures. Three \textit{Gaia} parameters were assessed to check for unresolved companions and are included in Table \ref{tab:SampleTable-astr}. First is the Renormalised Unit Weight Error (RUWE), where an elevated value may indicate an unresolved component. Ongoing RECONS work by LeBlanc et al.~(in prep) on the nearest $\sim$3000 M dwarf systems finds that RUWE $>$ 1.7 indicates an unseen companion, in line with the results of \cite{Vrijmoet_2020} which compared RECONS data to \textit{Gaia} DR2\footnote{While conventional criteria often use RUWE $>$ 1.4, we adopt a more conservative RUWE limit of 1.7 because our research finds this to be a more appropriate cutoff for true M dwarf binaries in the solar neighborhood \citep[][LeBlanc et al.~in prep]{Vrijmoet_2020}.}. All eight components considered here have RUWE $<$ 1.7, implying no companions that affect the astrometry over the 34-month timescale of the DR3 data. The second parameter is the error on RV$_{Gaia}$, where all are generally less than $\sim$3~km/s, appropriate for single stars with these magnitudes. Note, however, that KX~Com~B has an RV$_{Gaia}$ error value of nearly 5~km/s, implying a companion, which is in fact the case (see \S\ref{subsubsec:rv-results} and \S\ref{subsec:KXCOMAB}).

The third \textit{Gaia} parameter, \texttt{ipd\_frac\_multi\_peak} (IPDfmp), reports the fraction of \textit{Gaia} windows of the source for which a double peak is identified, possibly indicating an unresolved companion or contaminating source. For context, \cite{2023AJ....165..180T} demonstrated that source pairs closer than $\sim$2\farcs5 can generally display elevated IPDfmp values just due to proximity and not unseen bound companions. The components considered here all have IPDfmp$\leq$1\%, consistent with no unresolved sources, except NLTT~44989~B at 39\% presumably because of a very nearby \textit{Gaia} source 0\farcs84 away at the DR3 2016.0 epoch (0\farcs22 away at Ep=2021.0\footnote{We often use the notation ``Ep=year" throughout this paper to designation the Julian epoch of coordinates used to derive an angular separation value at a certain point in time.}). Our careful examination of archival DSS \citep{1996ASPC..101...88L, 2004AJ....128.3082G} and VPHAS \citep{2014MNRAS.440.2036D} images and \textit{Gaia} astrometry clearly shows this very nearby source --- along with a second nearby source 3\farcs23 away from B (3\farcs02 away at Ep=2021.0) --- are both physically unassociated fainter background stars that NLTT~44989~AB has approached over time via proper motion. These two contaminating sources are discussed further in \S\ref{subsec:contam}.

Our stars were searched for inclusion in the VizieR collection of \textit{Gaia} DR3 non-single stars catalogs \citep{2022yCat.1357....0G}, which report various assessments indicating likely unresolved multiples, but no matches were found. We also searched \textit{Gaia} DR3 for any potential additional wide companions within a 2d projected separation of 10,000~AU around each of our eight components, finding no sources in this radius with parallaxes within 10~mas of each associated twin star's parallax. A crossmatch found none of our components are present in the SB9 spectroscopic binary catalog as well \citep{2004A&A...424..727P}. Finally, the four systems were matched against the Washington Double Star Catalog \citep[WDS;][]{2001AJ....122.3466M}, where the only result of note was the entry of a supposed additional `C' component for NLTT~44989~AB. A careful investigation reveals this extra `C' source to be the aforementioned unassociated background star 3\farcs23 away from B, so it is not a real companion.

\section{Observations \& Data Processing} \label{sec:ObsDatRed}

Our twin targets have been observed with five observing campaigns: (1) long-term optical photometry with the CTIO/SMARTS 0.9m spanning several years to probe for stellar activity cycles, (2) short-term optical photometry with the CTIO/SMARTS 0.9m to capture rotation --- our rotation period determinations are also supported by archival data from \textit{TESS} \citep{TESS}, ZTF \citep{ZTF}, and ASAS-SN \citep{Shappee_2014, Jayasinghe_2019}, (3) multi-epoch high-resolution optical spectroscopy using the CHIRON echelle spectrograph on the CTIO/SMARTS 1.5m to determine radial velocities and H$\alpha$ equivalent widths, (4) \textit{Chandra} X-ray imaging observations to determine X-ray luminosities and coronal parameters, and (5) speckle imaging with HRCam on the SOAR 4.1m and QWSSI on the LDT 4.3m to search for hidden companions. We outline the methodology for each of these five observing campaigns in the following subsections: long-term photometry in \S\ref{subsec:longterm-methods}, rotation in \S\ref{subsec:rot-methods}, optical spectroscopy in \S\ref{subsec:chiron-methods}, X-rays in \S\ref{subsec:chandra-methods}, speckle in \S\ref{subsec:speckle-methods}, and a subsequent cumulative discussion about contamination in \S\ref{subsec:contam}.

\subsection{Stellar Cycles - CTIO/SMARTS 0.9m Long-term Campaign} \label{subsec:longterm-methods}

The CTIO/SMARTS 0.9m has been used to observe nearby M dwarfs as part of an ongoing RECONS long-term monitoring program since 1999 (see \cite{Henry_2018} for a recent summary). Past work has used the multi-decade photometry to investigate stellar variability \citep{Jao_2011, Hosey_2015, Clements_2017, Kar_2024}, with an ongoing project to reveal stellar activity cycles (Couperus et al.~in prep). GJ~1183~AB had fortuitously already been on the program since 2013, 2MA~0201+0117~AB and NLTT~44989~AB were added to the long-term program in 2019, and KX~Com~A-BC was added in 2021. KX~Com~A-BC is the only case we do not report long-term variability results for here, as the system still has insufficient coverage to be informative for long-term cycles.

Details of the differential photometry reduction and analysis procedures for the 0.9m long-term program are described in \cite{Jao_2011} and \cite{Hosey_2015}. To summarize, each target typically receives two visits per year with 5 frames taken per visit using the same optical $V$, $R$, or $I$ filter and positioned consistently in the 6\farcm8 square field to provide a set of 5--15 reference stars to be used as differential photometry calibrators. Measurements are made with SExtractor \citep{1996A&AS..117..393B}, specifically using the \texttt{MAG\_WIN} parameter; this routine obtains instrumental magnitudes by summing the source pixel counts falling within a circular Gaussian window function that is scaled to the light distribution of each source. Following the methodology of \cite{1992PASP..104..435H}, the instrumental magnitudes of all reference stars in all frames are simultaneously minimized from their individual mean brightnesses to yield corrective offsets for each frame due to changes in atmospheric transmission, instrumental efficiency, and exposure time. Any photometrically variable reference stars are identified by eye and removed to ensure that only constant calibrator stars are used. The offsets are then applied to the target science star magnitudes, giving the final relative light curve.

Results from the long-term program are discussed later in \S\ref{subsec:longterm-results}.

\subsection{Rotation - CTIO/SMARTS 0.9m, \textit{TESS}, ZTF, \& ASAS-SN} \label{subsec:rot-methods}

Archival data alone did not provide reliable periods identifiable for each star in a system in most cases, either because of blending or due to no rotation signal being evident in one component or the other. This situation motivated collecting our own observations with the 0.9m, as that telescope system can outperform the archival sources in vital ways. For example, the noise floor for the 0.9m is typically $\sim$7 mmag \citep{Jao_2011, Hosey_2015}, with the key advantage of high resolution with a 401 mas/pixel plate scale, which is the best of the four data sources utilized here. This is markedly better than for ZTF with $\sim$10--20 mmag precision at $r$=14--17 and 1\farcs01 pixels \citep{ZTF_data}, or ASAS-SN with $\sim$15--25 mmag at $V$=13--14 with 8\farcs0 pixels \citep{Jayasinghe_2019}. Whereas \textit{TESS} provides exquisite precision for the photometry, its 21\arcsec~pixels mean that all four of our systems are blended in \textit{TESS} measurements, thereby still requiring 0.9m measurements to assign rotations periods to individual components. For 2MA~0201+0117~AB, ZTF data were able to determine reliable periods for each star independently, but we still observed this system with the 0.9m to validate our rotation methodology.

The 0.9m observations targeting our twins' stellar rotation periods were carried out using NOIRLab time (ID 2023A-549259; PI Couperus). Observing cadences were tailored to each system based on likely or possible periods indicated by their H$\alpha$ activity and the archival data from \textit{TESS}, ZTF, or ASAS-SN. At the 0.9m, we made $\sim$50--70 visits to each target during two separate 20-night observing runs, with a few additional visits during adjacent long-term program (\S\ref{subsec:longterm-methods}) runs to extend baselines and coverage for GJ~1183~AB and NLTT~44989~AB. At each visit we routinely acquired four images with both components falling in a single detector field of view. Observations for all systems were made in the $V$ filter to provide enhanced spot contrast \citep{Hosey_2015} and to balance the brightnesses of the targets and reference stars. The light curves from the 0.9m rotation effort were derived following the same procedures as the long-term RECONS program (\S\ref{subsec:longterm-methods}).

For \textit{TESS}, we extracted Pre-search Data Conditioning Simple Aperture Photometry (PDCSAP) light curves using 20-second (20s) and 2-minute (2m) high-cadence data as well as 10-minute (10m) and 30-minute (30m) Full Frame Image (FFI) data, provided by the \textit{TESS}-SPOC pipeline \citep{2016SPIE.9913E..3EJ, 2020RNAAS...4..201C}. We used all available data products from all available \textit{TESS} sectors for each of our targets as follows: GJ~1183~AB has 2m and 10m data from sector 51, KX~Com~A-BC has a mix of 20s/2m/10m/30m data from sectors 23/49, 2MA~0201+0117~AB has a mix of 20s/2m/10m/30m data from sectors 4/42/43, and NLTT~44989~AB has a mix of 2m/10m/30m data from sectors 12/39. 

In addition to the standard \textit{TESS} results, we also generated FFI light curves with the \texttt{unpopular} package of \cite{unpopular}. This approach uses an alternative causal pixel model method that corrects for systematics by modeling trends common across many different sources in the field. A key facet of \texttt{unpopular} is the optional inclusion of a polynomial component that is simultaneously fit during this detrending process to better capture and preserve long-term astrophysical variations --- such as rotation signals with periods beyond half a \textit{TESS} sector baseline ($\sim$13.5d) --- thereby allowing us to search for longer duration signals in \textit{TESS}. We followed the approach outlined in \cite{Kar_2024}, where the polynomial component and any resulting long-term signal are only used and deemed reliable if the raw Simple Aperture Photometry (SAP) \textit{TESS} flux also shows a long-term signal. A long-term signal appearing in multiple sectors further validates a detection. Apertures used with \texttt{unpopular} were manually selected rectangles chosen to closely match the default \textit{TESS} pipeline apertures while minimizing blending and contamination where possible. The same sectors of data were used with \texttt{unpopular} for each target as noted above for the normal \textit{TESS} products.

Beyond the 0.9m and \textit{TESS}, two other sources of rotation data were utilized.  ZTF PSF-fit light curves were obtained from Data Release 18 via the IPAC/Caltech system \citep{ZTF_data}, using $zr$ and $zg$ filter data separately, with measurements from different ZTF fields and CCDs but for the same star all combined into a single light curve in each filter. ASAS-SN pre-computed light curves in the $V$-band were extracted via the photometry page\footnote{Available at \href{https://asas-sn.osu.edu/photometry}{https://asas-sn.osu.edu/photometry}.} \citep{Jayasinghe_2019}, or new $V$-band and $g$-band aperture photometry curves were generated with all cameras merged using the ASAS-SN Sky Patrol resource\footnote{Available at \href{https://asas-sn.osu.edu/}{https://asas-sn.osu.edu/}.} if the aforementioned pre-computed data were unavailable \citep{Kochanek_SkyPatrol}. No light curves or rotation data were available from \textit{Gaia} DR3 \citep{GaiaDR3_Var}, \textit{Kepler} \citep{Kepler}, K2 \citep{K2}, or MEarth \citep{Berta_2012} for the four systems discussed here, and a literature review found no other rotation results not already surpassed in quality by the 0.9m or archival sources.

To search for and measure rotation periods, we first addressed flares and outliers as follows. Any obviously strong flares in the 0.9m data were manually excluded. We removed poor quality measurements from the archival data (\textit{TESS}-SPOC, \textit{TESS}-unpopular, ZTF, and ASAS-SN) using provided quality flags if available, along with removal of any outlier points greater than 3$\sigma$ from the mean. Note that ASAS-SN Sky Patrol curves had stricter cuts for points at $>$2$\sigma$ and $>$50~mmag error owing to less curated starting data. For archival datasets, points indicating lingering mild flares not removed by our outlier cut were left untouched as we found the separate archival sources gave extremely congruent period measurements regardless.

Each light curve was inspected visually and analyzed with the Generalized Lomb-Scargle Periodogram \citep{Lomb_1976, Scargle_1982, Generalised_LS_2009} using the Astropy implementation \citep{2013A&A...558A..33A,Astropy2018}. Periodograms used 100,000 samples evenly spaced in frequency between 0.05--300 days. False Alarm Probabilities (FAPs) were computed using the approximate upper-limit Baluev method \citep{Baluev_2008}. The maximum power peak and its corresponding FAP value, 2/3/4 harmonic multiples, n = $-$3 to $+$3 one-day aliases (for ground-based observatories), and relative FAP lines were all considered in the determination of periods for each case. FAP values were de-emphasized in these reviews, and our manual assessment instead relied more on criteria such as the photometric amplitude of a signal relative to the noise level of the data itself, the visual robustness of candidate periods in raw and phase-folded light curves, a signal repeating over time or not, the periodogram power of a peak relative to the power of noise peaks, and the re-occurrence of trends in multiple independent data sources. That said, very small FAP values routinely accompanied our final choices. We point the interested reader to \cite{Understanding_LS} for a discussion of the subtleties involved with interpreting FAP values. 

The final rotation period chosen for each star was determined through a comprehensive review of all available light curves from the various sources outlined above, in conjunction with knowledge about the different amounts of blending, contamination, and photometric precision between the data sources. For example, our resolved 0.9m data might confidently suggest period X in star A but only gives a weak uncertain detection of period Y in star B, while blended \textit{TESS} data shows a combined signal of two robust periods also near X and Y, allowing us to confidently assess that the Y period is legitimate and belongs to B. Either data set alone may be inconclusive, but combined they confirm a period exists and to which star it belongs. The mix of data sources also allows us to better vet 1d sampling alias peaks in ground-based data by comparing to space-based data without such aliases. Generally, our resolved 0.9m results were able to either outright confirm or give indications towards a specific period in each star, with external blended data providing confirmation that such periods exist within the system as a whole to validate a weaker 0.9m detection. Note that we did not attempt to combine separate data sources into merged light curves or a global simultaneous analysis given the significantly different precision, systematics, cadences, filters, reduction procedures, timescales, blending, and contamination present across the many archival sources used individually here.

Results from the rotation analysis are discussed later in \S\ref{subsec:rot-results}.

\subsection{H$\alpha$ Equivalent Widths \& Radial Velocities - CTIO/SMARTS 1.5m \& CHIRON} \label{subsec:chiron-methods}

Optical spectra were obtained at the CTIO/SMARTS 1.5m with the CHIRON echelle spectrograph \citep{Tokovinin_2013, Paredes_2021}. Each of the four systems were observed at least 5 times spread over several months to determine H$\alpha$ equivalent widths (EWs) and radial velocities (RVs), with an additional sequence of 5 visits 5 nights in a row to search for close, potentially interacting unresolved companions via changes in RVs. For KX~Com~A and B only 3 of the 5 nightly sequence visits were secured due to poor weather. After preliminary RV analyses found a likely unresolved companion to KX~Com~B (the C component), we obtained another 23 single-spectrum visits on just B over one month to confirm or refute the companion. NLTT~44989~AB also received several additional visits to extend the time baseline beyond one year in order to further rule out any companion with an orbital period up to a few years.

Spectra were taken in fiber mode with 4$\times$4 binning, yielding R$\approx$27,000. Components were well resolved given the 2\farcs7 diameter fiber (see \S\ref{subsec:contam} for additional contamination details). A typical visit consisted of four total exposures, two on each binary component, along with ThAr wavelength calibration images at each pointing. For each system, spectra on each component were secured back-to-back, not at disjointed times from each other, so that our A-B comparisons are robust at consistent snapshots in time. Exposure times of 900--1800~sec were used and the two spectra at a given epoch were combined to yield a typical continuum signal-to-noise ratio (SNR\footnote{Our reported SNRs are the mean SNRs per pixel across both continuum regions using the per-pixel method in Eq.~(1) of \cite{Tokovinin_2013}.}) of $\sim$27.

Reduced data were received from the CHIRON pipeline as described in \cite{Paredes_2021}. The process includes routine bias and flat corrections, order extraction, and wavelength calibration using time-adjacent ThAr frames. We manually reviewed each spectrum to remove any cases with critical observing failures or strong cosmic rays on or near H$\alpha$. Spectra were then further processed using the procedures and code of J23, who used the same CHIRON configuration as our work and also targeted H$\alpha$ and RVs in M dwarfs. Briefly, the two back-to-back spectra in a single star's visit were barycenter corrected following \cite{2014PASP..126..838W}, combined to a mean spectrum to boost the SNR, blaze function normalized, and trimmed for cosmic rays along the way. RV and $v\sin(i)$ values were obtained via cross-correlation with the mid-M standard stars Barnard's star and GJ 273 in 6 echelle orders following the methodology of \cite{Irwin_2018} and \cite{Nisak_2022} (see J23). Each standard star gives 6 measures from the 6 respective orders, yielding a mean and standard deviation of the RV, as well as $v\sin(i)$. A weighted average of the two standard star results then gives our final RV and $v\sin(i)$ for that target star at that epoch.

Our H$\alpha$ EWs also follow the process outlined in J23. Spectra were shifted to rest-frame using the stellar RVs and manually reviewed to define three wavelength windows, one centered on the H$\alpha$ feature and two on either side to capture the mean continuum level. Our default regions for absorption cases are as given in Figure 2 and Table 2 of J23. In emission cases, the H$\alpha$ region was adjusted in width to capture line wings based on visual inspection using the template of J23. If an H$\alpha$ wing came close to or overlapped the default continuum regions, we shifted both continuum regions outward slightly to consistent `wide' positions designed to yield nearly the same mean continuum level as the default positions in order to avoid systematic offsets. Some poor SNR cases also used slightly wider continuum regions to better estimate the mean continuum levels. To avoid biasing our resulting measures and comparisons, specific care was taken to be as consistent as possible when defining all regions for two components in the same twin binary. 

Our EWs are measured following Eq.~(1) of J23, and we adopt the convention of negative EWs indicating H$\alpha$ emission. The EW uncertainties follow the procedure in \cite{Cayrel_1988} and use a Gaussian fit to H$\alpha$ to estimate the needed FWHM. We note that a handful of spectra with weak (often double-peaked) emission near the continuum were fit with overly wide Gaussians, falsely inflating the EW uncertainties from the typical $\sim$0.02~$\mathrm{\AA}$ to $\sim$0.10~$\mathrm{\AA}$ --- this has no meaningful impact on our results.

Results from the CHIRON spectral analysis are discussed later in \S\ref{subsec:chiron-results}.

\subsection{X-rays - \textit{Chandra} Observatory} \label{subsec:chandra-methods}

To evaluate the coronal behavior of the target stars, we obtained observations with the \textit{Chandra} X-ray Observatory for our four systems from 2020--2022 through the GO proposal \textit{Fraternal or Identical? The Magnetic Properties of M Dwarf Twins} (ID 22200260; PI Osten). The spatially-resolved ACIS-S imaging study resulted in eight exposures across the four systems --- three each for GJ~1183~AB and KX~Com~A-BC and one each for 2MA~0201+0117~AB and NLTT~44989~AB. The cases with multiple exposures are noted as TARGET--1, TARGET--2, etc., and all are outlined in Table \ref{tab:ChandraTable}. These observations allow us to produce X-ray light curves and non-grating spectra for subsequent analysis. Data were analyzed using the Chandra Interactive Analysis of Observations (CIAO) software package v4.14.3 and CALDB v4.9.8 \citep{Ciao2006}, with \texttt{chandra\_repro} used to apply the corresponding calibrations. Spectral fitting was carried out using the Sherpa package within CIAO \citep{Sherpa2001}.

Non-overlapping circular apertures were manually constructed for the A and B components in each observation, placed at corresponding source locations determined by the CIAO \texttt{wavdetect} algorithm. Source aperture radii were chosen based on the manual inspection of radial plots for each source, with selected radii of 3--6 pixels depending on the extent of each source's photon signals. Background apertures were $\sim$50-pixel radius circles encompassing the components, with enlarged regions excluded around each source to ensure the removal of all source photons. In a single case, NLTT~44989~A, we did not obtain a confident detection at the expected source location, the handling of which is detailed further in \S\ref{subsubsec:NLTT-A}.

\begin{figure}[t]
\centering
\includegraphics[scale=0.49]{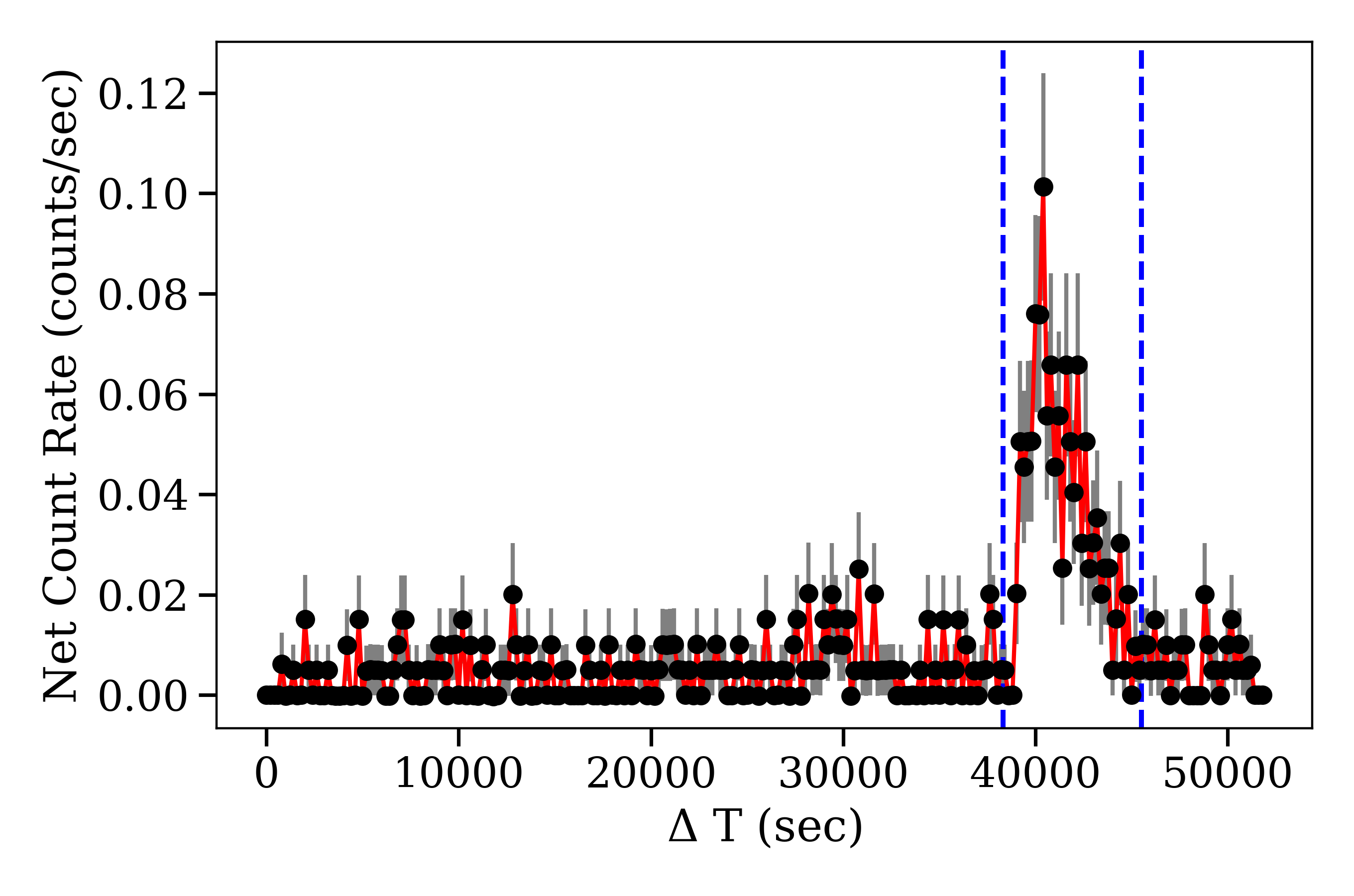}
\figcaption{A background-subtracted \textit{Chandra} X-ray light curve for NLTT~44989~B, showcasing a strong stellar flare during the exposure. Vertical dashed blue lines indicate the time period used for isolating the stellar flare. Counts are merged into 200-second bins. Light curves for all of the other X-ray datasets can be seen in Figure \ref{fig:all-xray-lc} in Appendix \ref{sec:appendix}. \label{fig:NLTT-B-LC}}
\end{figure}

For each source, X-ray light curves filtered to 0.3--10~keV, the nominal energy-calibrated range of ACIS, were inspected for noise background flares but none were found. We captured time-resolved stellar flares during four of the eight total exposures. An example light curve for the strongest flare, in this case for NLTT~44989~B, is shown in Figure \ref{fig:NLTT-B-LC} --- all other X-ray light curves can be seen in Figure \ref{fig:all-xray-lc} in Appendix \ref{sec:appendix}. In these flare cases, data were visually split into flaring and non-flaring time periods and analyzed separately; we report both the quiescent and flaring measurements separately in Table \ref{tab:ChandraTable}.
%
\begin{longrotatetable}
\movetabledown=17mm
\begin{deluxetable}{lccccc|rrrrrrrr}
\tablewidth{0pt}
\tablecaption{\textit{Chandra} X-ray Observations \& Results\label{tab:ChandraTable}}
\tablehead{
\colhead{Dataset} & \colhead{ObsID} & \colhead{ObsID-Start} & \colhead{Exp.} & \colhead{Counts} & \colhead{Pileup} & \colhead{$F_X$} & \colhead{$L_X$} & \colhead{$T_1$} & \colhead{$VEM_1$} & \colhead{$T_2$} & \colhead{$VEM_2$} & \colhead{Abund.} & \colhead{red.-$\chi^2$} \\
\colhead{} & \colhead{} & \colhead{} & \colhead{} & \colhead{} & \colhead{} & \colhead{$\times$10$^{-14}$} & \colhead{$\times$10$^{27}$} & \colhead{} & \colhead{$\times$10$^{50}$} & \colhead{} & \colhead{$\times$10$^{50}$} & \colhead{} & \colhead{} \\
\colhead{} & \colhead{} & \colhead{[yyyy-mm-dd.d]} & \colhead{[ks]} & \colhead{[cts]} & \colhead{[\%]} & \colhead{[erg s$^{-1}$cm$^{-2}$]} & \colhead{[erg/s]} & \colhead{[MK]} & \colhead{[cm$^{-3}$]} & \colhead{[MK]} & \colhead{[cm$^{-3}$]} & \colhead{[rel. solar]} & \colhead{}
}
\startdata
GJ 1183 A--1      & \dataset[24504]{https://doi.org/10.25574/24504} & 2020-12-15.4 & 14.7 & 475  & 2.2            & 48.8$_{-6.1}^{+3.7}$    & 18.0$_{-2.2}^{+1.4}$    & 10.91$_{-0.65}^{+0.65}$ & 20.1$_{-3.5}^{+4.2}$     & \nodata           & \nodata         & 0.159$_{-0.046}^{+0.057}$ & 1.00 \\
GJ 1183 A--2      & \dataset[24899]{https://doi.org/10.25574/24899} & 2020-12-15.8 & 13.3 & 345  & 1.7            & 36.9$_{-4.5}^{+3.2}$    & 13.6$_{-1.7}^{+1.2}$    & 10.49$_{-0.64}^{+0.66}$ & 13.8$_{-3.3}^{+3.6}$     & \nodata           & \nodata         & 0.198$_{-0.064}^{+0.101}$ & 1.62 \\
GJ 1183 A--3      & \dataset[23392]{https://doi.org/10.25574/23392} & 2022-04-12.4 & 28.3 & 994  & 2.4            & 59.8$_{-2.5}^{+2.3}$    & 22.0$_{-0.9}^{+0.8}$    & 11.41$_{-0.46}^{+0.37}$ & 23.3$_{-2.5}^{+2.9}$     & \nodata           & \nodata         & 0.176$_{-0.034}^{+0.039}$ & 1.25 \\
GJ 1183 B--1      & \dataset[24504]{https://doi.org/10.25574/24504} & 2020-12-15.4 & 14.7 & 340  & 1.6            & 39.6$_{-3.7}^{+3.2}$    & 14.6$_{-1.4}^{+1.2}$    & 9.21$_{-0.53}^{+0.53}$  & 14.1$_{-3.1}^{+3.4}$     & \nodata           & \nodata         & 0.216$_{-0.061}^{+0.089}$ & 0.87 \\
GJ 1183 B--2      & \dataset[24899]{https://doi.org/10.25574/24899} & 2020-12-15.8 & 14.9 & 196  & \nodata        & 23.0$_{-1.9}^{+2.0}$    & 8.5$_{-0.7}^{+0.7}$     & 8.66$_{-0.59}^{+0.58}$  & 7.9$_{-0.8}^{+0.8}$      & \nodata           & \nodata         & [0.216]                   & 1.17 \\
GJ 1183 B--3      & \dataset[23392]{https://doi.org/10.25574/23392} & 2022-04-12.4 & 28.3 & 777  & 1.9            & 47.3$_{-2.1}^{+2.0}$    & 17.4$_{-0.8}^{+0.7}$    & 11.41$_{-0.82}^{+0.55}$ & 22.2$_{-2.4}^{+3.9}$     & \nodata           & \nodata         & 0.105$_{-0.029}^{+0.028}$ & 1.10 \\
KX Com A--1       & \dataset[23393]{https://doi.org/10.25574/23393} & 2021-03-15.4 & 16.4 & 323  & 1.4            & 30.8$_{-2.8}^{+2.3}$    & 27.5$_{-2.5}^{+2.1}$    & 11.01$_{-0.70}^{+0.75}$ & 32.5$_{-6.1}^{+6.5}$     & \nodata           & \nodata         & 0.135$_{-0.042}^{+0.060}$ & 0.86 \\
KX Com A--2       & \dataset[24991]{https://doi.org/10.25574/24991} & 2021-03-15.8 &  9.3 & 140  & \nodata        & 24.6$_{-2.7}^{+2.7}$    & 22.0$_{-2.4}^{+2.4}$    & 9.64$_{-1.04}^{+0.88}$  & 25.8$_{-2.9}^{+3.5}$     & \nodata           & \nodata         & [0.135]                   & 1.12 \\
KX Com A--3       & \dataset[24503]{https://doi.org/10.25574/24503} & 2022-03-22.6 & 30.2 & 583  & 1.3            & 35.1$_{-2.8}^{+2.1}$    & 31.3$_{-2.5}^{+1.9}$    & 10.32$_{-0.55}^{+0.56}$ & 38.7$_{-5.5}^{+6.3}$     & \nodata           & \nodata         & 0.120$_{-0.030}^{+0.036}$ & 0.90 \\
KX Com BC--1      & \dataset[23393]{https://doi.org/10.25574/23393} & 2021-03-15.4 & 14.4 & 39   & \nodata        & 7.7$_{-3.6}^{+3.9}$     & 6.9$_{-3.2}^{+3.5}$     & 4.86$_{-1.04}^{+2.17}$  & 11.2$_{-5.1}^{+6.5}$     & \nodata           & \nodata         & [0.150]                   & 1.49 \\
KX Com BC--2      & \dataset[24991]{https://doi.org/10.25574/24991} & 2021-03-15.8 & 16.9 & 67   & \nodata        & 6.1$_{-1.0}^{+0.9}$     & 5.5$_{-0.9}^{+0.8}$     & 9.92$_{-1.09}^{+0.93}$  & 6.1$_{-0.9}^{+1.0}$      & \nodata           & \nodata         & [0.150]                   & 2.39 \\
KX Com BC--3      & \dataset[24503]{https://doi.org/10.25574/24503} & 2022-03-22.6 & 30.2 & 79   & \nodata        & 4.4$_{-0.5}^{+0.5}$     & 3.9$_{-0.4}^{+0.4}$     & 12.02$_{-1.33}^{+1.38}$ & 4.4$_{-0.5}^{+0.5}$      & \nodata           & \nodata         & [0.150]                   & 0.62 \\
2MA 0201+0117 A   & \dataset[23394]{https://doi.org/10.25574/23394} & 2020-10-10.2 & 13.9 & 605  & \nodata        & 62.5$_{-7.8}^{+4.2}$    & 181.5$_{-22.7}^{+12.2}$ & 10.87$_{-0.68}^{+0.59}$ & 207.6$_{-31.1}^{+41.8}$  & \nodata           & \nodata         & 0.149$_{-0.042}^{+0.046}$ & 1.04 \\
2MA 0201+0117 B   & \dataset[23394]{https://doi.org/10.25574/23394} & 2020-10-10.2 & 13.9 & 2071 & 1.7            & 225.8$_{-15.3}^{+11.1}$ & 654.8$_{-44.4}^{+32.3}$ & 9.75$_{-0.59}^{+0.71}$  & 370.3$_{-93.7}^{+228.8}$ & 20.9$_{-1.7}^{+5.5}$ & 331$_{-116}^{+52}$ & 0.170$_{-0.065}^{+0.062}$ & 1.05 \\
NLTT 44989 A     & \dataset[23395]{https://doi.org/10.25574/23395} & 2022-07-28.6 & 49.8 & $\le$~4.3  & \nodata  & $\le$~0.2                & $\le$~0.1                & [10.00]                 & $\le$~0.1                 & \nodata                   & \nodata                 & [0.200]                   & \nodata   \\
NLTT 44989 B     & \dataset[23395]{https://doi.org/10.25574/23395} & 2022-07-28.6 & 42.8 & 223  & \nodata        & 9.7$_{-0.9}^{+0.9}$     & 3.9$_{-0.4}^{+0.4}$     & 9.09$_{-0.67}^{+0.64}$  & 4.4$_{-0.4}^{+0.5}$      & \nodata           & \nodata         & [0.150]                   & 1.32 \\
\hline
GJ 1183 A--2 Flare    & \dataset[24899]{https://doi.org/10.25574/24899} & 2020-12-15.8 &  1.6 & 96   & 4.7  & 111.8$_{-14.9}^{+15.0}$ & 41.2$_{-5.5}^{+5.5}$    & 10.10$_{-1.14}^{+1.11}$ & 46.2$_{-6.5}^{+7.5}$     & \nodata           & \nodata         & [0.150]                   & 0.57 \\
KX Com A--2 Flare     & \dataset[24991]{https://doi.org/10.25574/24991} & 2021-03-15.8 &  7.6 & 327  & 2.8  & 69.0$_{-10.0}^{+6.4}$   & 61.6$_{-8.9}^{+5.7}$    & 6.56$_{-2.95}^{+1.67}$  & 35.4$_{-11.2}^{+24.5}$   & 18.3$_{-2.3}^{+5.3}$ & 39$_{-10}^{+11}$   & [0.150]                   & 1.65 \\
KX Com BC--1 Flare    & \dataset[23393]{https://doi.org/10.25574/23393} & 2021-03-15.4 &  2.0 & 43   & 1.5  & 34.0$_{-5.6}^{+5.8}$    & 30.5$_{-5.0}^{+5.2}$    & 12.97$_{-1.88}^{+2.22}$ & 34.6$_{-5.7}^{+5.8}$     & \nodata           & \nodata         & [0.150]                   & 0.71 \\
NLTT 44989 B Flare   & \dataset[23395]{https://doi.org/10.25574/23395} & 2022-07-28.6 &  7.1 & 252  & 2.3  & 52.3$_{-5.1}^{+4.2}$    & 21.0$_{-2.0}^{+1.7}$    & 13.94$_{-1.22}^{+1.74}$ & 27.4$_{-5.2}^{+5.3}$     & \nodata           & \nodata         & 0.092$_{-0.044}^{+0.074}$ & 1.60 \\
\enddata
\tablecomments{Exposure times are what is left after separating the stellar flare segments, where relevant. Counts are for 0.3--10~keV after background subtraction. Cases with a reported pileup fraction used a pileup component in their spectral modeling while those without a reported value did not. Values in square brackets were fixed in the coronal models. The uncertainties are all 1-$\sigma$ (68\% confidence interval) values (see \S\ref{subsec:chandra-methods} and \S\ref{subsec:chandra-results} for details). NLTT~44989~A was a weak- or non-detection, with values provided here indicating estimated upper limits only (see \S\ref{subsubsec:NLTT-A} for details).}
\end{deluxetable}
\end{longrotatetable}

Non-grating Pulse Height Amplitude (PHA) spectra were extracted for each detected source, background subtracted, and filtered to 0.3--10~keV. We grouped data to 9 counts per bin in all cases for consistency --- in spectra with total counts $\lesssim$200 we tested 6 counts per bin as well, but it did not substantively change our resulting measurements. The X-ray coronal spectra were fit with Astrophysical Plasma Emission Code (APEC) models within Sherpa that parameterize the plasma temperature (kT in keV units), joint coronal abundances relative to solar of 13 atomic species other than hydrogen, redshift (fixed to 0 for our nearby stars), and a normalization parameter tied to the emission measure.  We used a forward-folding technique typical for this application that takes into account the instrumental response function and utilized the \texttt{chi2xspecvar} reduced-$\chi^2$ statistic. Fits were carefully tested in every case with the Sherpa \texttt{levmar} optimizer and \texttt{moncar} MCMC optimizer, along with several different initial parameter values, to validate consistent convergence to the final selected solutions.

A model component to account for interstellar medium absorption was tested using hydrogen column density estimates from the local interstellar cloud model of \cite{Redfield-and-Linsky-2000}\footnote{We used the column density web calculator available at \href{http://lism.wesleyan.edu/ColoradoLIC.html}{http://lism.wesleyan.edu/ColoradoLIC.html}.}, which gave relatively small values of 10$^{16}$--10$^{18}$~cm$^{-2}$ because the model only extends out to a few parsecs. Using these low densities, including the absorption model component had functionally no impact on the resulting coronal parameters --- the minuscule changes were significantly smaller than the underlying parameter uncertainties. Given our stars can be found out to nearly 50~pc, we also tested with column densities of $\sim$10$^{20}$~cm$^{-2}$ for the highest and lowest SNR datasets; this indicated the resulting coronal parameters and fluxes would deviate by no more than $\sim$0.1--0.5~$\sigma$, insufficient to change the interpretation of our results. Based on this and the low densities from \cite{Redfield-and-Linsky-2000} we removed the absorption component for simplicity.

Another concern was pileup, the coincident arrival of two or more X-ray photons in the same pixel region within a single frame time. We used the Sherpa \texttt{jdpileup} implementation of the \cite{Davis2001-pileup} pileup model and followed suggestions in the \textit{Chandra ABC Guide to Pileup} documentation\footnote{The \textit{Chandra} pileup guide is found at \href{https://cxc.harvard.edu/ciao/download/doc/pileup\_abc.pdf}{https://cxc.harvard.edu/ciao/download/doc/pileup\_abc.pdf}.}. Test fits indicated the pileup parameters \texttt{alpha} and \texttt{psfrac} were poorly constrained even in our best cases, so we adopted a fixed pileup model with parameters set to typical values advised by the documentation. When the fit-measured pileup fractions were $\lesssim$1\% the overall impact was negligible and changed the resulting coronal parameters by markedly less than the underlying uncertainties --- the pileup component was removed in these cases for simplicity. When $\gtrsim$1\%, values sometimes changed by more than the uncertainties, especially in cases with a second hotter coronal component --- the pileup model was kept in these $\gtrsim$1\% cases and typically slightly improved the reduced-$\chi^2$. Table \ref{tab:ChandraTable} reports a pileup fraction if it was included in the spectral model for a dataset.

\begin{figure}[t]
\centering
\gridline{\fig{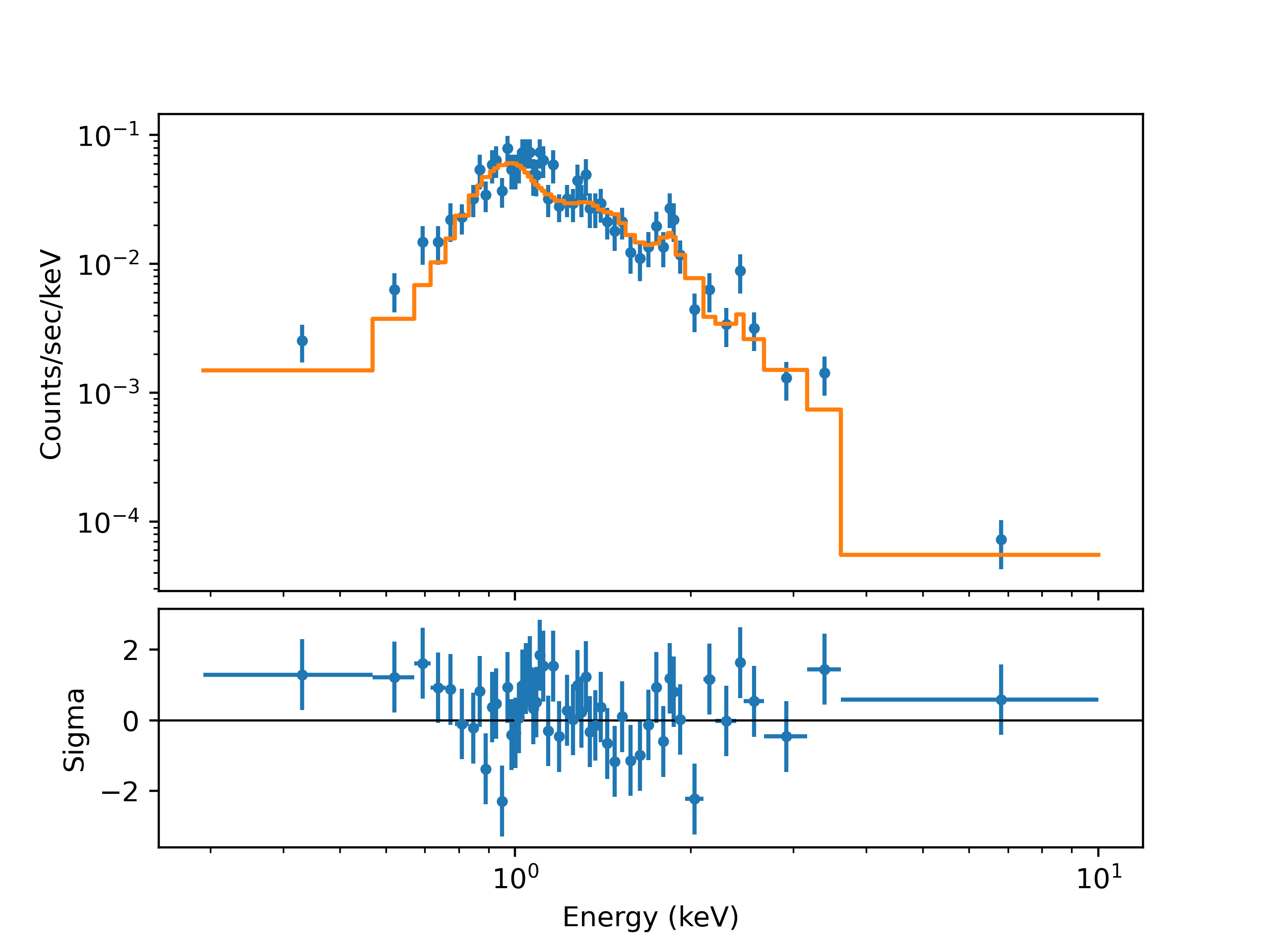}{0.49\textwidth}{(2MA 0201+0117 A)}
          \fig{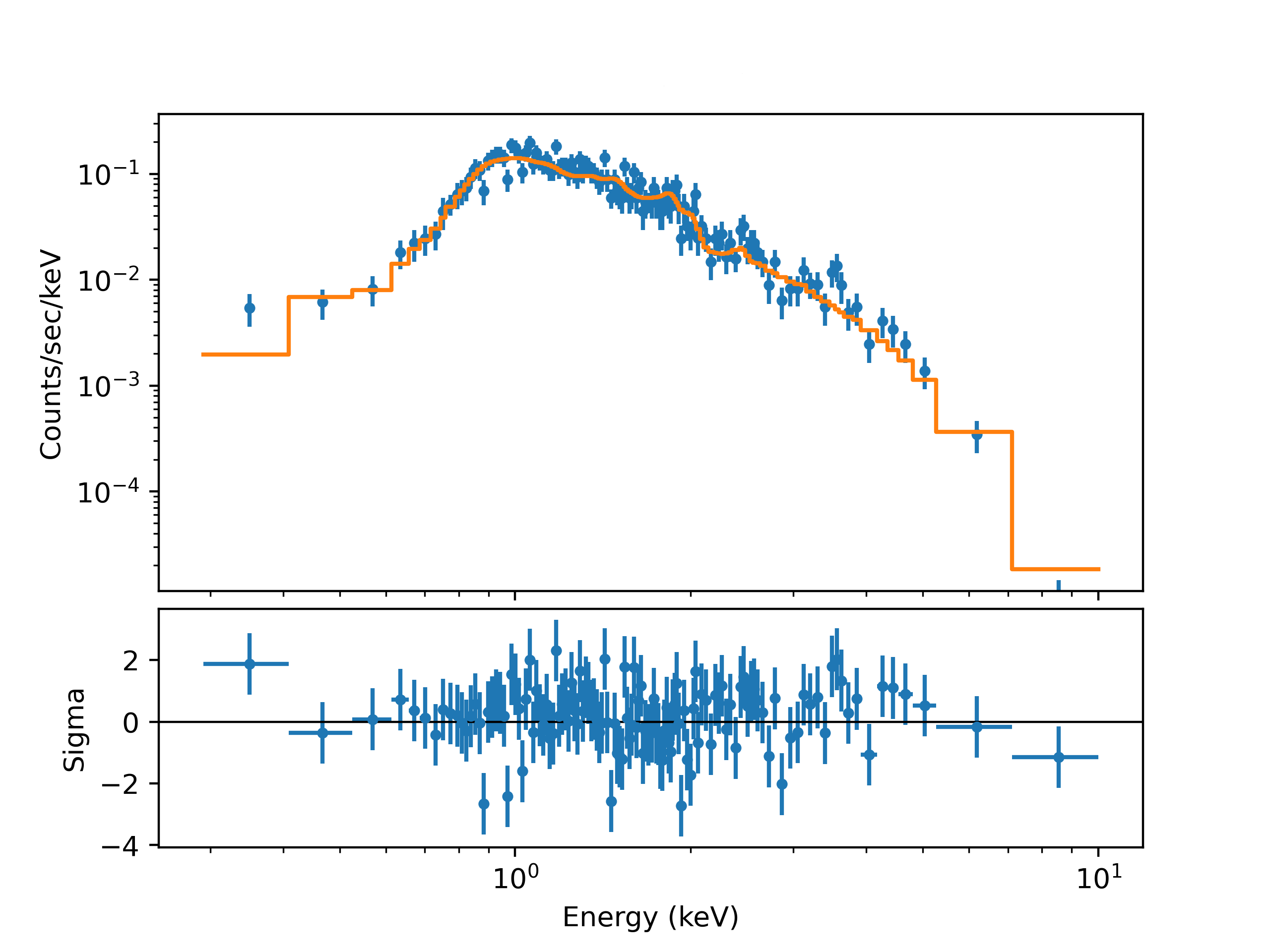}{0.49\textwidth}{(2MA 0201+0117 B)}}
\figcaption{Blue data points show the observed \textit{Chandra} coronal X-ray spectrum from 0.3--10~keV for the $\beta$ Pic Moving Group members 2MA~0201+0117~A and B, grouped to 9 counts per bin. Overplotted in orange are the forward-folded best-fit APEC models; A uses a single temperature component whereas B uses two temperature components in the coronal model. The bottom panels show the residual values divided by the errors, with horizontal bars indicating the energy bin widths. Spectral fits for all of the other quiescent X-ray datasets can be seen in Figure \ref{fig:all-xray-spectra} in Appendix \ref{sec:appendix}. \label{fig:2MA0201-SpectraFit}}
\end{figure}

Each spectrum was tested with gradually increasing model complexity, including 1-temperature and 2-temperature coronae, pileup inclusion or exclusion as outlined above, varying or fixed global coronal abundances, and alternate VAPEC models which use APEC models just with different combinations of fixed and varying individual elemental abundances. Such thorough testing was motivated by the range of signal-to-noise values and different features present in various datasets. Final model selections were informed foremost by the reduced-$\chi^2$ proximity to unity, the presence of any poorly constrained or unconstrained parameters in the solution, the visual quality of the fit to the data, and in some cases F-test comparisons between competing models. We favored simpler models and consistent choices when there was an ambiguity in the best choice. Example spectral fits can be seen in Figure \ref{fig:2MA0201-SpectraFit} for the $\beta$ Pic Moving Group members 2MA~0201+0117~A and B, with all other quiescent spectral fits shown in Figure \ref{fig:all-xray-spectra} in Appendix \ref{sec:appendix}. For most cases, 1-temperature corona models provided reasonable fits to the data. A second temperature component was indicated but not fully constrained for datasets GJ~1183~A--2, GJ~1183~A--3, and GJ~1183~B--3.

A few notes about the global coronal abundances are in order. In two cases, GJ~1183~B--2 and KX~Com~A--2, the abundances were not constrained, so were instead fixed to quiescent values obtained from separate exposures on the same stars taken earlier in each day. Global abundances were otherwise fixed to a representative sub-solar value of 0.15 if unconstrained in other fits, with 0.15 determined from the average measured abundance we see in the quiescent datasets. We tested our fixed sub-solar abundance cases using a solar abundance instead, but most $F_X$ and coronal temperature results deviated by less than $\sim$1--2~$\sigma$. We ultimately chose to exclude solar-abundance-fixed models given that \textit{all} of our measured abundances indicated firmly sub-solar values around 0.1--0.2, in agreement with the typically sub-solar coronal abundances found in other M dwarfs \citep{2005A&A...435.1073R}. For the four flaring events, we used the same fixed 0.15 sub-solar coronal abundance where needed despite knowing that abundances can change during X-ray flares \citep[e.g.,][]{2000A&A...353..987F} because our one measured abundance during a flare was still sub-solar at 0.092 in NLTT~44989~B. Flares are not our principle science focus here; a more careful analysis of the flares is possible, but beyond the scope of this work. Finally, we note that sub-solar abundances were indicated but not fully constrained for the GJ~1183~A--2--Flare, GJ~1183~B--2, KX~Com~A--2, and KX~Com~BC--3 datasets.

We report 68\% confidence interval (1-$\sigma$) asymmetric uncertainties for the APEC coronal parameters, computed with the \texttt{conf} Sherpa sampling method. X-ray fluxes were determined between 0.3--10~keV using the \texttt{sample\_energy\_flux} Sherpa method, which repeatedly draws parameter values and sums over the model to calculate a flux at each iteration. We adopt the median of the resulting distribution of 10,000 flux samples as our chosen flux value, with the asymmetric 1-$\sigma$ bounds of the distribution as our flux uncertainties.

Results from the \textit{Chandra} X-ray analysis are discussed later in \S\ref{subsec:chandra-results}.

\subsubsection{NLTT 44989 A Detection} \label{subsubsec:NLTT-A}

The expected source location for NLTT~44989~A at the epoch of the \textit{Chandra} observation, based on \textit{Gaia} DR3 coordinates and proper motions, did not show a clear detection above the background noise, nor did the \texttt{wavdetect} source-detection method identify any sources within several arcseconds. The projected separation of the AB pair gives an orbital period $>$1000 years, eliminating orbital motion as a possible explanation. Furthermore, the \textit{Gaia} astrometry over 2014--2017 yielded a proper motion for A that is consistent with the RECONS proper motion fit using data over 2019--2024\footnote{The RECONS data undergo a full astrometric analysis alongside each photometric analysis, yielding the RECONS proper motion mentioned here. See \cite{Henry_2018} for a recent summary of the RECONS long-term program.}, indicating no deviation in the star's path between \textit{Gaia} DR3 and the \textit{Chandra} observations in 2022.

There is a weak grouping of roughly 5--10 counts across all energies over several pixels within $\sim$1$\arcsec$ of the expected location, but this is qualitatively comparable to many other regions of noise in the image. That said, we cannot strictly rule out the possibility of some detected source photons from A, so we used the few counts at its expected location to derive an upper limit on its X-ray luminosity and emission measure.

We used a 2-pixel radius source aperture centered at the expected location of NLTT~44989~A to capture the small grouping of nearby counts. After filtering to 0.3--10~keV and subtracting the background, only 4.3 counts remain. The associated count rate was then used with the \textit{Chandra} PIMMS\footnote{PIMMS is found at \href{https://cxc.harvard.edu/toolkit/pimms.jsp}{https://cxc.harvard.edu/toolkit/pimms.jsp}.} calculator to determine an X-ray flux assuming an APEC source model (T=10$^7$~K, Abundance=0.2, Redshift=0, nH=0), returning a Norm value of $2\times10^{-6}$ as well. The final limiting luminosity and emission measure values are reported in Table \ref{tab:ChandraTable}. If the 4.3 counts are true source photons and not a coincidental clustering of background noise, then this offers a rough estimate for the NLTT~44989~A X-ray flux under our model assumptions; it otherwise gives an approximate upper limit only.

\subsection{Speckle Imaging - SOAR \& LDT} \label{subsec:speckle-methods}

GJ~1183~AB and NLTT~44989~AB were observed using the High-Resolution Camera \citep[HRCam;][]{Tokovinin_2018} with the SOAR Adaptive Module \citep[SAM;][]{Tokovinin_2016} on SOAR through a separate RECONS project led by coauthor Vrijmoet and summarized in \cite{Vrijmoet_2022}. Observations occurred during 2019--2020, with one visit to the GJ~1183~AB system and two visits to the NLTT~44989~AB system. Data were taken in the \textit{I}-band, and otherwise used procedures typical for the observing program as outlined in \cite{Vrijmoet_2022}. Data were processed using the methodology of \cite{Tokovinin_2010} and \cite{Tokovinin_2018}, yielding measures of the angular separation and magnitude difference either as detections or limits.

2MA~0201+0117~AB and KX~Com~A-BC were observed with the Quad-camera Wave-front-sensing Six-wavelength-channel Speckle Interferometer instrument \citep[QWSSI;][]{QWSSI_2020} on the Lowell Discovery Telescope as part of an ongoing RECONS speckle effort led by coauthor Henry. Stars were observed once each during 2021 in each of the four 40~nm wide channels at 577, 658, 808, and 880~nm. Data were processed following the procedures typical for QWSSI, which are similar to that of its predecessor DSSI as outlined in \cite{2009AJ....137.5057H, 2015AJ....150..151H}. As with the SOAR data, the results are parameters for detections or limits for non-detections.

Results from the speckle analysis are discussed later in \S\ref{subsec:speckle-results}.

\subsection{Blending and Contamination} \label{subsec:contam}

Our new data from all five observing campaigns spatially resolve the A and B components in a twin pair in all four cases in reasonably good seeing.  As shown in Table \ref{tab:SampleTable-astr}, GJ~1183~AB, KX~Com~A-BC, and 2MA~0201+0117~AB are well separated by 7--13$\arcsec$, and all three systems are also free from contamination because the nearest background \textit{Gaia} DR3 sources are $>$15$\arcsec$ away at Ep=2021.0.

For the closest of the four pairs, NLTT~44989~AB with a separation of 4\farcs75, care was taken to only observe and use data with suitably good seeing to prevent AB blending. Background contamination is more complex for this system given its location in a dense field, with sources closer than roughly 4\farcs5 possibly contaminating the new measurements. Here we examine potentially contaminating sources, although overall we deem all of the new observations of NLTT~44989~A and B to be suitably free from contamination at any meaningful level.

NLTT~44989~A has two \textit{Gaia} sources within 4\farcs5 away but each is $\sim$6.5 mag fainter than A at $G$, making them negligible in the optical where four of our five campaigns observe. In X-rays, we see no confident detection at A's location (\S\ref{subsubsec:NLTT-A}) nor at its $<$4\farcs5 nearby neighbors, so we consider the star to be uncontaminated in the \textit{Chandra} measurements as well.

\begin{figure}[t]
\centering
\gridline{\fig{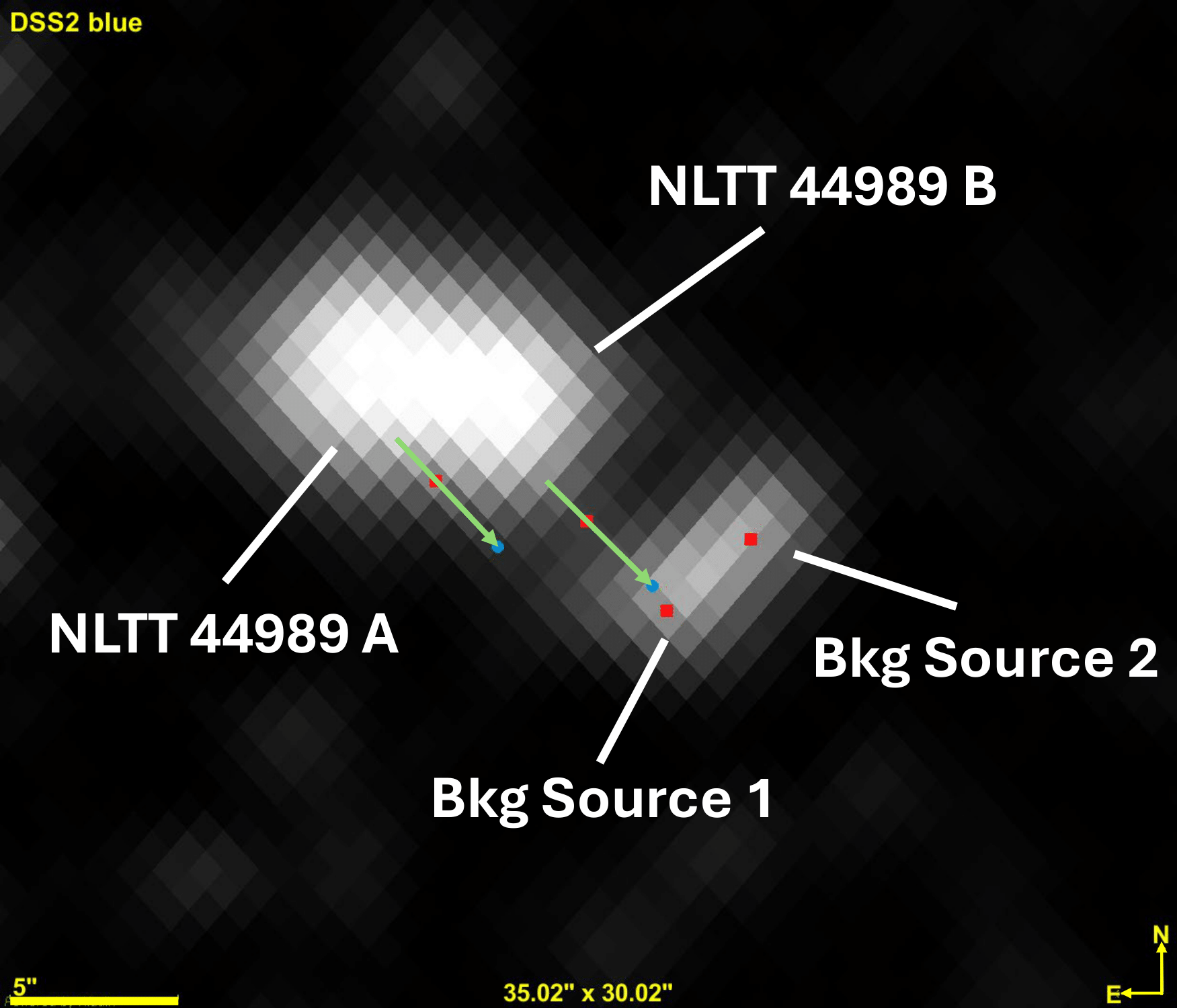}{0.49\textwidth}{(DSS2-Blue)}
          \fig{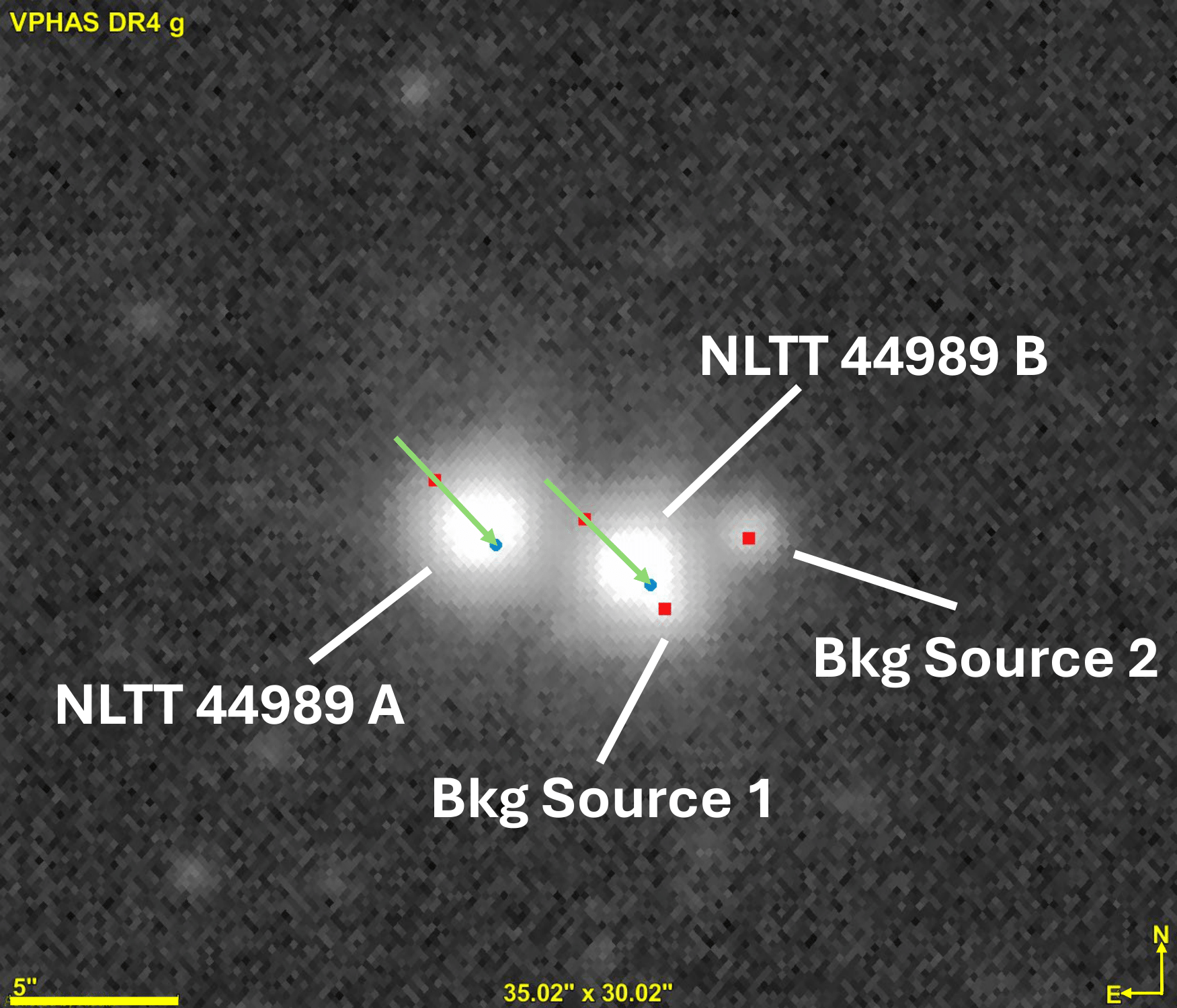}{0.49\textwidth}{(VPHAS DR4 g-band)}}
\figcaption{Optical images of the NLTT~44989~AB system, showing the relative positions of the brighter target M dwarfs and fainter background contaminating sources. The left panel shows a DSS2-Blue image taken roughly half a century ago \citep[Ep.~$\approx$~1975;][]{1996ASPC..101...88L, 2004AJ....128.3082G} and the right panel shows a VPHAS DR4 g-band image taken more recently about a decade ago \citep[Ep.~$\approx$~2012;][]{2014MNRAS.440.2036D}. Red squares and blue circles show the relevant source positions from \textit{Gaia} DR3 at J2000 and J2016 respectively, with green arrows along the proper motion vectors of the twin stars --- the background sources have negligible proper motions. Our new observations of the system span 2019 to 2024, when proper motion has moved B farther on top of background source 1. Several other background sources in \textit{Gaia} DR3 are not shown in these images for visual clarity, but all are negligibly faint and/or resolved from our target stars. See \S\ref{subsec:contam} and \S\ref{subsubsec:contam-mags} for a discussion of the small contamination impacts in the various observations we use. \label{fig:NLTT-field}}
\end{figure}

NLTT~44989~B has several background sources within 4\farcs5; all are negligible at 6.4--7.4~mag fainter in $G$ except for two that warrant further consideration. The first (source 1) is separated from B by only 0\farcs22 at 2021.0 (0\farcs84 away at Ep=2016.0), is 3.75~mag fainter in $G$, and has no parallax information available. The second (source 2) is 3\farcs02 away at 2021.0 (3\farcs23 away at Ep=2016.0), is 3.97~mag fainter in $G$, and is possibly an evolved giant star based on the \textit{Gaia} DR3 parallax with large error. Both are shown relative to the twin system over time in Figure \ref{fig:NLTT-field}. In the 0.9m photometry of NLTT~44989~B we utilize relative brightness changes, so while source 1 always adds $\sim$3\% contaminating flux, its impact is negligible within our uncertainties --- this assumes source 1 does not vary by large fractions of its entire brightness at timescales (or morphologies) matching our observed signal. Source 2 is trickier for the 0.9m because seeing changes that vary the contaminating flux could easily mimic a weak variability signal. We took extreme care to only observe with excellent seeing $\leq$1\farcs4 and manually reviewed radial source profiles for \textit{all} 0.9m frames to remove any cases with unacceptable overlap between the B star and source 2 light distributions. \textit{TESS} data, which always have both background sources entirely blended with both NLTT~44989~A and B, yield period measures from two sectors that are consistent with the 6.55d signal we see in the 0.9m photometry of B (discussed later in \S\ref{subsec:rot-results}), confirming that variable blending with source 2 is not markedly influencing our 0.9m results for B. In addition, a fully convective M dwarf with this period would be expected to display activity that is generally consistent with what we observe in H$\alpha$ and $L_X$ for B (\S\ref{sec:rot-activity-results}). In the CHIRON spectra of B, source 1 adds $\sim$3\% contaminating optical flux while source 2 may occasionally contribute minimal contamination from its wings depending on the seeing. However, the H$\alpha$ EWs are total brightness measures (as opposed to differential measurements), and B's observed H$\alpha$ variability is well beyond 3\% (\S\ref{subsec:chiron-results}), so these sources don't meaningfully impact the H$\alpha$ results. For RVs, our measures from the CHIRON spectra for NLTT~44989~A and B are consistent and in agreement with resolved \textit{Gaia} DR3 RV values as well. Finally, in the X-ray observations, we do not see a clear indication of B's X-ray source being a merged or multi-source profile with source 1 and we also see no signal beyond the noise at the location of source 2.

Considering the rotation archival data sources, for all systems for which data are available in ZTF, the components are resolved. \textit{TESS} and ASAS-SN data always blend A and B together in each system, sometimes with background sources too, but their resulting blended periods agree with the periods we find in the resolved 0.9m data so we do not elaborate further on their contamination here, with the exception of NLTT~44989~A discussed further in \S\ref{subsec:rot-results}.

\subsubsection{Blending and Contamination in the \textit{Gaia} and 2MASS Apparent Magnitudes} \label{subsubsec:contam-mags}

The final data we assess for contamination are the $BP$, $RP$, \textit{J}, \textit{H}, and $K_s$ magnitudes we used for selecting equal-mass components, where disruptions could make non-twins appear falsely twin-like and influence the interpretation of our results with even small deviations. Focusing first on \textit{Gaia} $BP$ and $RP$, values are obtained from spectral extraction windows 3\farcs5 by 2\farcs1 in size \citep{2023A&A...674A...2D}, subsequently sampling a region roughly 3\farcs5 by 3\farcs5 wide around a given source as \textit{Gaia} scans along different angles --- two sources closer than roughly 3\farcs5 would therefore contaminate each other. We can estimate the contaminating flux in $BP$ and $RP$ using \textit{Gaia} DR3 $G$ because $G$ magnitudes come from spatially smaller windows with profile-fitting and are much less susceptible to nearby contamination \citep{2016A&A...595A...3F, 2021A&A...649A..11R}. All four systems here have AB separations $>$4$\arcsec$ so do not blend A and B, and GJ~1183~AB, KX~Com~A-BC, and 2MA~0201+0117~AB all lack background sources within at least 15$\arcsec$ of each star so are entirely uncontaminated. NLTT~44989~A has a single source within 3\farcs5 but it only adds $\sim$0.2\% flux in $G$ so A is also functionally uncontaminated. However, NLTT~44989~B has five background sources 0\farcs84--3\farcs26 away at Ep=2016.0 totaling about 6.3\% extra optical flux based on $G$, possibly falsely brightening $BP$ and $RP$ for the component by up to that much depending on the exact extraction regions and spectral energy distributions of the background sources.

We also assessed the \textit{Gaia} DR3 \texttt{phot\_bp\_rp\_excess\_factor}, which compares combined $BP$ and $RP$ fluxes relative to $G$ fluxes to find inconsistencies between the measures that can indicate contamination or other issues \citep{2021A&A...649A...3R}. We calculated the corrected C* excess factors for this using the relations provided in Table 2 of \cite{2021A&A...649A...3R}, finding seven of our eight components have minimal excess $BP+RP$ flux $\le$3.1\%, but the eighth case of NLTT~44989~B again finds a 6.3\% excess, congruent with our more approximate manual estimations of contaminating fluxes above. Furthermore, the background source 0\farcs84 away from NLTT~44989~B at Ep=2016.0 and 3.75~mag fainter in $G$ (source 1 above) has the \textit{Gaia} parameter \texttt{ipd\_frac\_odd\_win} (IPDfow) significantly elevated at 80\%, meaning most scans have this background source's astrometric $G$-band windows truncated or otherwise disrupted; this implies source 1 and B are somewhat blended in their individual $G$ measurements as well, so further contaminating deviations of order $\sim$1--3\% might exist in $G$ and the subsequent $BP+RP$ excess for NLTT~44989~B but are hard to constrain.

Overall, this indicates the $BP$ and $RP$ measures are not markedly impacted by blending or contamination for seven of our eight components, with just NLTT~44989~B possibly having $BP$ and $RP$ falsely brightened by up to approximately 6.3\% ($\sim$66~mmag). We used $BP$ to estimate our masses (see \S\ref{sec:Sample}), so NLTT~44989~B may therefore actually have a mass very slightly smaller than we calculated. The important implications of this for our results are discussed later in \S\ref{subsubsec:discussion-explanations-masses}.

Now considering the 2MASS catalog used in our target selection criteria, it hosts 2\farcs0 pixels and a typical 2\farcs4--2\farcs7 FWHM \citep{2006AJ....131.1163S}. Per the 2MASS catalog, our sources' 2MASS magnitudes were derived from either PSF profile fitting or 4$\arcsec$ radii apertures, with 14$\arcsec$--20$\arcsec$ background annuli --- sources within 5$\arcsec$ of each other also used simultaneous multi-component fitting. GJ~1183~AB, KX~Com~A-BC, and 2MA~0201+0117~AB are therefore resolved and uncontaminated in all three bands given their 7\farcs72--13\farcs07 AB separations and aforementioned lack of background sources within 15$\arcsec$ --- KX~Com~A-BC at 7\farcs72 may have minimal blending on the wings, but its 2MASS magnitudes are ignored given we already know it has a third companion blended into its photometry. NLTT~44989~A and B show some overlap between their wings, with nearby contaminating background sources almost certainly mildly disrupting the target star measures and sky background corrections. That said, our nearby M dwarf stars are much brighter in the 2MASS near-IR bands than the distant background stars. Minor issues impacting both components similarly would still retain similar relative measures as well. The 2MASS magnitude uncertainties for our four systems span 0.016--0.061~mag, which compared to our $<$0.10~mag selection cutoff indicates minor contamination of up to several percent would also not markedly impact our results. Overall, our mass estimates derived from the higher resolution $BP$ mags are in excellent agreement between components, showing $\Delta M \le 2.3\%$ for the pairs here, so we do not expect these 2MASS contamination factors to change our general takeaways despite using the magnitudes in our sample construction.

\section{Results} \label{sec:Results}

Results from each of the five observing campaigns are described next. For the long-term variability (\S\ref{subsec:longterm-results}), rotation (\S\ref{subsec:rot-results}), H$\alpha$ activity (\S\ref{subsubsec:ha-results}), and X-ray properties (\S\ref{subsec:chandra-results}), our primary interest is whether or not the two stars in a given twin pair display similar or dissimilar activity and rotation behaviors. For the radial velocities (\S\ref{subsubsec:rv-results}) and speckle imaging (\S\ref{subsec:speckle-results}), our goal is to validate the twin natures of our targets by searching for and ruling out unseen companions.

\subsection{Stellar Cycles - CTIO/SMARTS 0.9m Long-term Campaign} \label{subsec:longterm-results}

\begin{figure}[!t]
\centering
\gridline{\fig{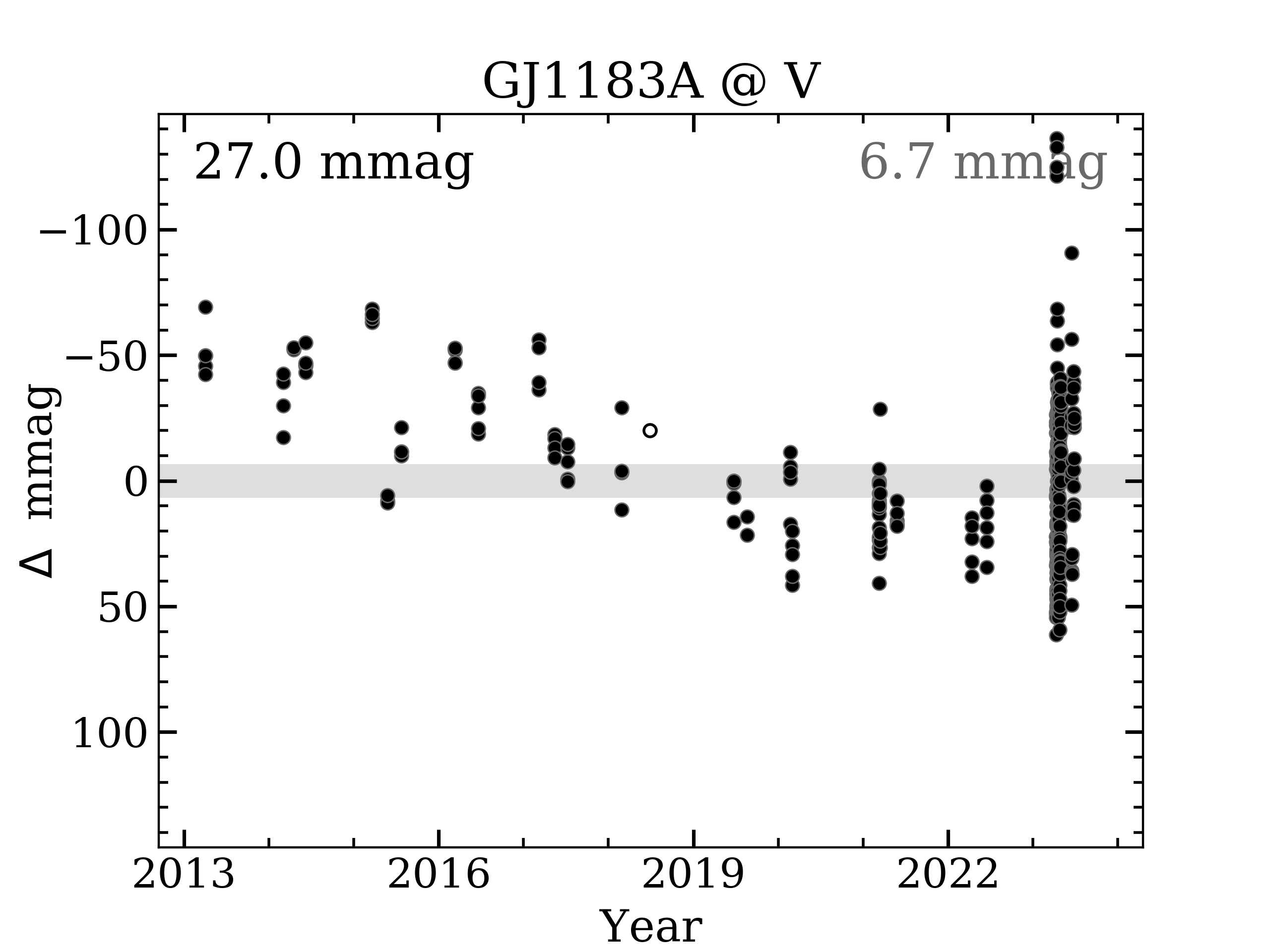}{0.33\textwidth}{}
          \fig{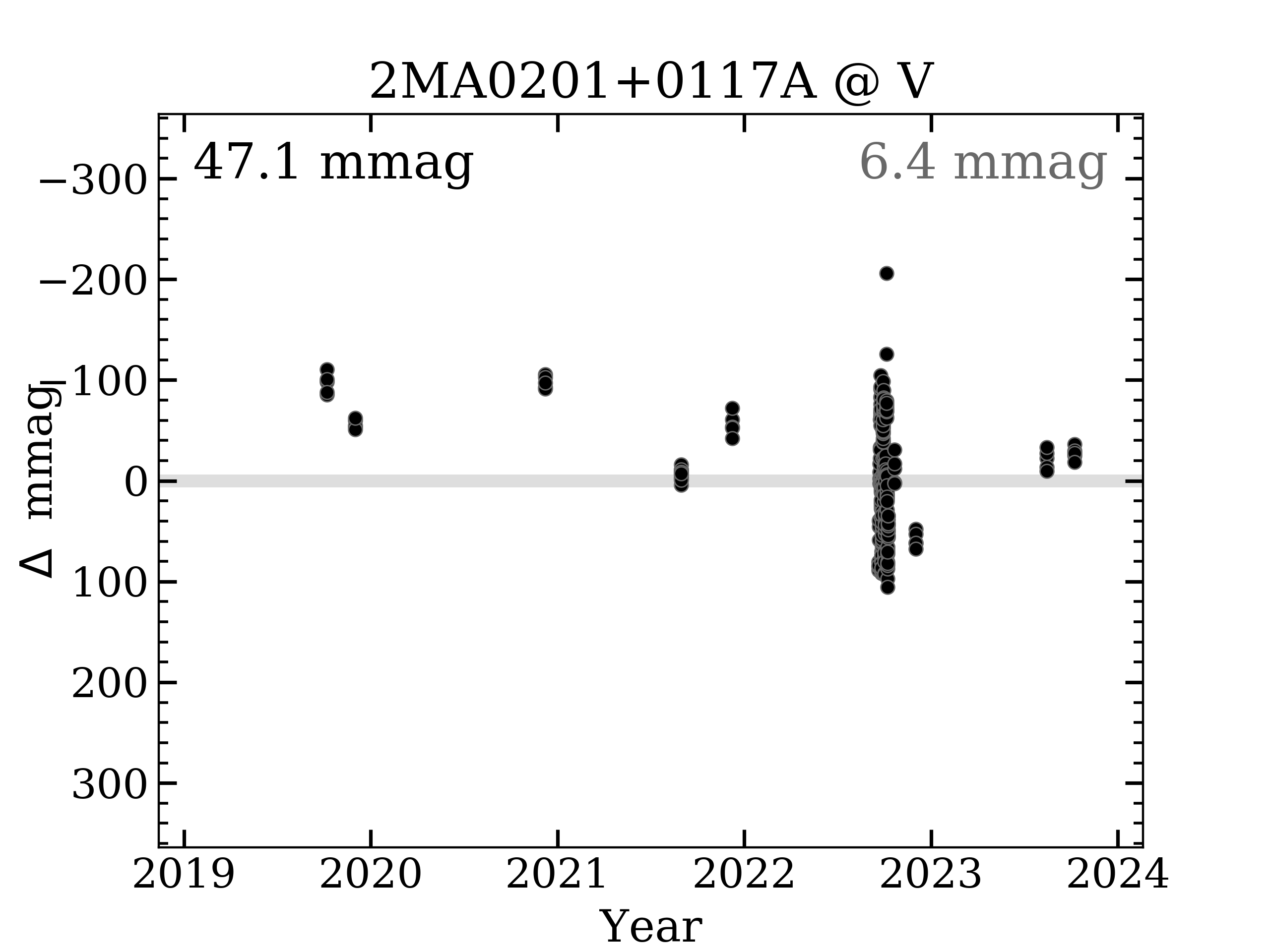}{0.33\textwidth}{}
          \fig{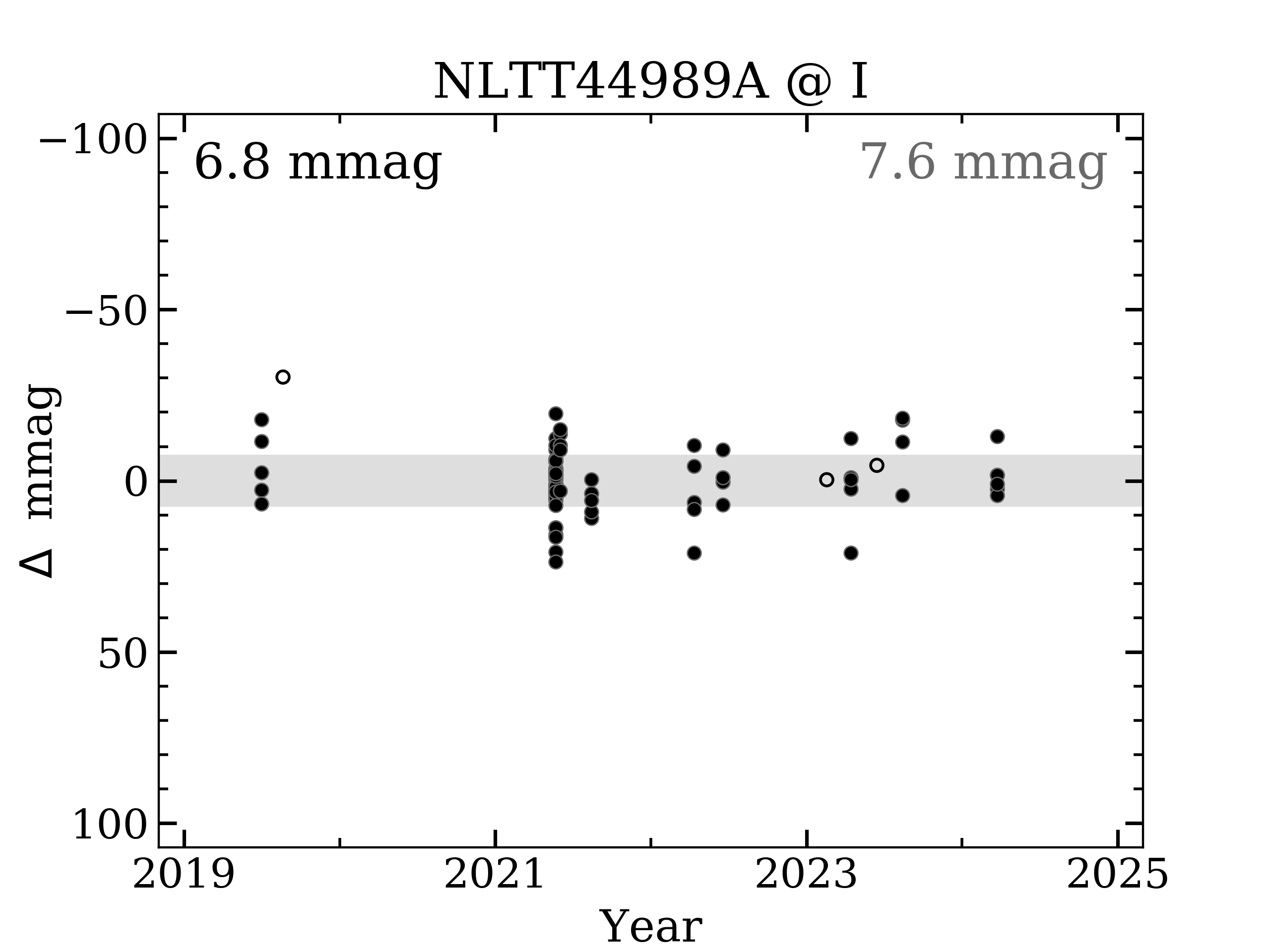}{0.33\textwidth}{}}
\vspace*{-8mm}
\gridline{\fig{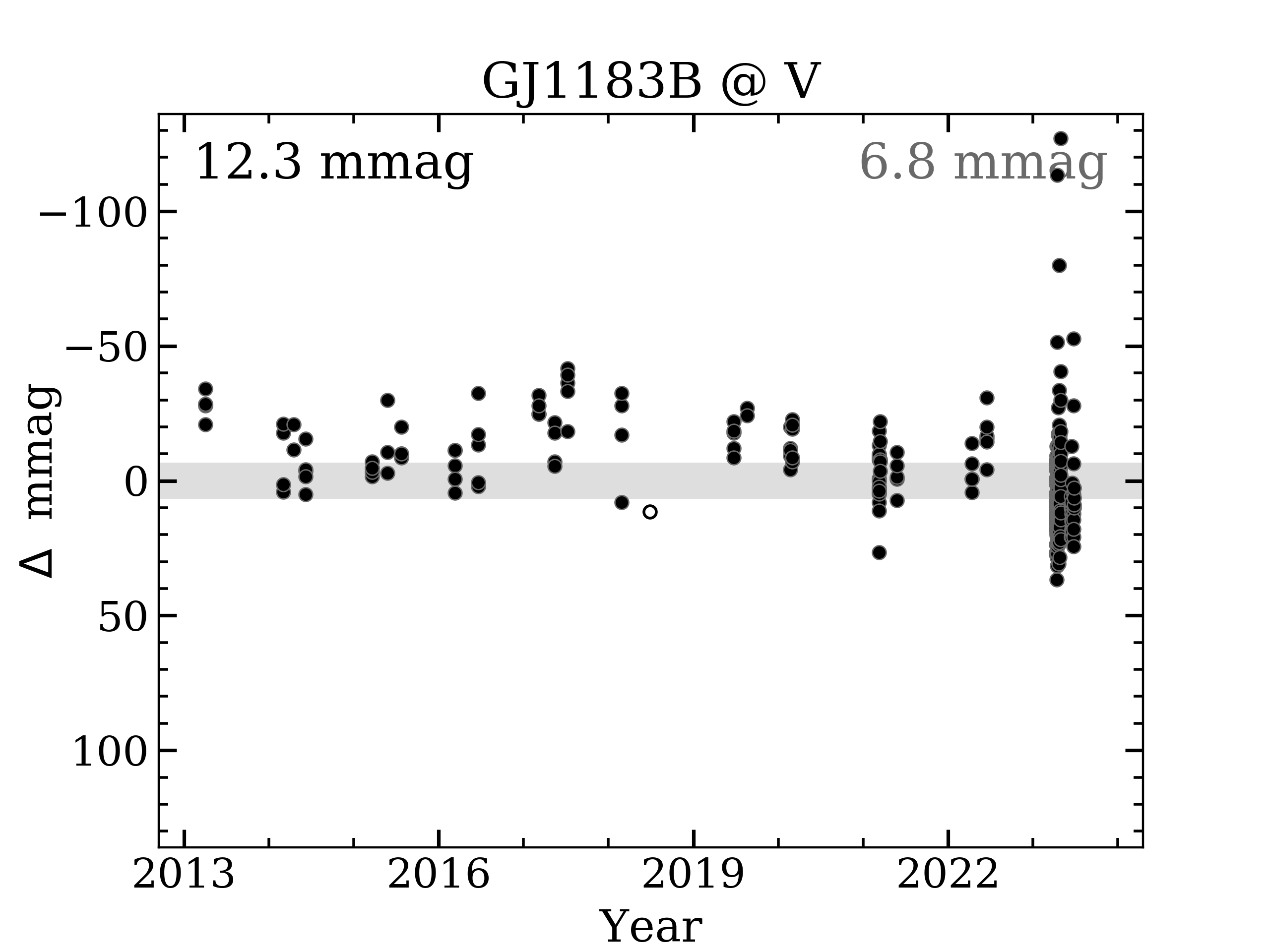}{0.33\textwidth}{(GJ 1183 A \& B)}
          \fig{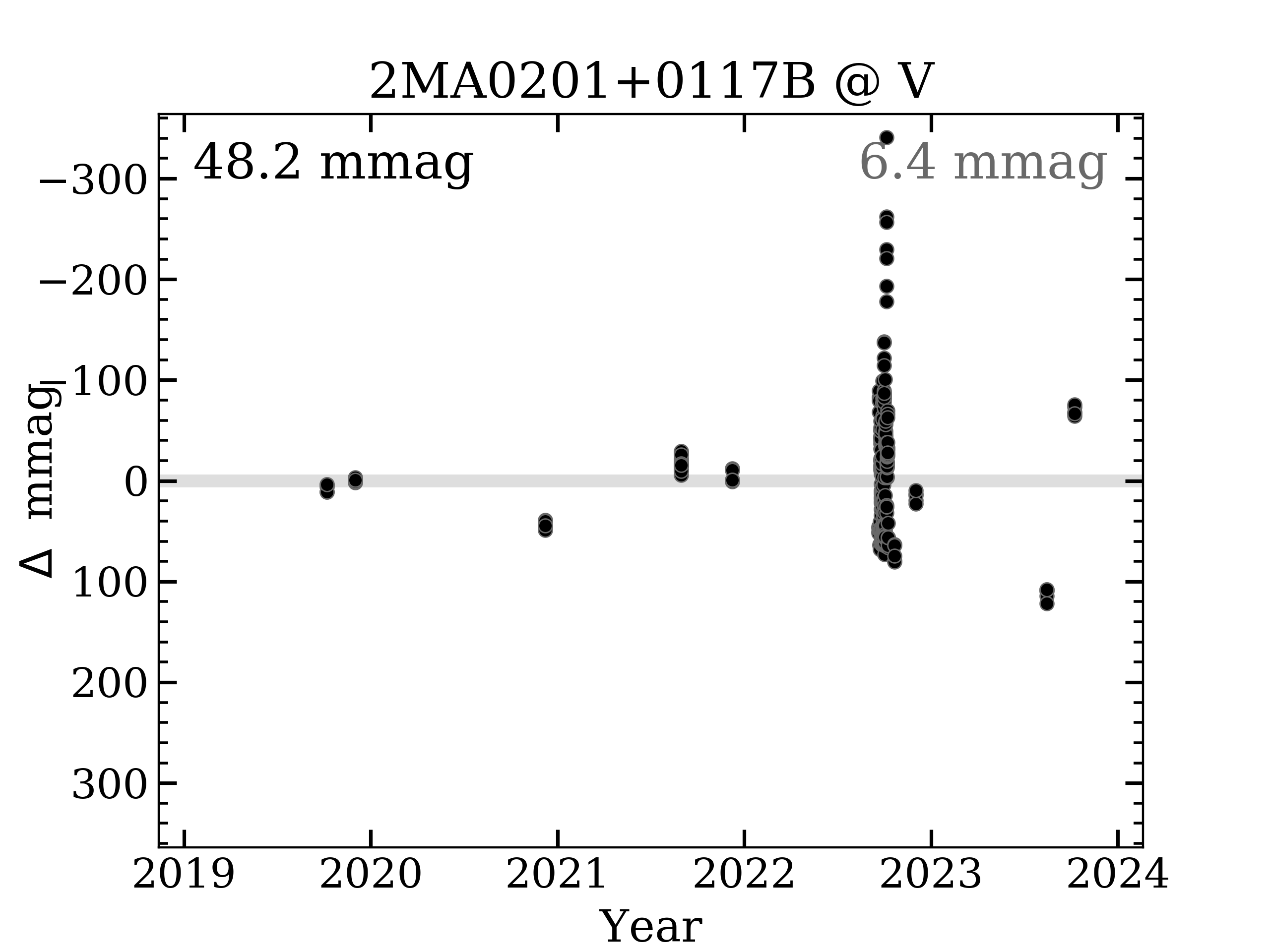}{0.33\textwidth}{(2MA 0201+0117 A \& B)}
          \fig{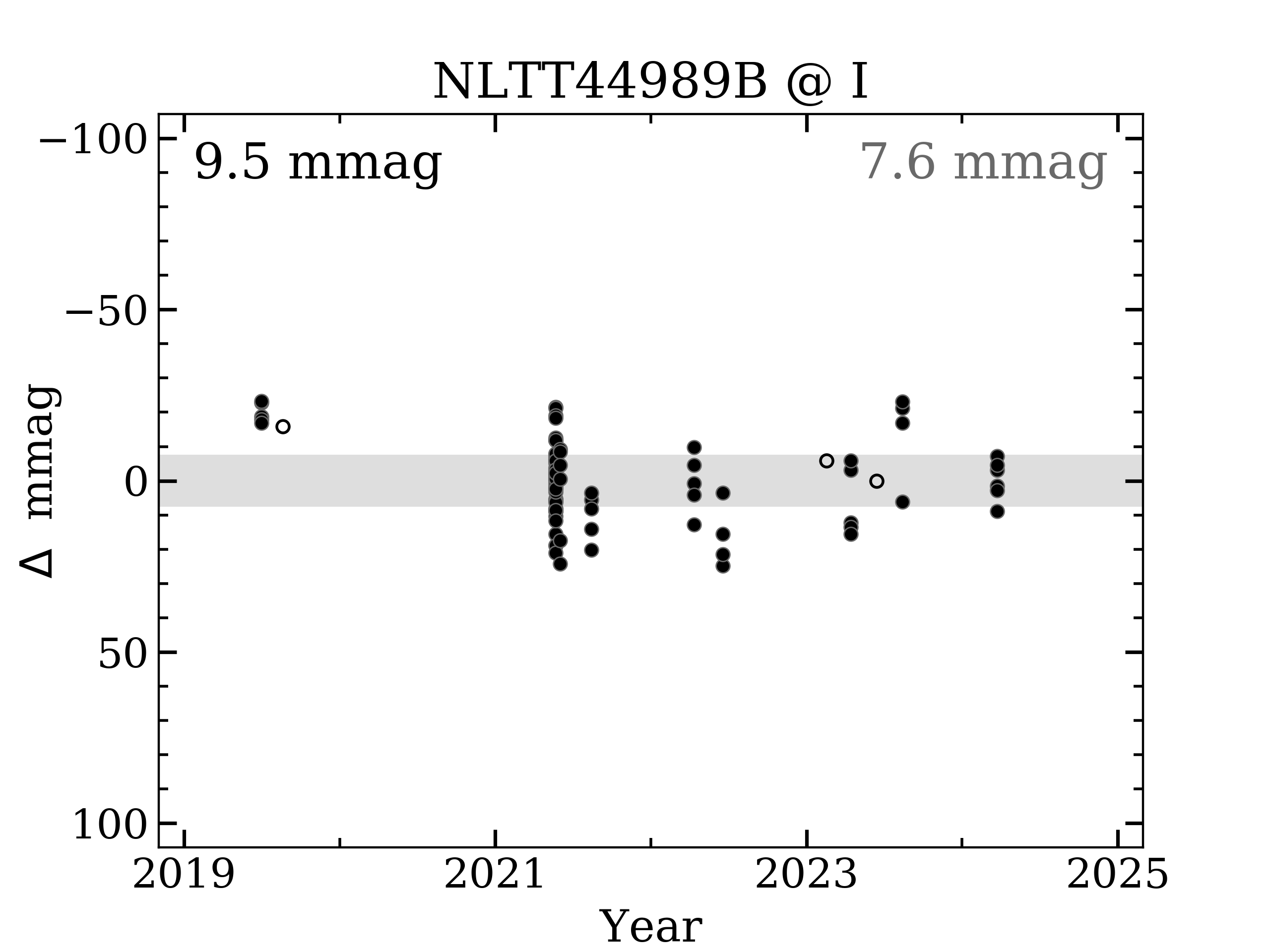}{0.33\textwidth}{(NLTT 44989 A \& B)}}
\figcaption{Long-term 0.9m light curves for the three true twin systems. A (top) and B (bottom) components are shown for each pair, with a brightening trend indicated by a rise in each plot. Observation filters are noted in each subplot title. GJ~1183~A and B are shown spanning over a decade of coverage, while the other two systems capture several years. Note the very different y-axis scale for 2MA~0201+0117~A and B. Year times are the Julian epoch, i.e., J2000 plus the number of Julian years since then. Large clusters of points in 2022 for 2MA~0201+0117~AB and 2023 for GJ~1183~AB are from the additional rotation visits analyzed in Figure \ref{fig:09m-rot}, with moderate flaring points left included here as examples. Open circles represent visits when only a single frame was secured; these are excluded from all quantitative analyses. Black numbers in the upper left of each panel are the Mean Absolute Deviation (MAD) values (average of absolute value offsets from the mean) for the solid points for each component. Grey numbers in the top right of each panel are the average noise levels of the non-varying reference stars used in the differential photometry analyses, again calculated as MAD values; these noise levels are indicated visually as grey shaded regions above and below 0 in each panel. The light curve data are available as Data behind the Figure (DbF) products in the online journal. \label{fig:longterm-lcs}}
\end{figure}

One of the four systems targeted here has substantial long-term data from the SMARTS 0.9m --- the decade-long light curves of GJ~1183~A and B are shown in Figure \ref{fig:longterm-lcs}. The same set of comparison reference stars is used for both target stars, so the significantly higher level of variability for A compared to B is secure. Given the long-term nature of these light curves, we strongly favor the explanation of overall enhanced spot activity in A compared to B, rather than a coincidental mismatch in spot contrast levels. Indications of stellar activity cycles are possibly evident in these light curves, but moderate rotational modulation is also present and the true cycle periods are likely close to or longer than a decade if they exist. Nonetheless, these results indicate that the difference in spot activity is sustained over at least a decade in GJ~1183~AB. Figure \ref{fig:longterm-lcs} also shows the few years of data presently available for the other two true twin systems; 2MA~0201+0117~A and B both show substantial scatter presumably from their strong rotational modulation shown later in Figure \ref{fig:09m-rot}, whereas NLTT~44989\footnote{NLTT~44989~AB was observed in I-band for the long-term visits shown in Figure \ref{fig:longterm-lcs} but V-band for the rotation observations shown later in Figure \ref{fig:09m-rot}.} A and B both show minimal photometric variation close to the noise limit.

The long-term 0.9m program also analyzes differential astrometry in search of positional photocenter perturbations that might indicate an unresolved orbiting companion (see \cite{Vrijmoet_2020} for a recent discussion of this process with these data). No such perturbations are seen in the 10 years of astrometry for GJ~1183~A and B or in the more limited $\sim$2--5~yrs of data available for the other three systems, consistent with no additional companions within the measurement limitations (roughly 7~mas photocentric displacement). Proper motion has also not moved any of the systems over background stars during the $\sim$2--10 years of long-term 0.9m observations for these targets, except for NLTT~44989~AB where minimal motion of $\sim$0.8$\arcsec$ over roughly five years has occurred around the nearby fainter background stars discussed in \S\ref{subsec:contam}.

\subsection{Rotation - CTIO/SMARTS 0.9m, \textit{TESS}, ZTF, \& ASAS-SN} \label{subsec:rot-results}

\begin{figure}[!hp]
\centering
\gridline{\fig{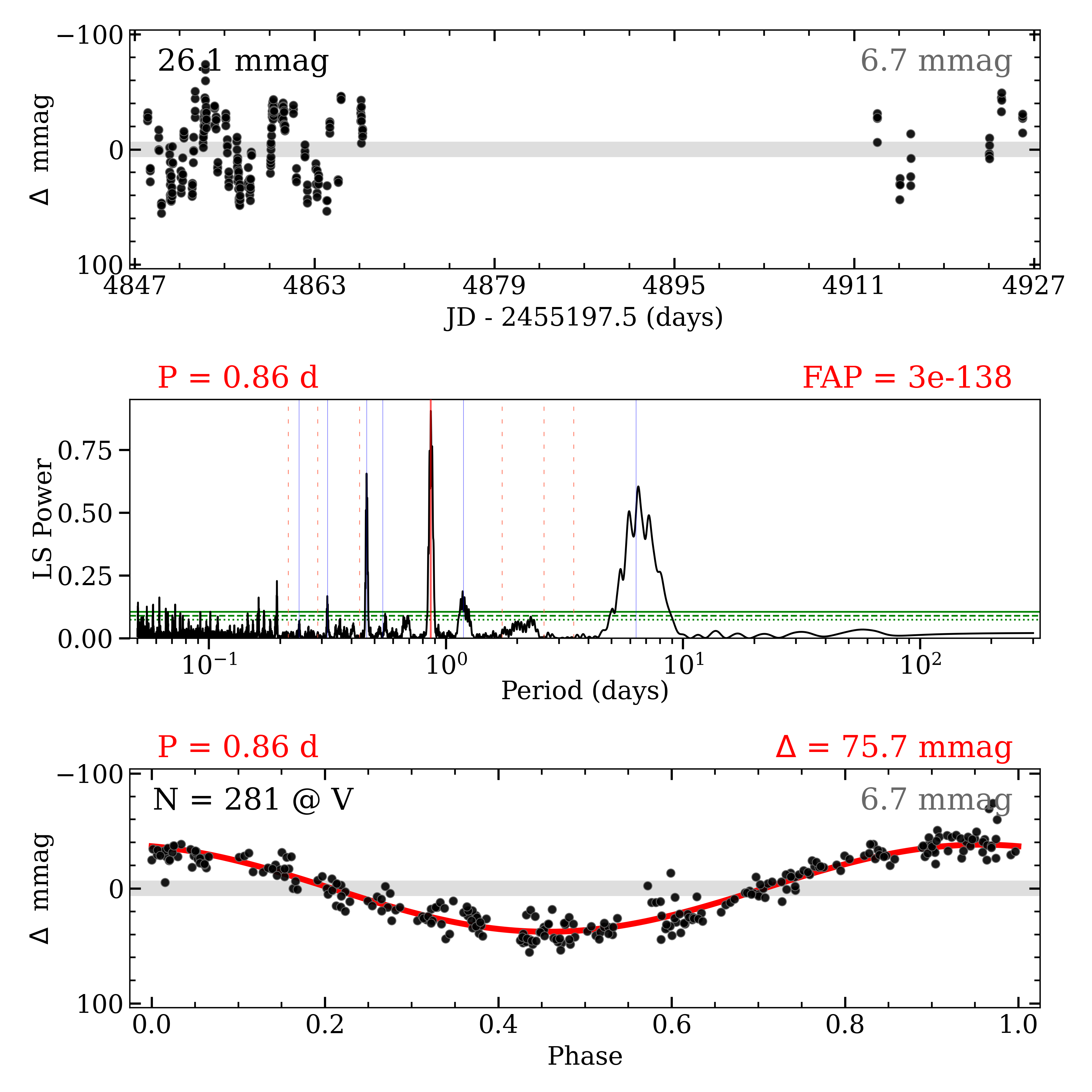}{0.47\textwidth}{$\uparrow$  GJ 1183 A  $\uparrow$}
          \fig{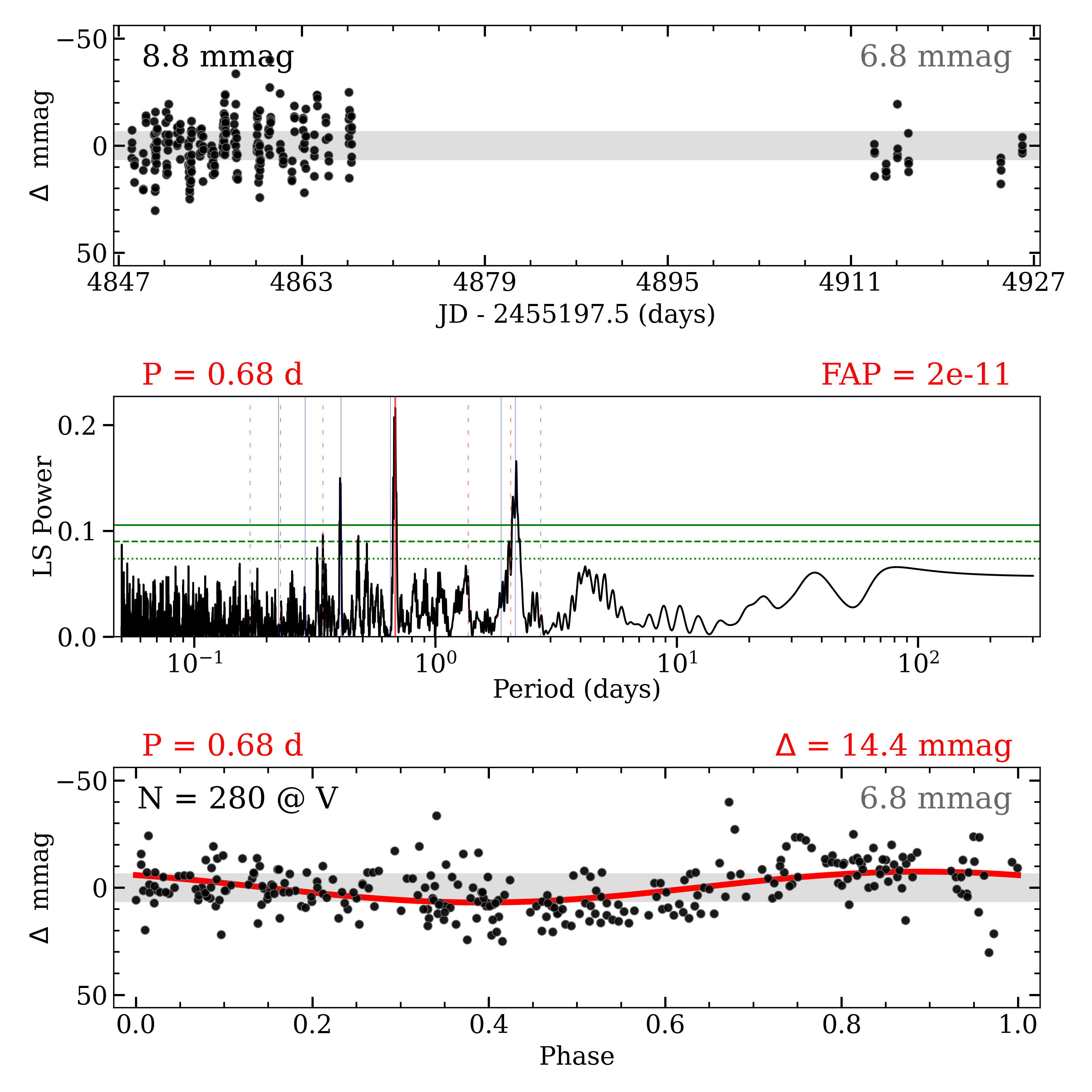}{0.47\textwidth}{$\uparrow$  GJ 1183 B  $\uparrow$}}
\gridline{\fig{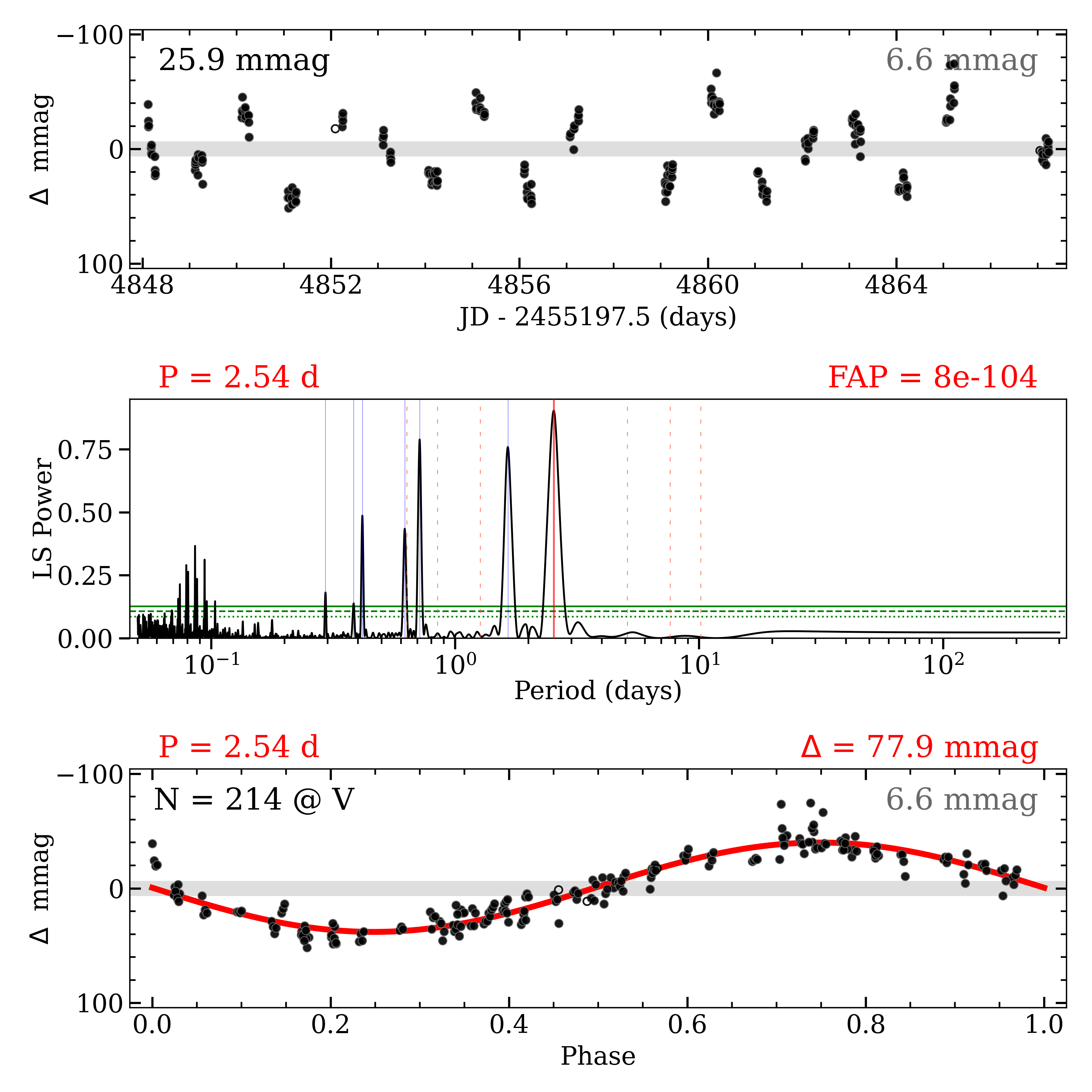}{0.47\textwidth}{$\uparrow$  KX Com A  $\uparrow$}
          \fig{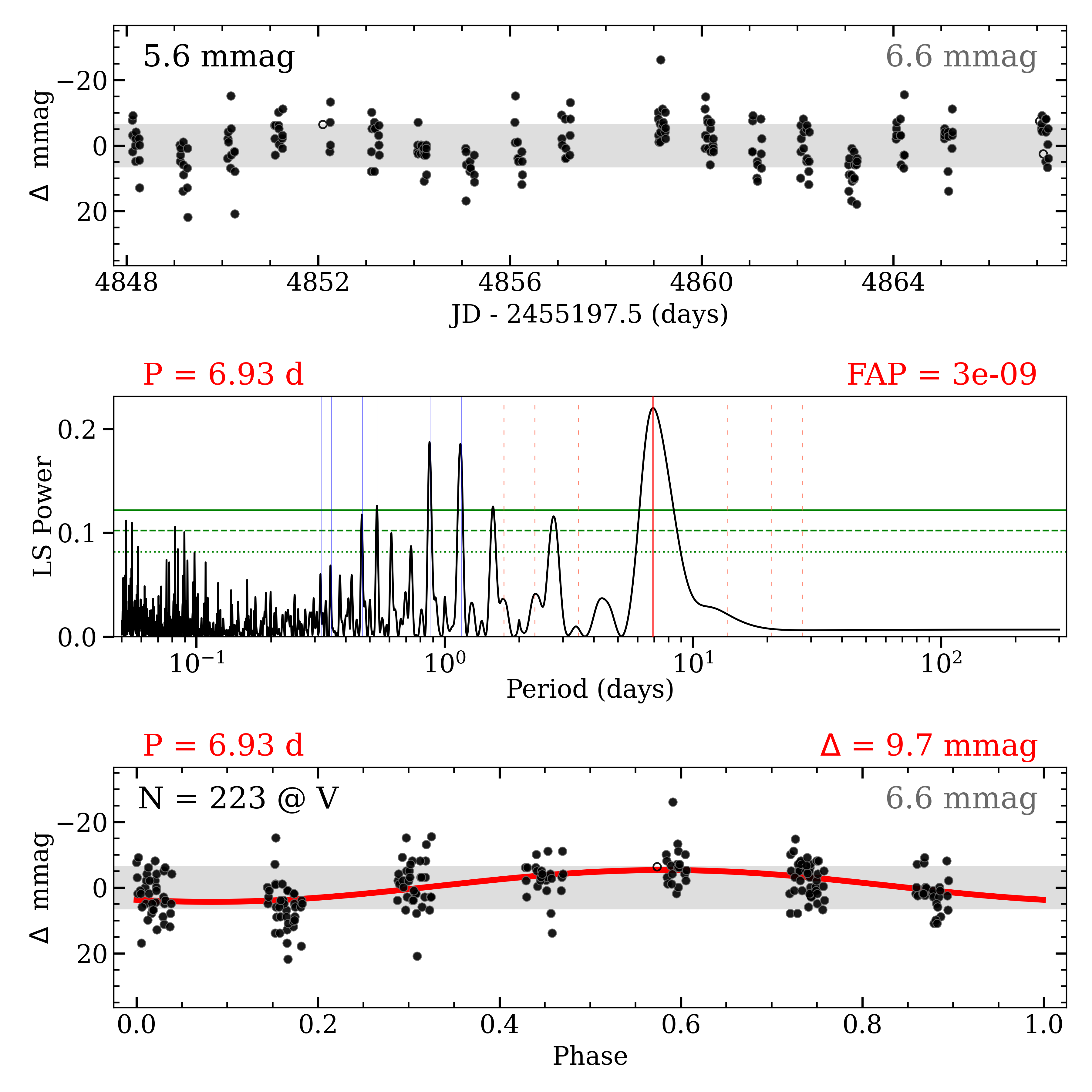}{0.47\textwidth}{$\uparrow$  KX Com BC  $\uparrow$}}
\figcaption{Each set of three vertical panels shows 0.9m $V$-band rotation results for a single component in a twin system. The top panel is a relative light curve in the same general format as Figure \ref{fig:longterm-lcs}, the middle panel is a Lomb-Scargle Periodogram of the same data, and the bottom panel is a phase-folded light curve of the same data based on the measured period. The Julian time of 2455197.5 corresponds to the 2010.0 epoch, a convenient reference time preceding all new observations in this paper. Note the different vertical scales between light curves. The periodogram shows horizontal green lines at the 10\% (dotted) / 1\% (dashed) / 0.1\% (solid) Baluev False Alarm Probability (FAP) values, the selected maximum power period peak with a vertical solid red line, vertical dashed orange lines at the 2/3/4 harmonic multiples, and vertical solid purple lines at the integer $n=[-3,...,3]$ 1/day sampling aliases. Phase-folding begins at the epoch of the first point, with a red sine wave corresponding to the selected Lomb-Scargle model result. Values in red are given for the rotation period, FAP, and peak-to-peak amplitude ($\Delta$) in mmag. The number of filled data points (N) and the observation filter ($V$ for all these stars) are given in the top left of the phase-folded subplot --- open points correspond to incomplete visits that only obtained one frame. The un-phased light curve data are available as Data behind the Figure (DbF) products in the online journal. \textit{Figure continued on the next page.} \label{fig:09m-rot}}
\end{figure}

The rotation results are summarized in Table \ref{tab:RotTable}, providing the measured periods from each data source along with the final period we adopt for each star. In \textit{TESS} and ASAS-SN data the systems are unresolved so the periods are noted with `bl' for blended. All 0.9m light curves are shown in Figure \ref{fig:09m-rot}, and example light curves from the archival data sources can be seen in Figures \ref{fig:external-rot} and \ref{fig:NLTT-AB-tess}. We note that the \textit{TESS} and \textit{TESS}-unpopular periodogram peaks can often correspond to FAPs numerically comparable to 0, hence their non-values for some cases in Figures \ref{fig:external-rot} and \ref{fig:NLTT-AB-tess}. Here we provide details for each system.
%
\begin{deluxetable}{lccccc|cc}[!t]
\tablewidth{0pt}
\tablecaption{Rotation Periods and Amplitudes\label{tab:RotTable}}
\tablehead{
\colhead{Name} & \colhead{0.9m $P_{rot}$} & \colhead{ZTF $P_{rot}$} & \colhead{\textit{TESS} $P_{rot}$} & \colhead{\textit{TESS}-unpop $P_{rot}$} & \colhead{ASAS-SN $P_{rot}$} & \colhead{Final $P_{rot}$} & \colhead{0.9m $\Delta$}\\
\colhead{} & \colhead{[d]} & \colhead{[d]} & \colhead{[d]} & \colhead{[d]} & \colhead{[d]} & \colhead{[d]} & \colhead{[mmag]}
}
\startdata
GJ 1183 A      & 0.86  & 0.86 or 6.49 & bl: 0.86         & bl: 0.69 \& 0.86    & bl: 0.86 & 0.86 & 75.7 \\
GJ 1183 B      & 0.68  & \nodata      & bl: 0.86         & bl: 0.69 \& 0.86    & bl: 0.86 & 0.68 & 14.4 \\
KX Com A       & 2.54  & 2.55         & bl: 2.55         & bl: 2.55            & bl: 2.55 & 2.55 & 77.9 \\
KX Com BC      & 6.93  & \nodata      & bl: 2.55         & bl: 2.55            & bl: 2.55 & 6.93 & 9.7 \\
2MA 0201+0117 A & 5.96  & 6.01         & bl: [5.92--6.09] & bl: 6.33            & bl: 6.01 & 6.01 & 152.8 \\
2MA 0201+0117 B & 3.30  & 3.30         & bl: [5.91--6.03] & bl: 6.32            & bl: 6.01 & 3.30 & 129.9 \\
NLTT 44989 A   & 38.27 & \nodata      & bl: [6.36--6.62] & bl: $\gtrsim$ 22 & \nodata  & 38   & 12.6 \\
NLTT 44989 B   & 6.55  & \nodata      & bl: [6.42--6.71] & bl: $\gtrsim$ 22 & \nodata  & 6.55 & 8.8 \\
\enddata
\tablecomments{Rotation periods from four sources (two treatments for \textit{TESS}) are shown for the components in the four targeted systems. An unreported result for a given archival source indicates data were either unavailable or showed no confident rotation signal. Entries with a leading `bl' flag are derived from photometry with the components either partially or entirely blended. Adopted rotation periods are given in the next to last column and the peak-to-peak model amplitude ($\Delta$) in the $V$ filter from the 0.9m is given in the final column. This 0.9m amplitude value is chosen because the four systems are all spatially resolved, rotation was detected in all cases, and a consistent filter was used. The 6.93d result for KX~Com~BC is from photometry blending the B and C stars but resolved from A. See also $\S$\ref{subsec:rot-results}.}
\end{deluxetable}

{\bf GJ~1183~A} has a period of 0.86d in 0.9m data, consistent with values from all other sources. The alternative alias peak around 6.5d in the ZTF light curve is not found in the \textit{TESS} data, confirming 0.86d as the correct period. {\bf GJ~1183~B} shows a low-amplitude 0.68d signal from the 0.9m that is supported by the presence of an asymmetric secondary peak at 0.69d in the AB-blended \textit{TESS}-unpopular results (see bottom right of Fig.~\ref{fig:external-rot}) distinct from the 0.86d peak.

{\bf KX~Com~A} shows a period of 2.54d in 0.9m data, in excellent agreement with the 2.55d period in the resolved ZTF and blended \textit{TESS} and ASAS-SN datasets. The various strong alias peaks present in ground-based data for A do not appear in the \textit{TESS} results, confirming 2.55d as the true signal. \cite{Chen2020} found a 2.55d rotation period from zr-band ZTF data for KX~Com~A and \cite{Magaudda_2022} found 2.55d from ABC blended \textit{TESS} data, both exactly matching the results of our own rotation analyses of these data. {\bf KX~Com~BC} exhibits a clear period of 6.93d in the 0.9m data, seen best in the raw light curve with two clear repeats (bottom right of Fig.~\ref{fig:09m-rot}). This period is not evident in any other dataset. Because B and C are unresolved, this could be the rotation period for either star, but is likely from B based on H$\alpha$ information discussed later in \S\ref{subsec:KXCOMAB}.

{\bf 2MA~0201+0117~A and B} have periods from the 0.9m that are in excellent agreement with the resolved ZTF periods. The longer period of $\sim$6d for A is seen in all datasets. \textit{TESS} blended data visually show intermixing of the stronger $\sim$6d pattern and a second weaker signal at shorter period, presumably the 3.30d period for the B component. Beyond what we report in Table \ref{tab:RotTable}, \cite{2017A&A...600A..83M} also report rotation periods of 6.00d/3.41d from ASAS \citep{1997AcA....47..467P}, 5.87d from INTEGRAL/OMC \citep{2010ASSP...14..493D}, and 5.98d/3.30d from NSVS \citep{2004AJ....127.2436W}, all using blended photometry. \cite{2012AcA....62...67K} report 6.01d, also from blended ASAS data. Observations by Evryscope found 6.00d with AB blended \citep{2020ApJ...895..140H}. Separate from our own ASAS-SN rotation analysis, \cite{2019MNRAS.486.1907J} report 6.00d from blended ASAS-SN data. All are consistent with our results for the stars.

\begin{figure}[!htbp]
\figurenum{6}
\centering
\gridline{\fig{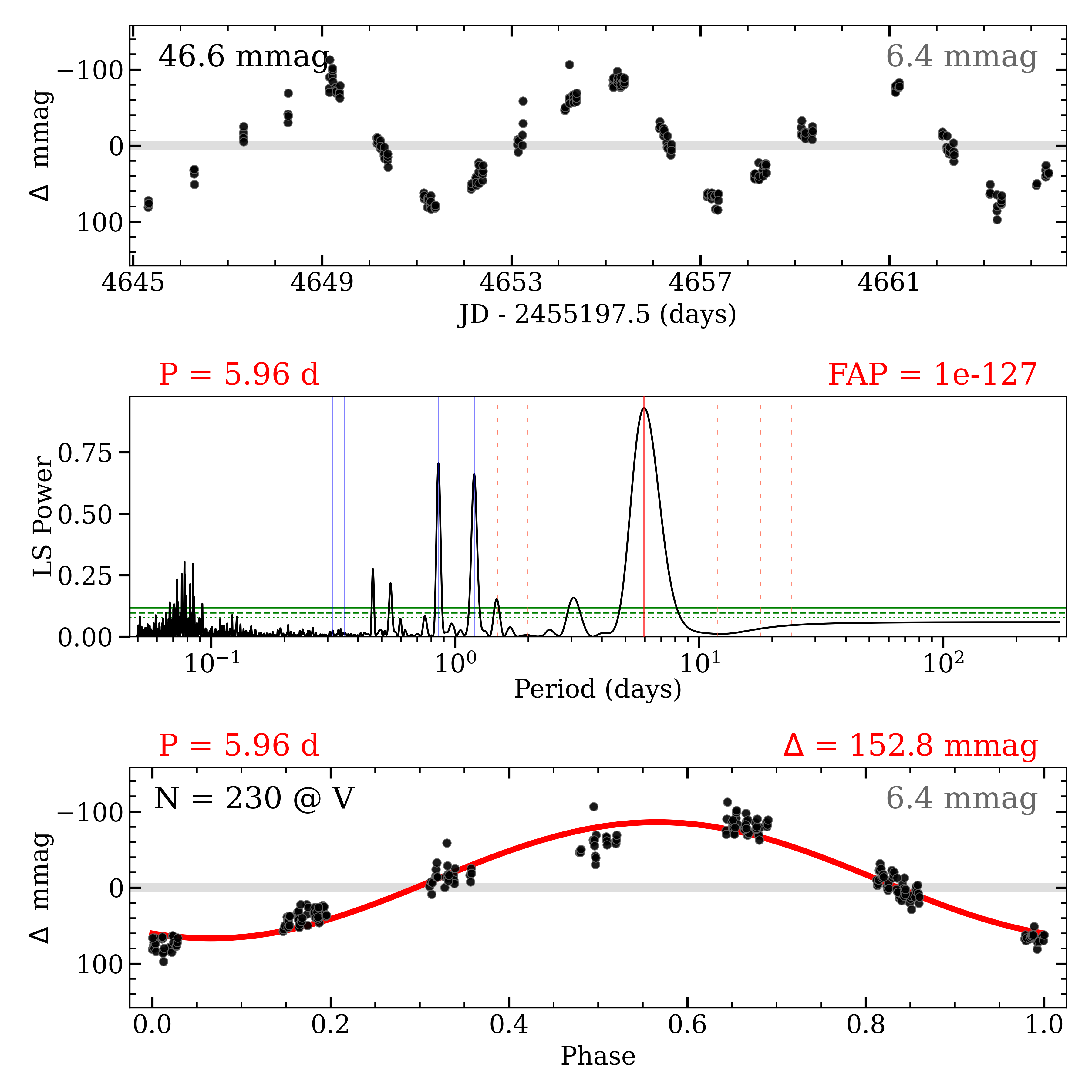}{0.47\textwidth}{$\uparrow$  2MA 0201+0117 A  $\uparrow$}
          \fig{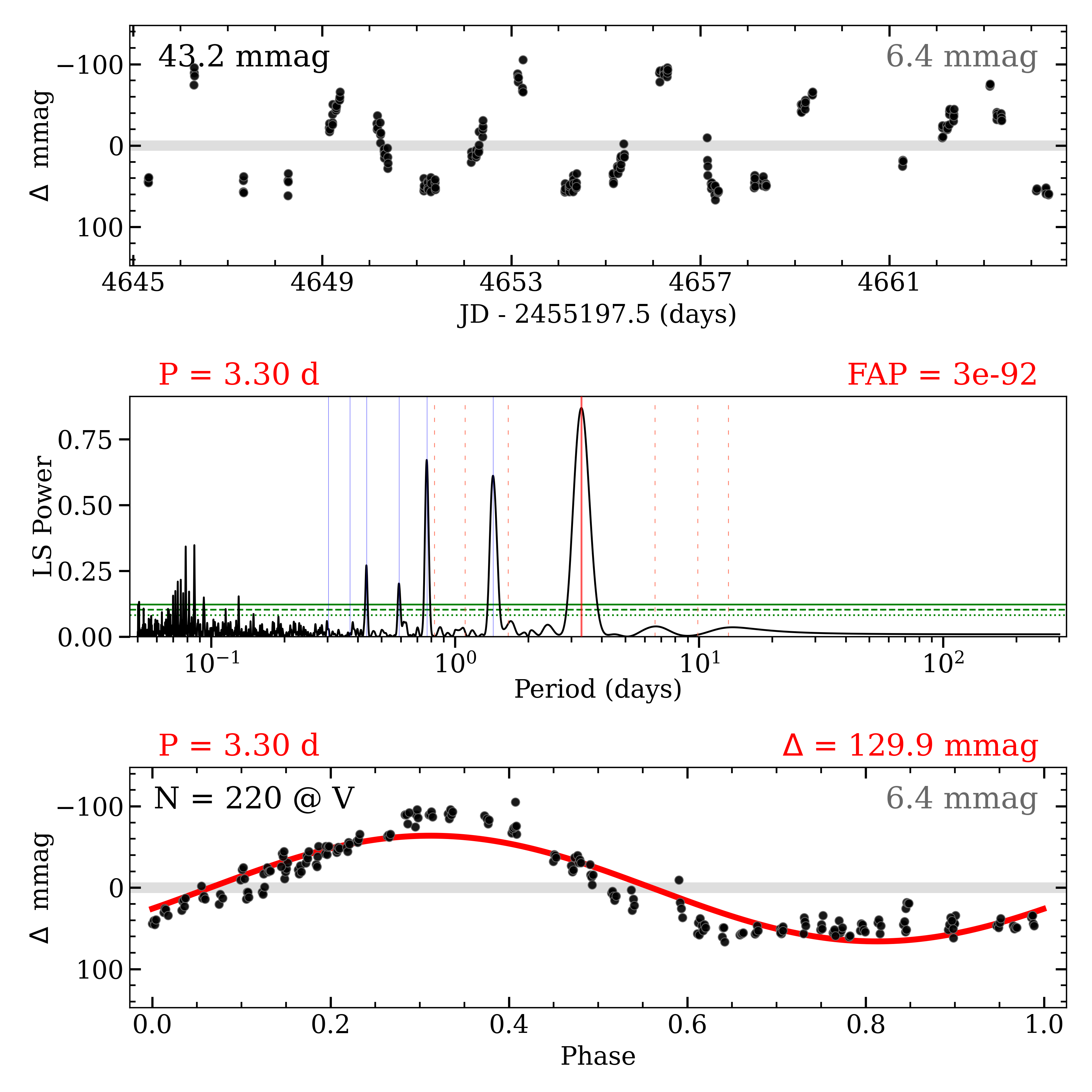}{0.47\textwidth}{$\uparrow$  2MA 0201+0117 B  $\uparrow$}}
\gridline{\fig{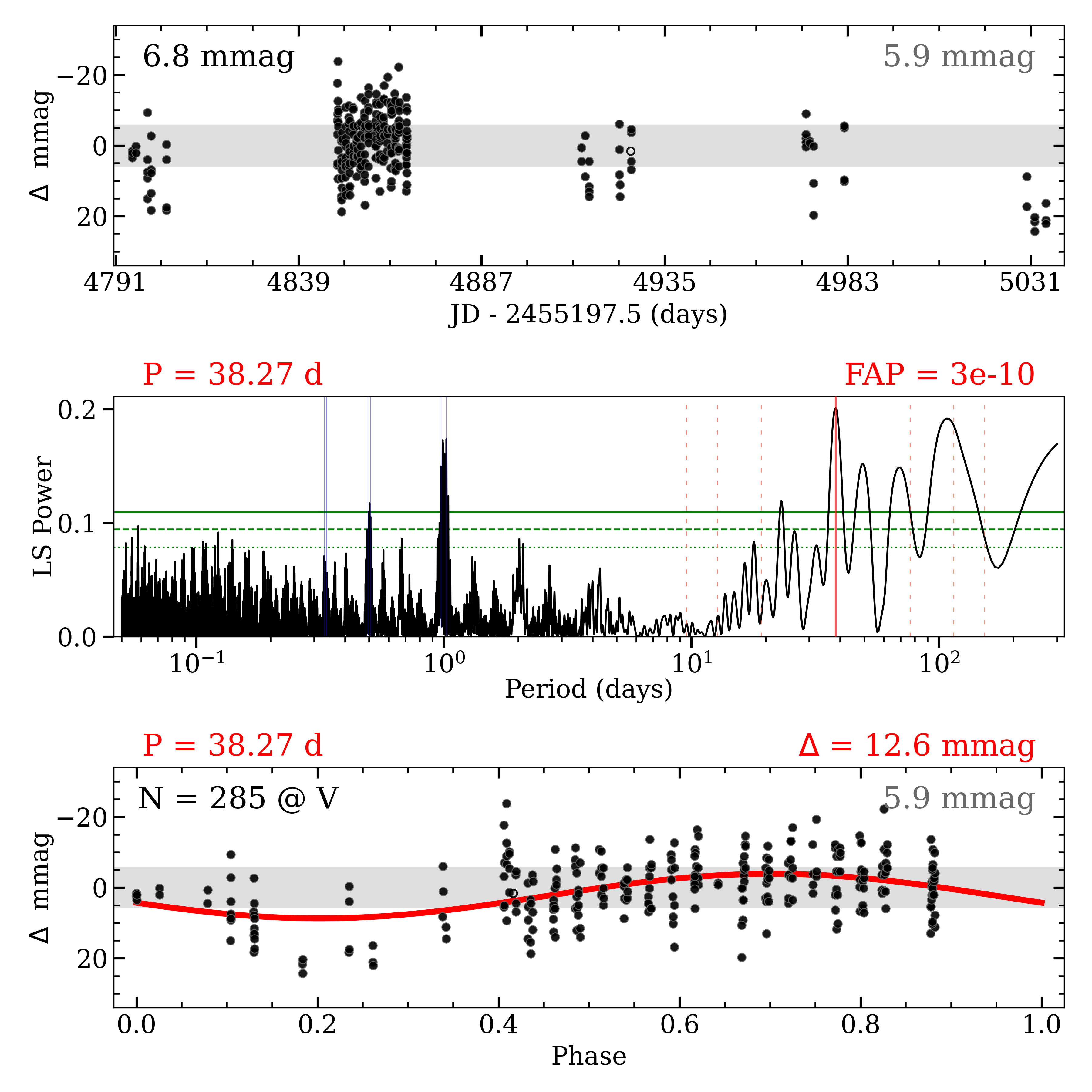}{0.47\textwidth}{$\uparrow$  NLTT 44989 A  $\uparrow$}
          \fig{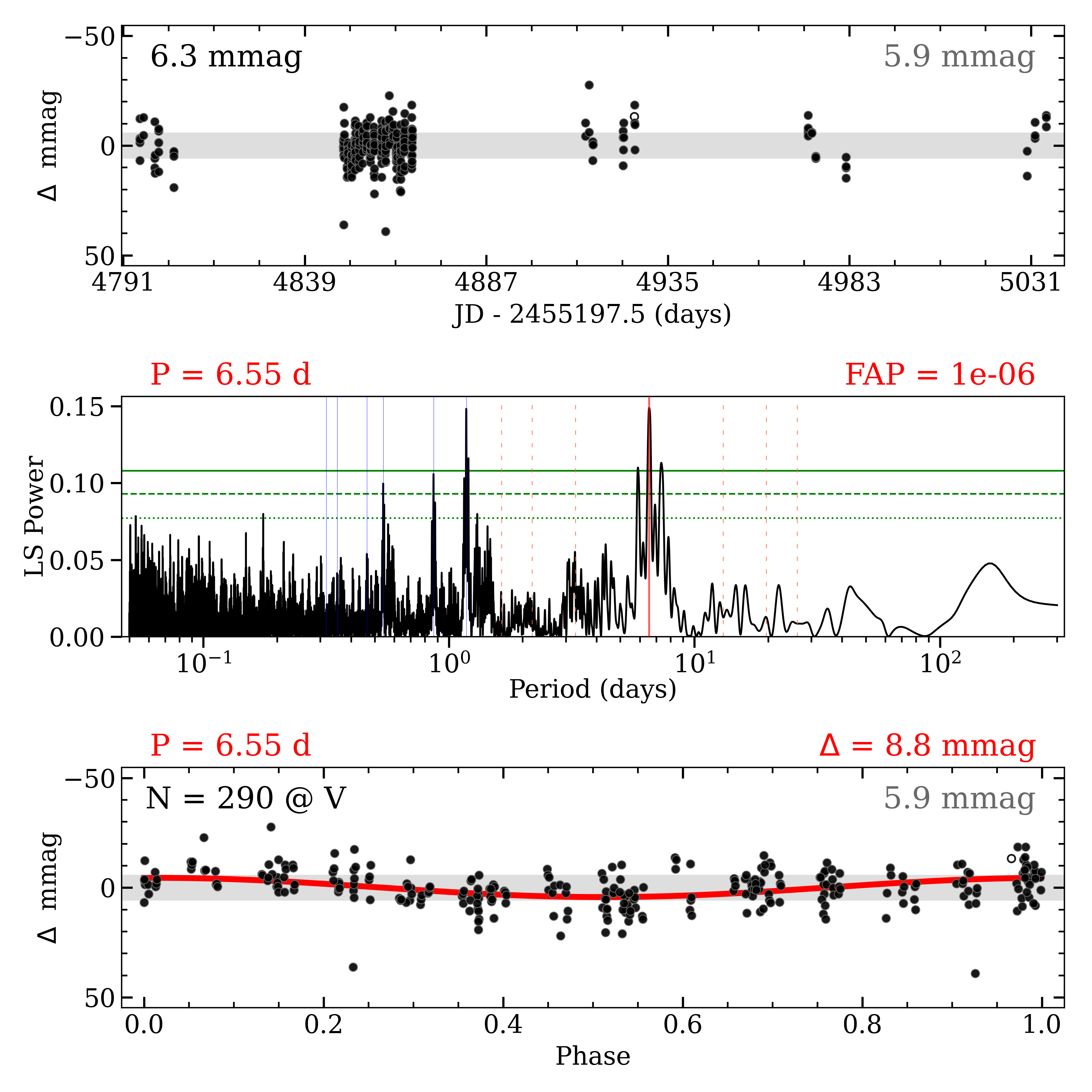}{0.47\textwidth}{$\uparrow$  NLTT 44989 B  $\uparrow$}}
\figcaption{\textit{(contd.)}}
\end{figure}

\begin{figure}[!htb]
\centering
\gridline{\fig{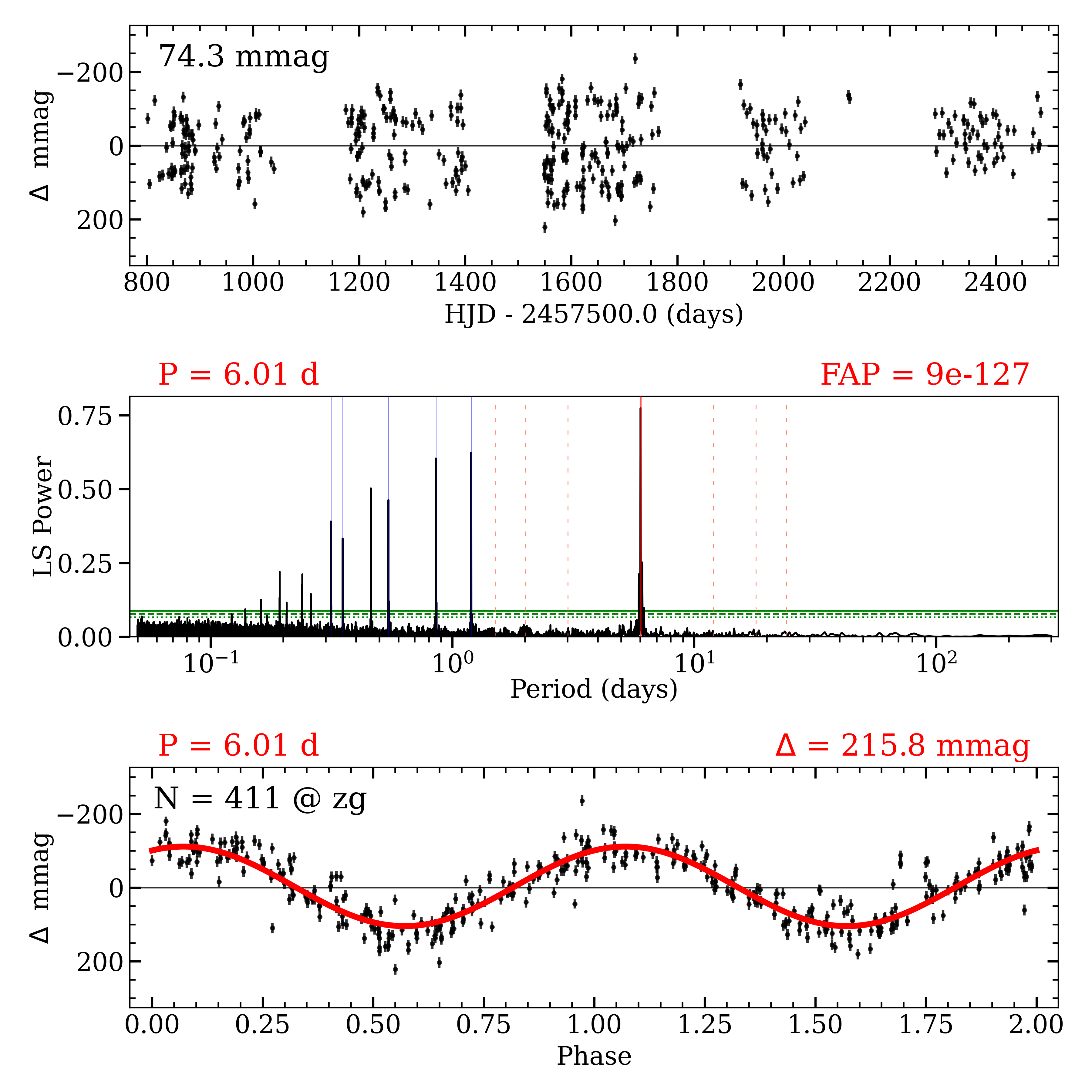}{0.49\textwidth}{$\uparrow$  2MA 0201+0117 A : ZTF  $\uparrow$}
          \fig{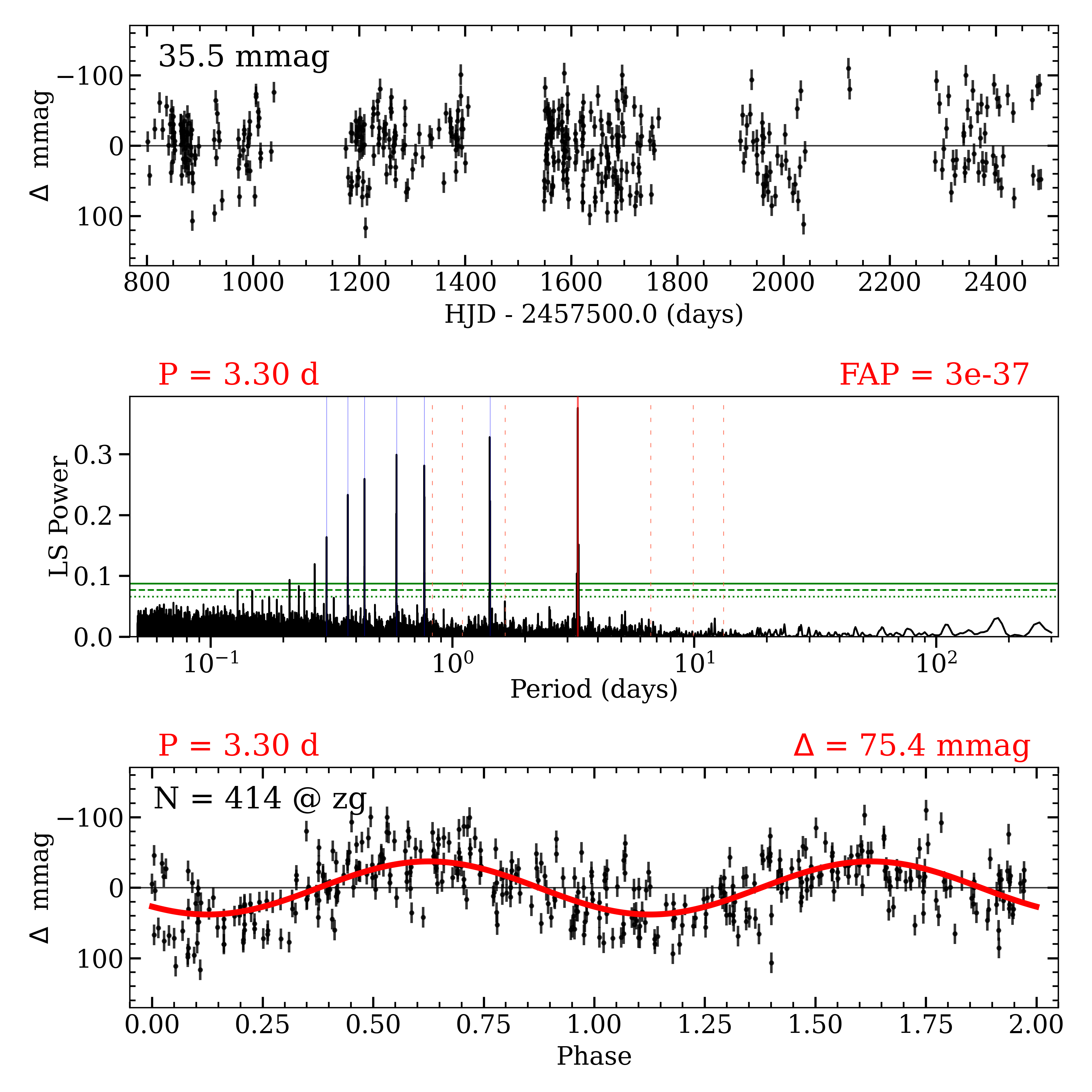}{0.49\textwidth}{$\uparrow$  2MA 0201+0117 B : ZTF  $\uparrow$}}
\gridline{\fig{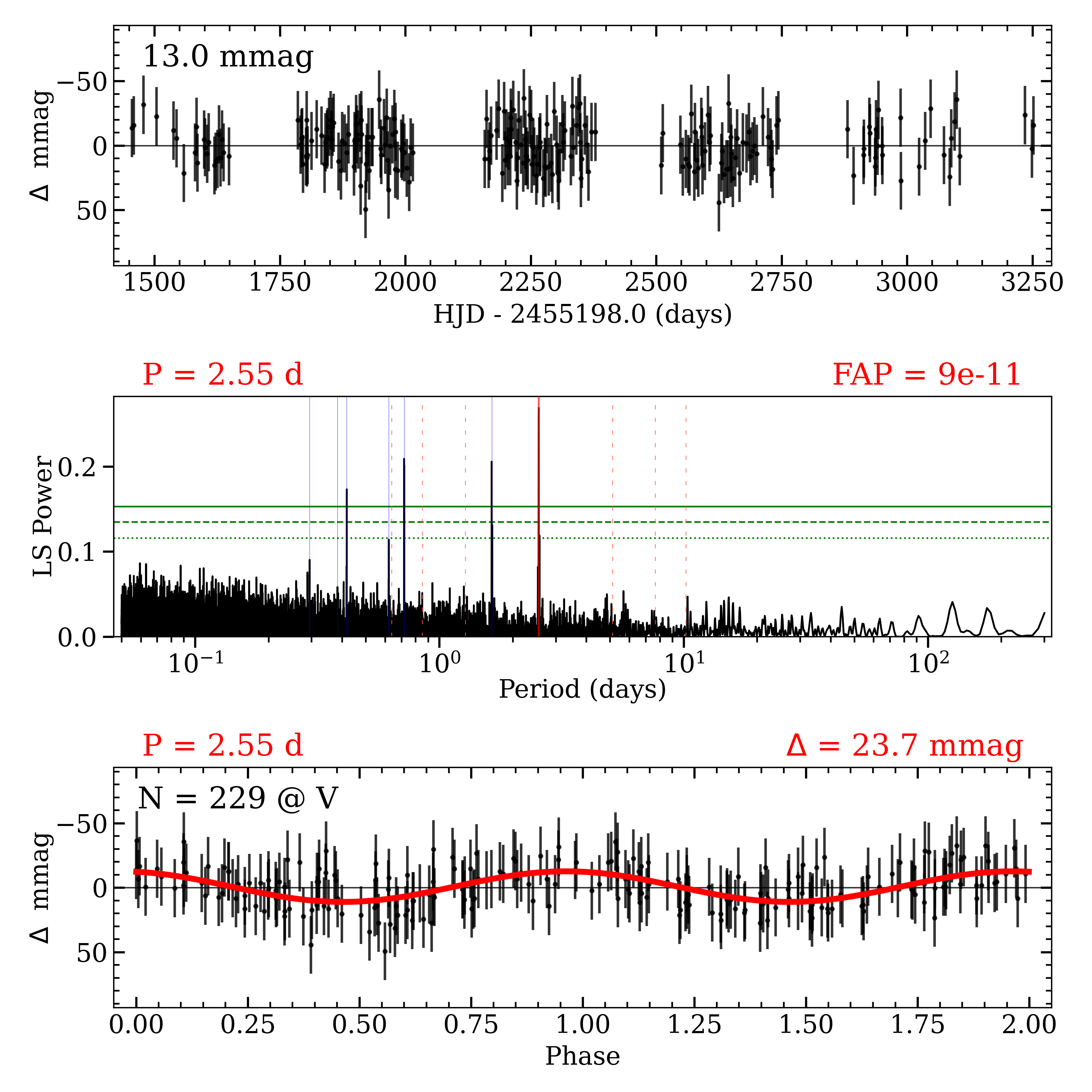}{0.49\textwidth}{$\uparrow$  KX Com ABC : ASAS-SN  $\uparrow$}
          \fig{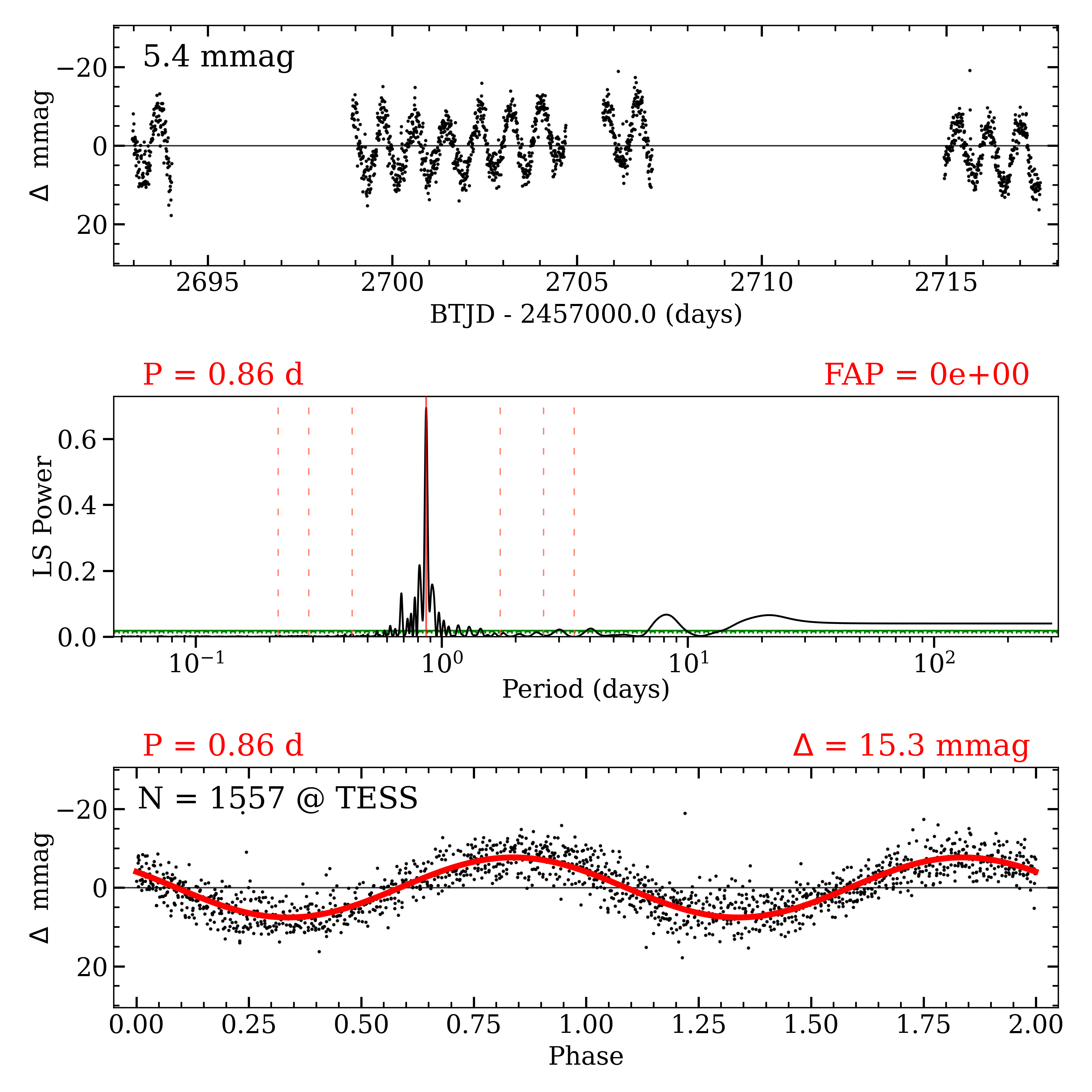}{0.49\textwidth}{$\uparrow$  GJ 1183 AB : \textit{TESS}-unpopular  $\uparrow$}}
\figcaption{Format details are largely the same as for Figure \ref{fig:09m-rot}, now showing example results using archival data from ZTF (top left and top right), ASAS-SN (bottom left), and \textit{TESS}-unpopular (bottom right). The phased light curves show data folded every two phases in time instead of one phase for visual clarity. \textit{TESS} is also space-based without 1-day sampling, so does not show the corresponding alias lines. ZTF spatially resolves 2MA~0201+0117~A and B, while the other two examples have the A-B(C) components blended. \label{fig:external-rot}}
\end{figure}

\begin{figure}[!htbp]
\centering
\gridline{\fig{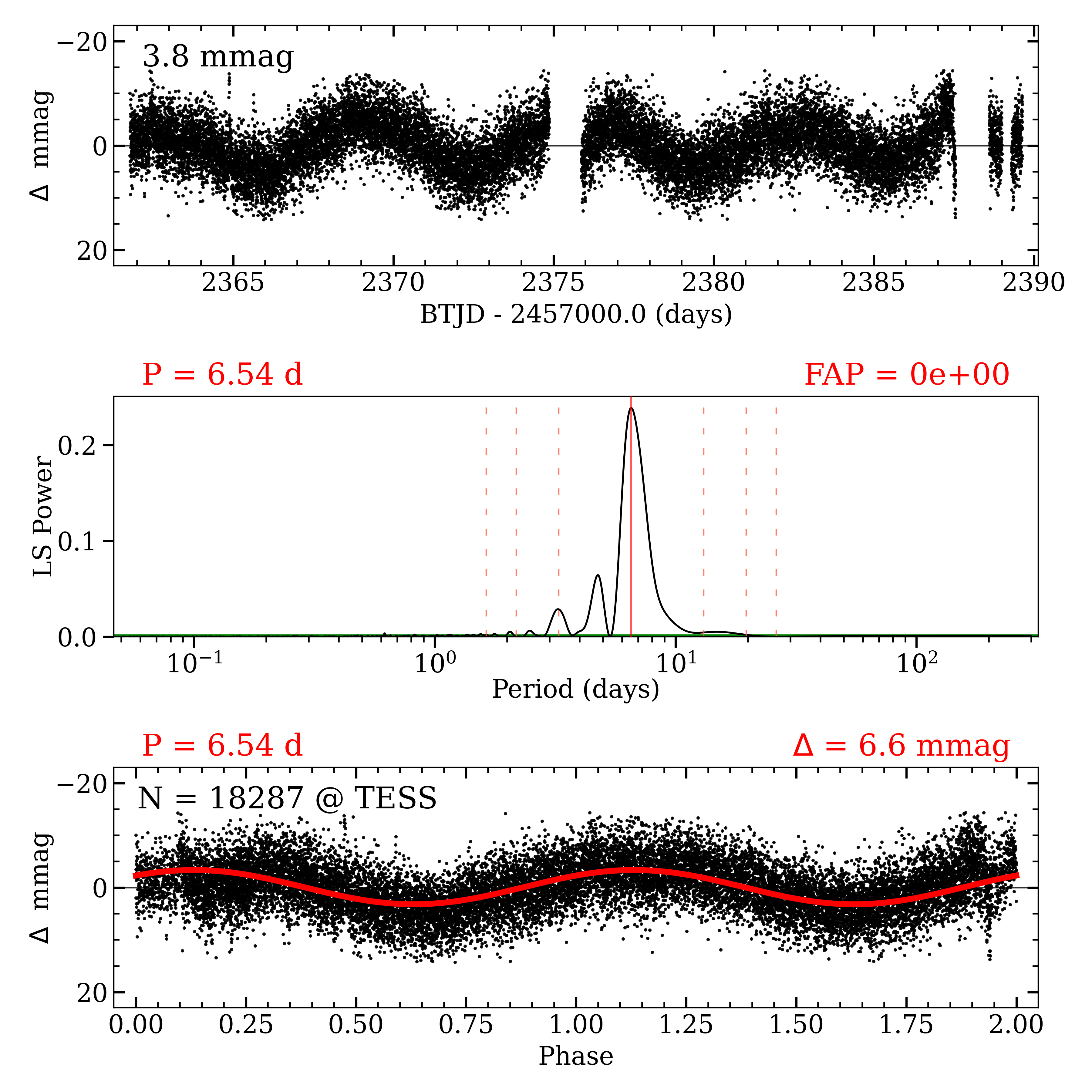}{0.49\textwidth}{$\uparrow$  NLTT 44989 AB : \textit{TESS}-2min  $\uparrow$}
          \fig{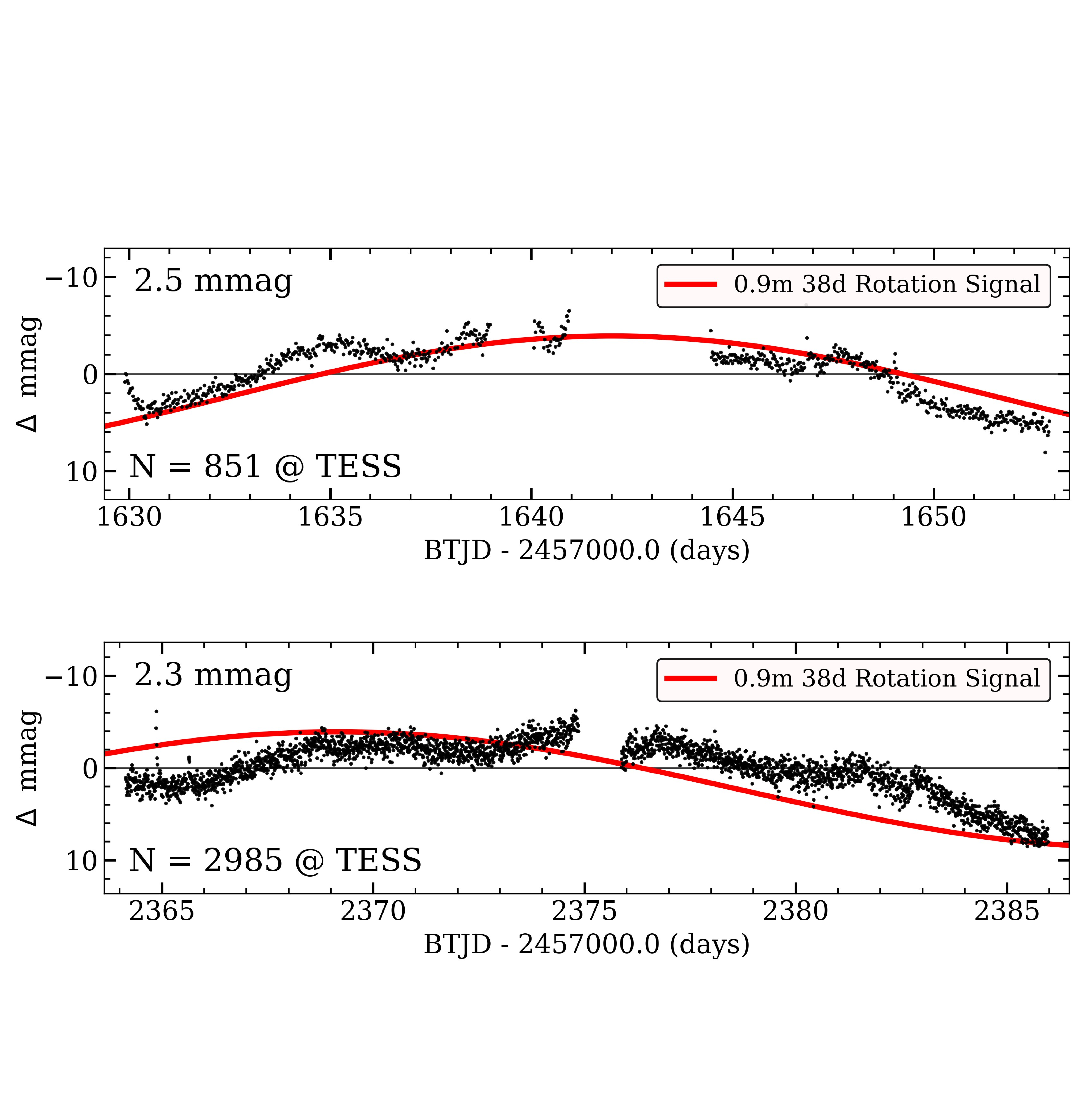}{0.49\textwidth}{$\uparrow$  NLTT 44989 AB : \textit{TESS}-unpopular  $\uparrow$}}
\figcaption{Blended light curves for the NLTT~44989~AB system. (Left) --- \textit{TESS} 2-minute cadence PDCSAP pipeline data from sector 39, shown in the same general format as Figure \ref{fig:09m-rot} and Figure \ref{fig:external-rot}. The $\sim$6.55d rotation period from the B component is plainly evident, while the A component's longer signal is absent due to the default \textit{TESS} pipeline processing. (Right) --- Alternative \textit{TESS}-unpopular light curves from sectors 12 (top) and 39 (bottom), with the MADs given in the top left and numbers of data points (N) in the bottom left. The shorter periodic rotation pattern of $\sim$6.55d from the B component is weakly visible intermixed with the more obvious long-term rotation signal of the A component. Red curves overlay the 38.27d Lomb-Scargle sine wave result from the 0.9m data (Fig.~\ref{fig:09m-rot}) but extended in time to compare the phase alignment with the signal in each \textit{TESS}-unpopular sector; the small shift in phase in sector 39 is likely explained by a small offset between the adopted and true period, or possibly differential rotation. See \S\ref{subsec:rot-results} for further discussion. \label{fig:NLTT-AB-tess}}
\end{figure}

{\bf NLTT~44989~A} has an uncertain low-amplitude period near 38d (or longer) in the resolved 0.9m data. The blended PDCSAP \textit{TESS} data don't show the long period trend (see Fig.~\ref{fig:NLTT-AB-tess}), presumably because it is removed by the default \textit{TESS} pipeline, but we note a long-term signal does appear in both sectors' raw SAP light curves (not shown here). The \textit{TESS}-unpopular data, which better preserve long-term astrophysical trends, are shown on the right in Figure \ref{fig:NLTT-AB-tess} --- a candidate long-term signal $\gtrsim$22d is seen mixed with the B component's shorter 6.55d variations across two non-consecutive sectors. A Lomb-Scargle analysis of the two sectors' unpopular data merged indicates a possible period of roughly 20--100d. Our \textit{TESS}-unpopular aperture for NLTT~44989~A (with B blended) was a 3$\times$3 pixel grid (63$\times$63 arcseconds), capturing many faint background stars from the dense field, although A and B are still the brightest sources in the aperture. To support the trend's connection to the A component instead of a background star, we re-analyzed the \textit{TESS}-unpopular data using 16 different aperture configurations drawn from the same set of 3$\times$3 pixels in various arrangements; the long-term signal remained evident in both sectors for all 16 cases. In Figure \ref{fig:NLTT-AB-tess} we overlay the 38d 0.9m signal extended in time to assess the phase alignment with each sector's turnover trend, finding generally good agreement despite the two sectors and 0.9m data all having been observed several years apart from each other --- the small phase mismatch in sector 39 is possibly due to our measured period deviating slightly from the true value, or differential rotation could be occurring. We also note that as we acquired additional 0.9m data for NLTT~44989~A the periodogram peak around $\sim$38d generally strengthened relative to other peaks. Altogether, while the spatially-resolved 0.9m period detection is weak, it is congruent in period and phase with the \textit{TESS}-unpopular signal and with the low activity levels we observe for the star in H$\alpha$ and $L_X$ (\S\ref{sec:rot-activity-results}), so we adopt 38d as our final period for NLTT~44989~A. A rotation period shorter than $\sim$38d would move the star to a unique position in the rotation-activity plane for both H$\alpha$ and $L_X$ (see Figure \ref{fig:Prot-activity}, discussed in \S\ref{sec:rot-activity-results}), where no other field stars are observed to exist, further supporting the long period we adopt. A period longer than 38d may be possible but would actually compound the A/B mismatch and strengthen our overall findings.

{\bf NLTT~44989~B} shows a candidate low-amplitude 6.55d period from the 0.9m with a similar-strength alias peak of 1--1.5 days, but blended \textit{TESS} data reveal a confident signal around $\sim$6.5d with no significant peaks in the 1--1.5d region, confirming 6.55d as the true signal for B.

\subsection{H$\alpha$ Equivalent Widths \& Radial Velocities - CTIO/SMARTS 1.5m \& CHIRON} \label{subsec:chiron-results}

Results from the optical spectroscopy effort are summarized in Table \ref{tab:MeanSpecTable}, with individual epochal values outlined in Table \ref{tab:AllSpecTable} and available in the online journal.  For the four systems considered here, we measure H$\alpha$ EWs spanning $-$15.03 (a flare epoch) to 0.31~$\mathrm{\AA}$ and reach typical single-visit RV precision of 210~m/s and averages of 20--60 m/s. Our CHIRON RVs are in good agreement with the \textit{Gaia} DR3 RVs (provided in Table \ref{tab:SampleTable-astr}) when available. Our $v\sin(i)$ measurements span 1.41--15.84~km/s, but we caution that any $v\sin(i)$ values below $\sim$10~km/s should be considered less reliable, per J23, and these are enclosed in parentheses in Table \ref{tab:MeanSpecTable}.
%
\begin{deluxetable}{lccrcrcrc}[!t]
\tablewidth{0pt}
\tablecaption{Optical Spectra --- Mean CHIRON Measurements\label{tab:MeanSpecTable}}
\tablehead{
\colhead{Name} & \colhead{N$_{All}$} & \colhead{N$_{H\alpha}$} & \colhead{$\overline{EW_{H\alpha}}$} & \colhead{$EW_{H\alpha}$ Lo-Hi} & \colhead{$\overline{RV}$} & \colhead{$\sigma_{\overline{RV}}$} & \colhead{$\overline{v\sin(i)}$} & \colhead{$\sigma_{\overline{v\sin(i)}}$} \\
\colhead{} & \colhead{} & \colhead{} & \colhead{[$\mathrm{\AA}$]} & \colhead{[$\mathrm{\AA}$]} & \colhead{[km/s]} & \colhead{[km/s]} & \colhead{[km/s]} & \colhead{[km/s]}
}
\startdata
     GJ 1183 A & 10 &  7 & $-$8.05 & [$-$7.65,$-$8.57] & $-$11.23 & 0.05 & 13.08  & 0.11 \\
     GJ 1183 B & 10 &  7 & $-$6.37 & [$-$5.61,$-$6.84] & $-$11.26 & 0.05 & 15.52  & 0.13 \\
      KX Com A &  8 &  7 & $-$3.71 & [$-$2.90,$-$4.61] &  $-$7.61 & 0.04 & (4.98) & \nodata \\
     KX Com BC & 30 &  7 & $-$0.53 & [$-$0.43,$-$0.82] &      var &  var &   var  &  var \\
2MA 0201+0117 A & 11 &  9 & $-$3.67 & [$-$3.18,$-$4.09] &     5.61 & 0.04 & (4.53) & \nodata \\
2MA 0201+0117 B & 10 &  9 & $-$5.60 & [$-$4.93,$-$6.49] &     6.36 & 0.06 & 10.52  & 0.24 \\
  NLTT 44989 A & 15 & 15 & $+$0.25 & [$+$0.31,$+$0.12] &    42.05 & 0.02 & (1.85) & \nodata \\
  NLTT 44989 B & 15 & 15 & $-$0.88 & [$-$0.54,$-$1.42] &    43.69 & 0.02 & (2.31) & \nodata \\
\enddata
\tablecomments{The H$\alpha$ EW mean and range values consider only the epochs (N$_{H\alpha}$) with both A and B observed back-to-back successfully and with neither flaring. RV and $v\sin(i)$ values use all available visits for each star (N$_{All}$) and give the weighted means and associated uncertainties. Square brackets indicate the range of observed H$\alpha$ values. $v\sin(i)$ values \textless 10~km/s are less reliable measurements, indicated with parentheses and exclusion of the unreliable uncertainties.}
\end{deluxetable}

At some epochs spectra were only successfully acquired for one component, either due to inclement weather or because one star had to be thrown out due to strong cosmic rays on H$\alpha$ or critical observational mistakes. The successful component's data are still useful for RV and $v\sin(i)$ analyses, but we entirely remove such epochs for any A-B H$\alpha$ activity comparisons to ensure we only ever compare H$\alpha$ data taken at consistent snapshots in time for both stars.
%
\begin{deluxetable}{cccccccccc}[!t]
\tablewidth{0pt}
\tablecaption{Optical Spectra --- All CHIRON Measurements\label{tab:AllSpecTable}}
\tablehead{
\colhead{Name} & \colhead{J. Epoch} & \colhead{Flare?} & \colhead{EW$_{H\alpha}$} & \colhead{$\sigma$-EW$_{H\alpha}$} & \colhead{RV} & \colhead{$\sigma$-RV} & \colhead{$v\sin(i)$} & \colhead{$\sigma$-$v\sin(i)$} & \colhead{SNR} \\
\colhead{} & \colhead{[YYYY.YYYY]} & \colhead{[y/n]} & \colhead{[$\mathrm{\AA}$]} & \colhead{[$\mathrm{\AA}$]} & \colhead{[km/s]} & \colhead{[km/s]} & \colhead{[km/s]} & \colhead{[km/s]} & \colhead{}
}
\startdata
      GJ1183A & 2022.2497 &  y & $-$15.03 &   0.01 & $-$11.03 & 0.14 & 13.28 &  0.38 &  32.8 \\
      GJ1183B & 2022.2497 &  n &  $-$6.91 &   0.02 & $-$11.23 & 0.08 & 15.57 &  0.38 &  33.0 \\
       KXComA & 2021.3732 &  n &  $-$3.63 &   0.02 &  $-$7.60 & 0.13 &  4.95 &  1.46 &  25.4 \\
      KXComBC & 2023.2708 &  y &  $-$5.36 &   0.03 &  $-$8.79 & 0.17 &  7.12 &  0.68 &  35.0 \\
2MA0201+0117A & 2019.9441 &  n &  $-$3.66 &   0.02 &     5.50 & 0.18 &  4.17 &  1.01 &  25.4 \\
2MA0201+0117B & 2019.9441 &  n &  $-$4.93 &   0.01 &     6.52 & 0.21 & 10.39 &  0.98 &  21.6 \\
   NLTT44989A & 2022.2473 &  n &  $+$0.23 &   0.02 &    42.08 & 0.05 &  1.85 &  0.52 &  35.2 \\
   NLTT44989B & 2022.2474 &  n &  $-$0.69 &   0.11 &    43.68 & 0.07 &  2.18 &  0.46 &  36.1 \\
\enddata
\tablecomments{One example set of CHIRON measurements is shown here for each component. The full table of all measurements is available in machine-readable form in the online Journal. The `Flare?' column indicates visits with an H$\alpha$ flare observed, with `y' for a flare and `n' for no flare. Any $v\sin(i)$ values less than 10~km/s should be treated as potentially unreliable, with weaker confidence as the value decreases. SNRs are for the continuum near H$\alpha$ and not the H$\alpha$ line itself.}
\end{deluxetable}

\begin{figure}[!t]
\centering
\gridline{\fig{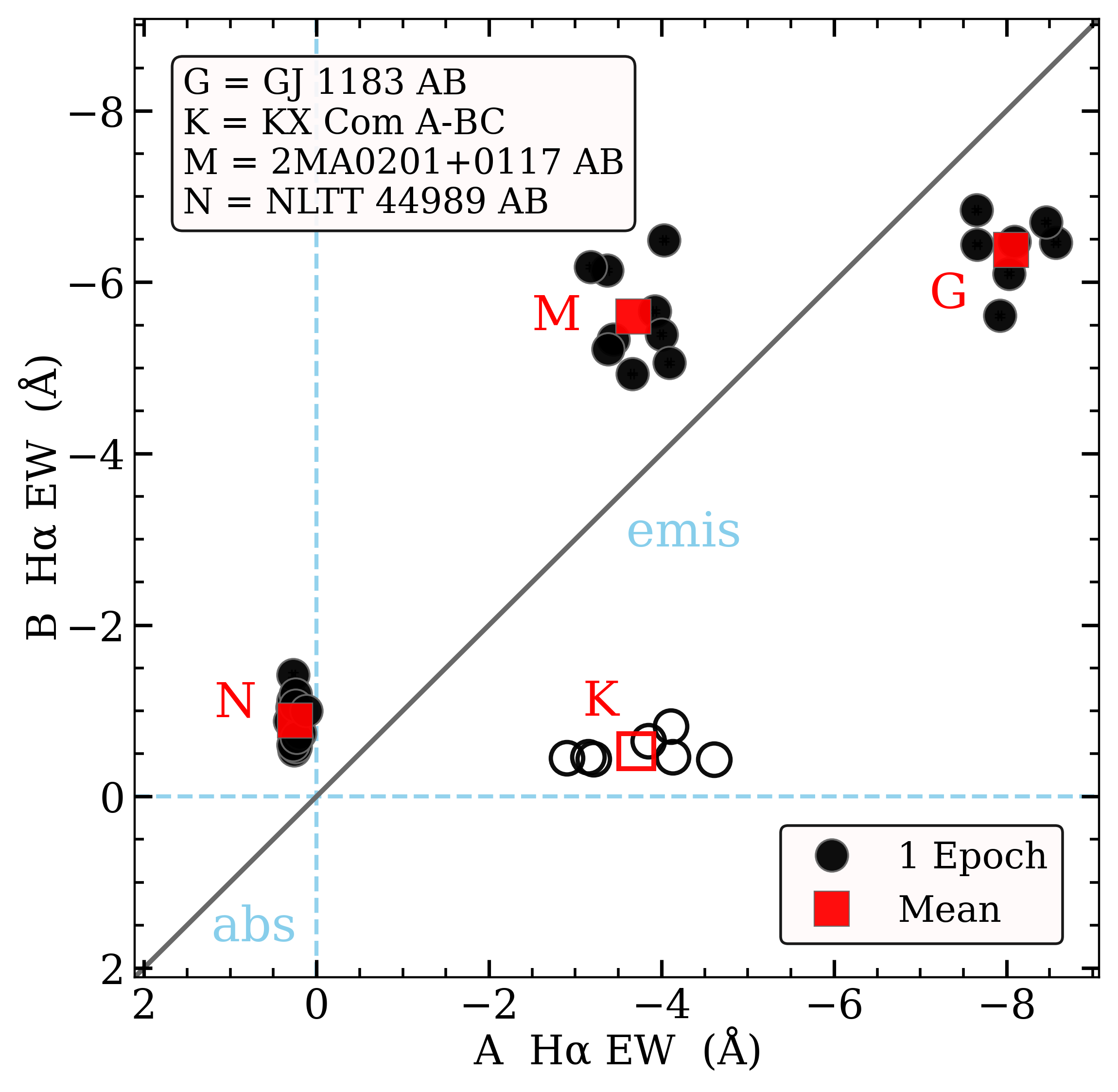}{0.47\textwidth}{}}
\vspace*{-7mm}
\figcaption{A one-to-one equivalency plot comparing H$\alpha$ EWs between components in each system. Individual epochal measurements are black circles while the means of these are red squares. Open symbols indicate the non-twin triple KX~Com~A-BC. Absorption and emission regions are indicated with blue lines and labels. We only include non-flaring epochs with both stars successfully observed back-to-back. \label{fig:Ha-equal}}
\end{figure}

We searched for Li 6708 doublet features in our spectra but found no confident detections, even for the 2MA~0201+0117~AB binary that is a member of the $\beta$ Pictoris association, nor for the GJ~1183~AB system that has components slightly elevated in the Hertzsprung–Russell Diagram (Fig.~\ref{fig:HRD}).

\subsubsection{H$\alpha$ Activity} \label{subsubsec:ha-results}

Before comparing H$\alpha$ in binary components, we first consider observations when any component is flaring.  A few visits captured prominent H$\alpha$ flares (indicated in Table \ref{tab:AllSpecTable}), which we identify as epochal EWs satisfying the following: (1) show emission ($EW < 0$), (2) have stronger emission at that epoch than the median EW for the star ($EW < EW_{Median}$), and (3) the EW changes by more than 3 times the typical scatter in EW for that star ($Abs(EW - EW_{Median}) > 3 \times MAD_{Median}$). These cutoffs were derived in part via manual inspection of all the spectra and EW timeseries to identify brightened outliers and flaring H$\alpha$ line profiles compared to non-flaring epochs for the same star. When comparing the H$\alpha$ activity between components, we exclude any epochs for both A and B if either star is flaring at an epoch.

\begin{figure}[!t]
\centering
\gridline{\fig{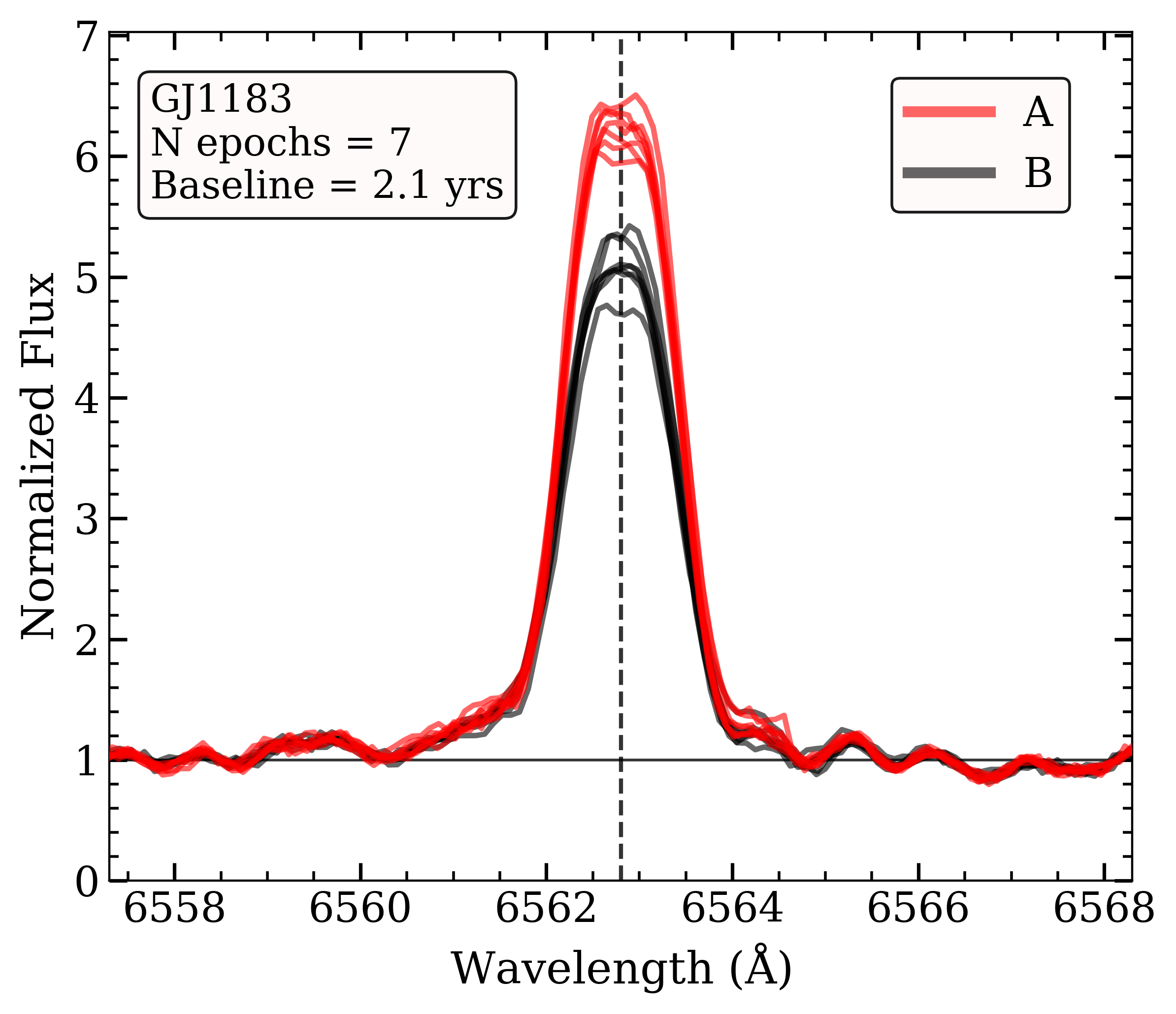}{0.49\textwidth}{}
          \fig{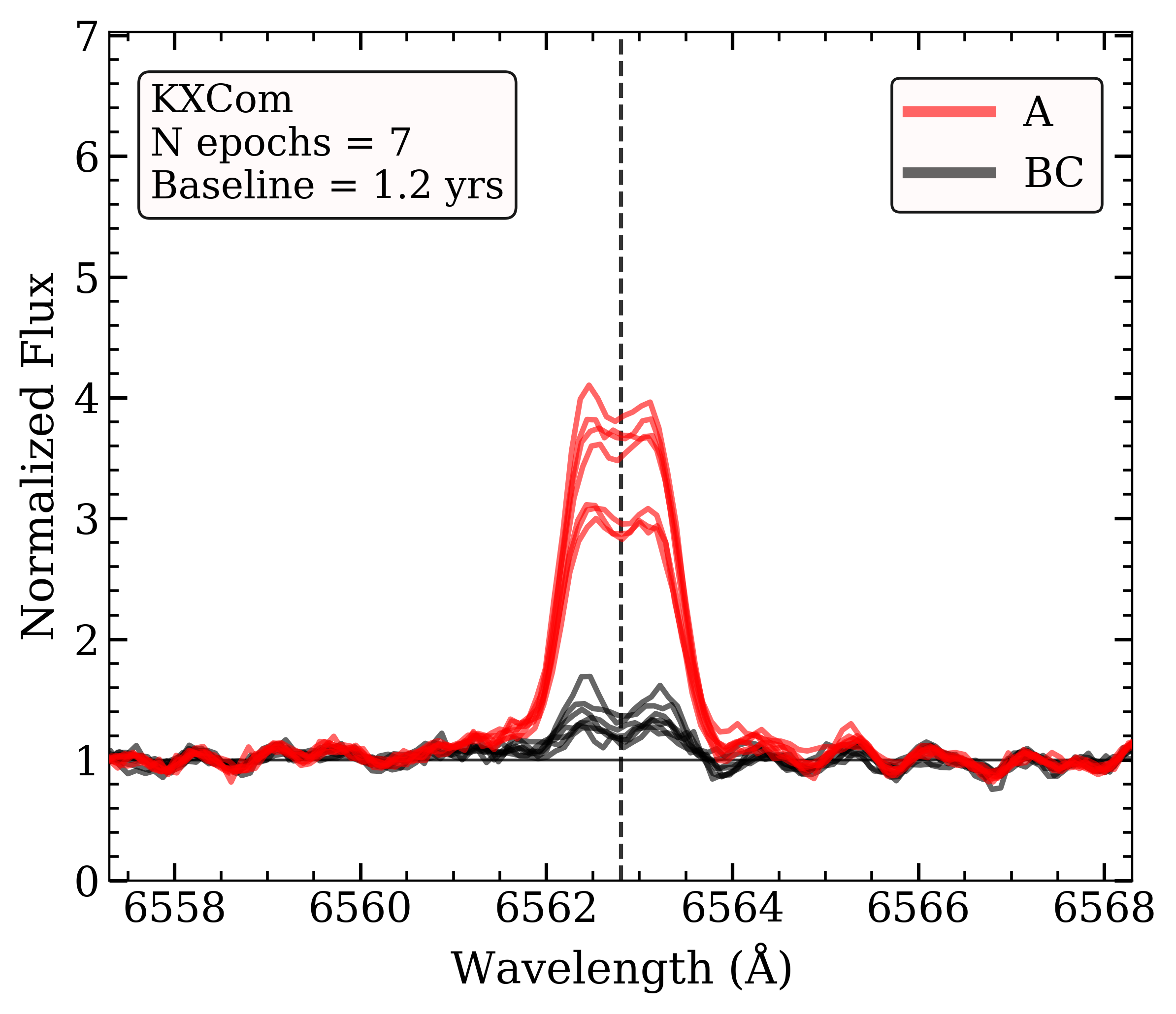}{0.49\textwidth}{}}
\vspace*{-9mm}
\gridline{\fig{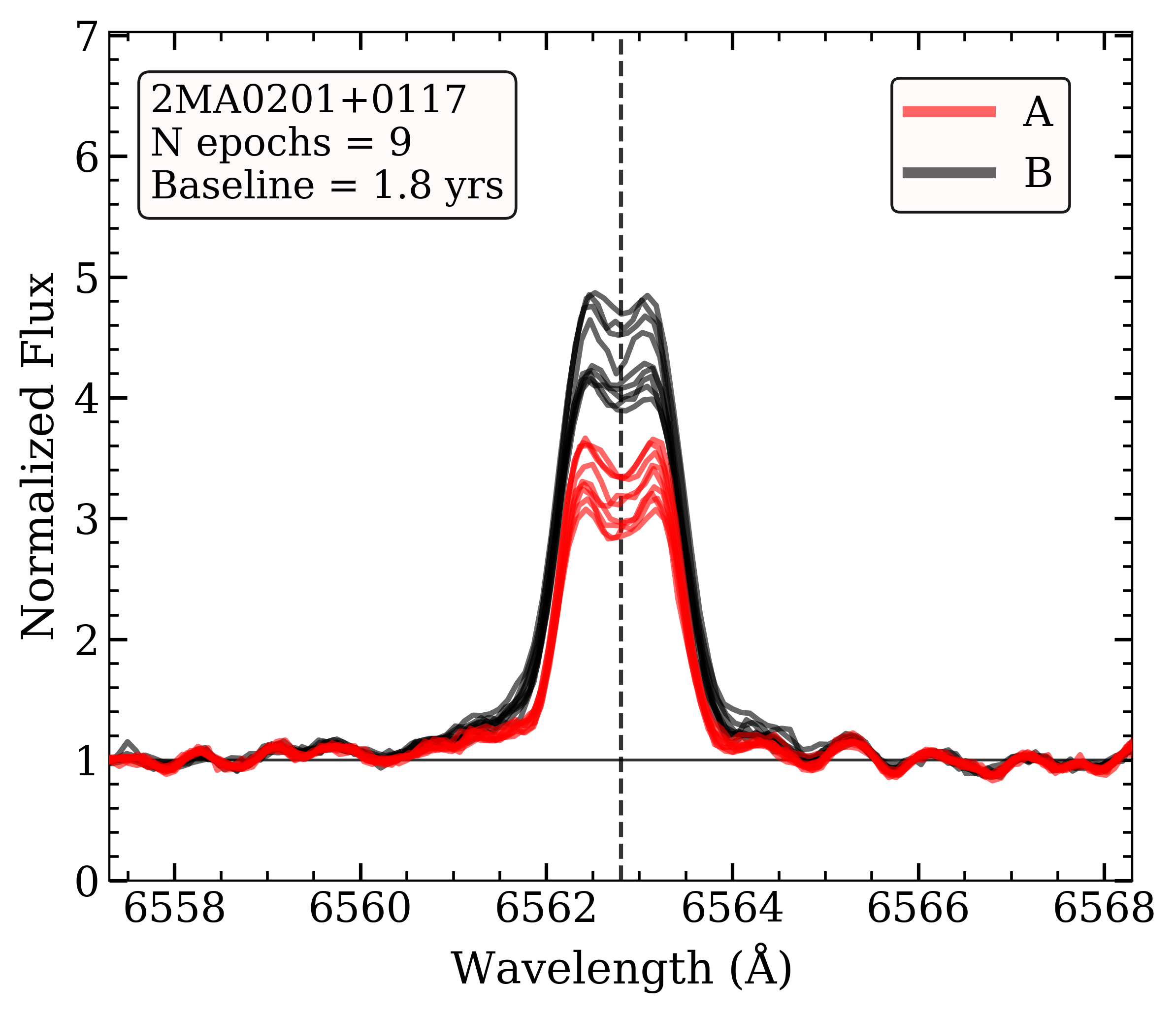}{0.49\textwidth}{}
          \fig{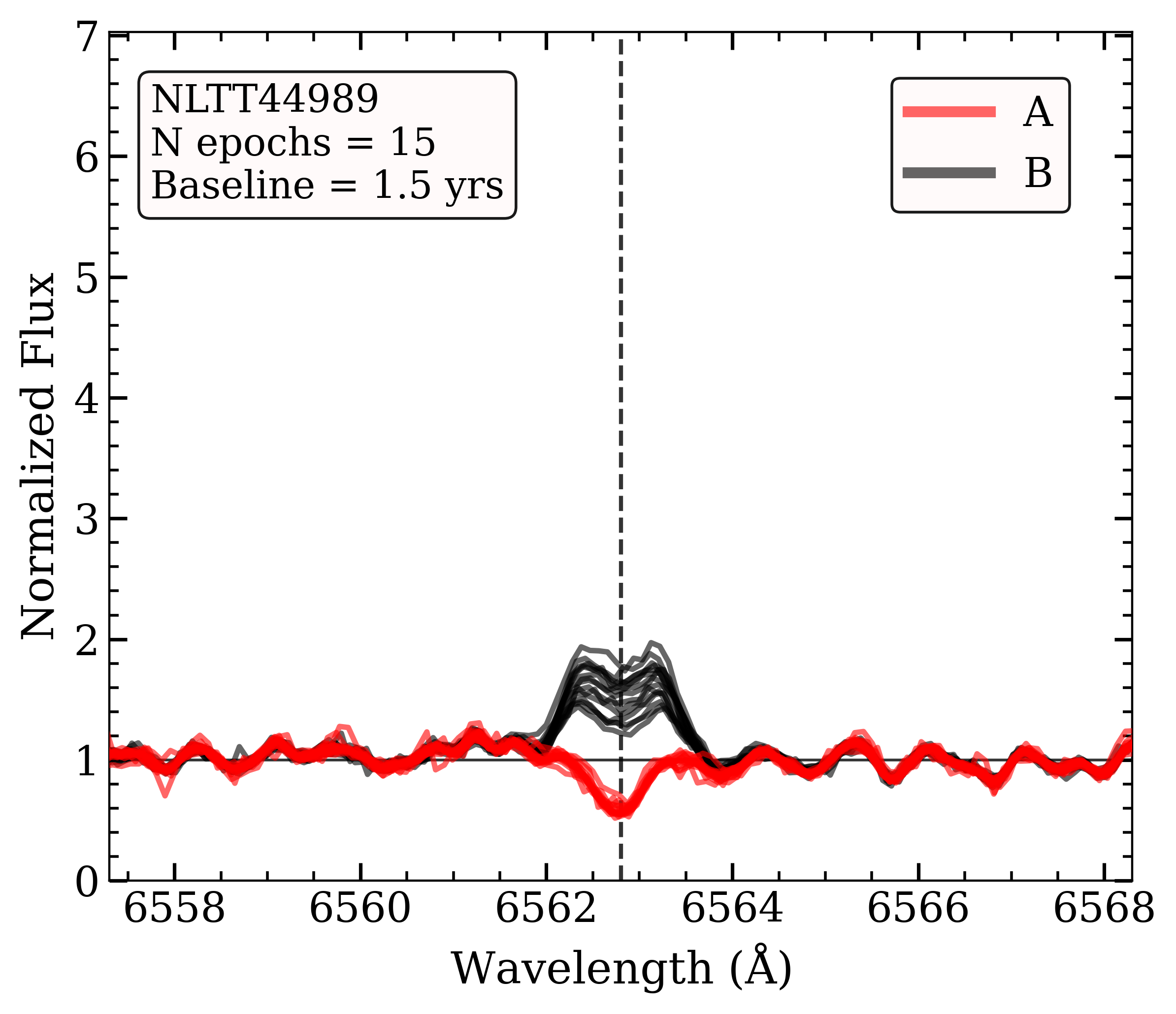}{0.49\textwidth}{}}
\vspace*{-7mm}
\figcaption{CHIRON spectra of the H$\alpha$ line region --- shifted to zero RV and blaze corrected --- stacked for multiple epochs to visually compare the A and B components in each system. We only include non-flaring epochs with both stars successfully observed back-to-back; the legends indicate the final number of epochs shown and the timespan of those spectra. Work by \cite{Pass_2024_ApJ} (P24) indicates the emission seen in the blended spectra for KX~Com~BC is primarily from the B star while C is flat (see \S\ref{subsec:KXCOMAB}). \label{fig:Ha-spec}}
\end{figure}

H$\alpha$ EWs are compared between A and B components with an equivalency plot in Figure \ref{fig:Ha-equal}. The scatter for any individual star indicates the intrinsic H$\alpha$ variability of the source. Even accounting for this scatter, it is clear that sustained differences exist between components in all four systems examined, i.e., none of the points lie on the one-to-one line. The corresponding H$\alpha$ line profiles are compared between components in Figure \ref{fig:Ha-spec}, where the same conclusion is evident. NLTT~44989~AB offers a significant result, with one component in emission and the other in absorption. This disparity in H$\alpha$ implies very different levels of magnetically-induced chromospheric heating, representing a total mismatch in magnetic activity between twin stars.

Several stars in Figure \ref{fig:Ha-spec} display split-horned H$\alpha$ emission profiles. Such patterns are expected theoretically, as discussed in \cite{Cram_Mullan_1979}, due to non-LTE and optical depth effects in the heated chromospheres of active M dwarfs. Figure \ref{fig:Ha-spec} also demonstrates a remarkable overlap in the continuum `wiggles' between components in each system. The spectrum for a given star and epoch was RV shifted and blaze corrected in isolation, i.e., the analysis gave no consideration to the RV or spectral behavior at other epochs or in the other component. This means the overlap in continuum features is truly astrophysical and consistent over many visits (even in the RV-variable case of KX~Com~BC, see below), validating and underscoring the twin-like natures of our pairs.

\subsubsection{Radial Velocities} \label{subsubsec:rv-results}

\begin{figure}[!t]
\centering
\gridline{\fig{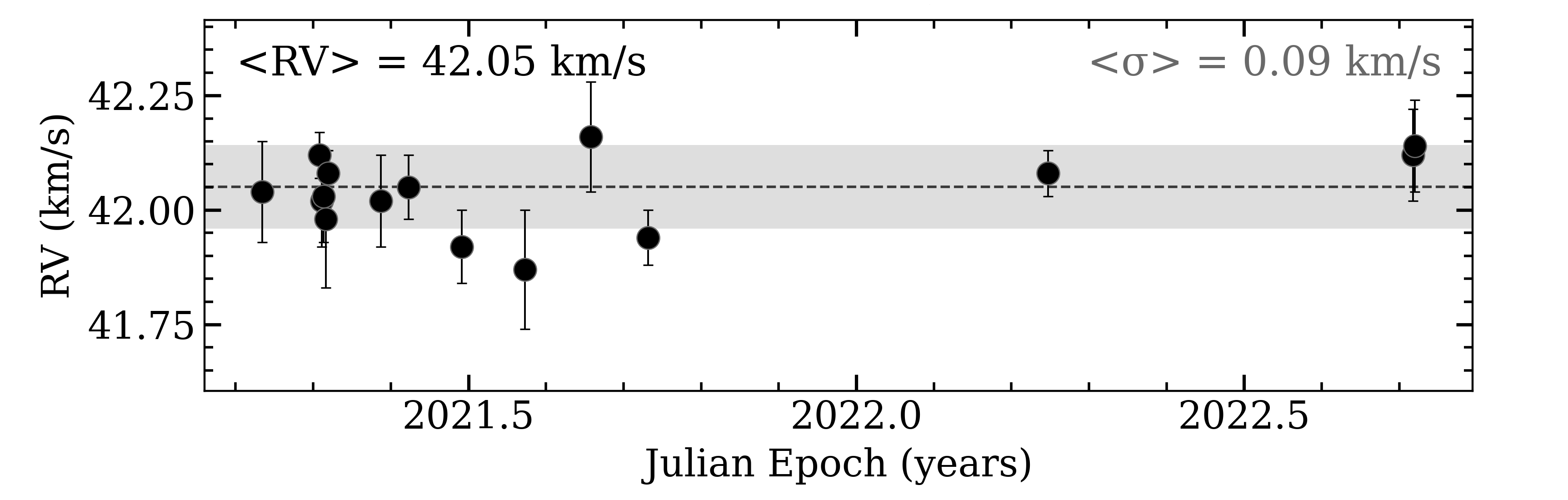}{0.49\textwidth}{(NLTT 44989 A)}
          \fig{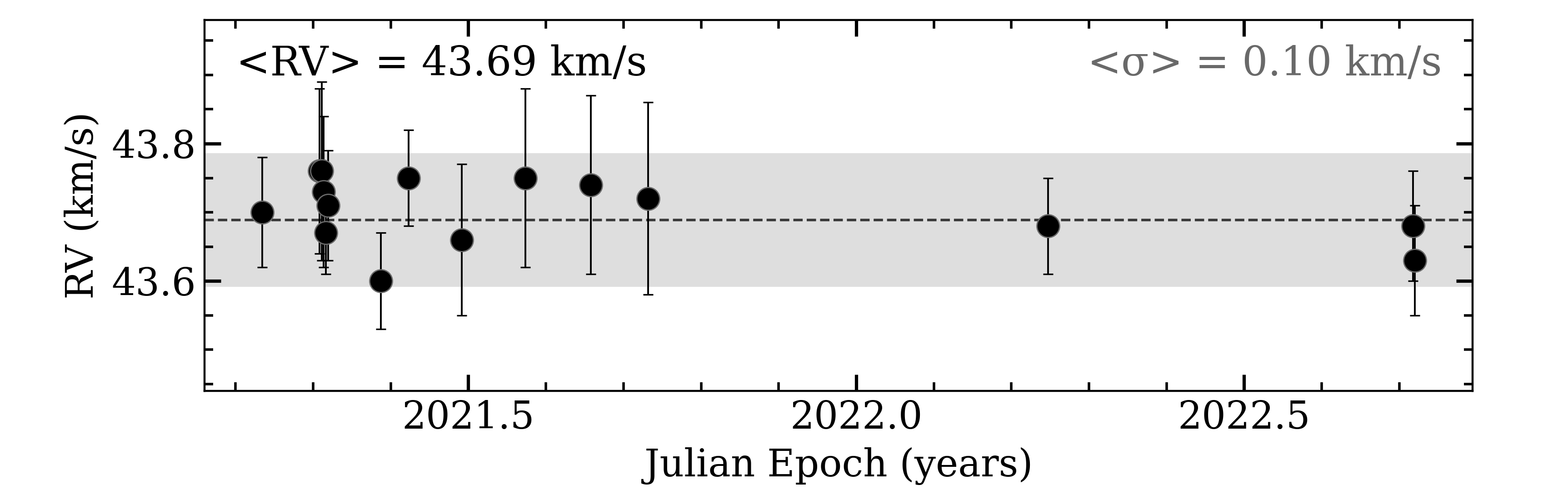}{0.49\textwidth}{(NLTT 44989 B)}}
\gridline{\fig{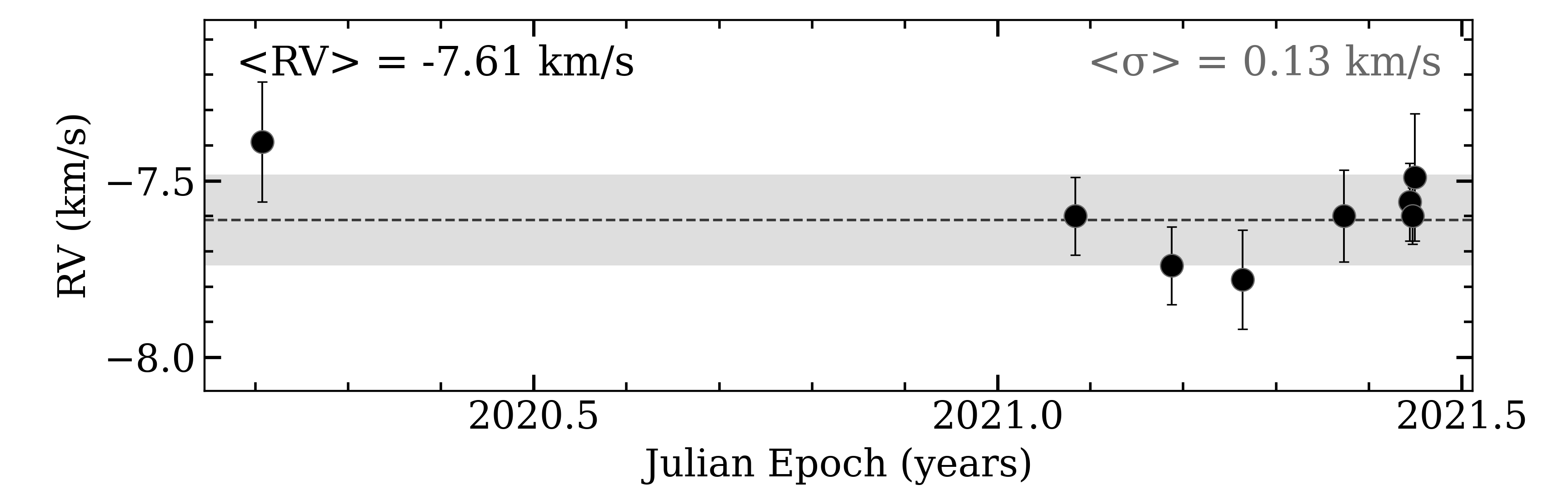}{0.49\textwidth}{(KX Com A)}
          \fig{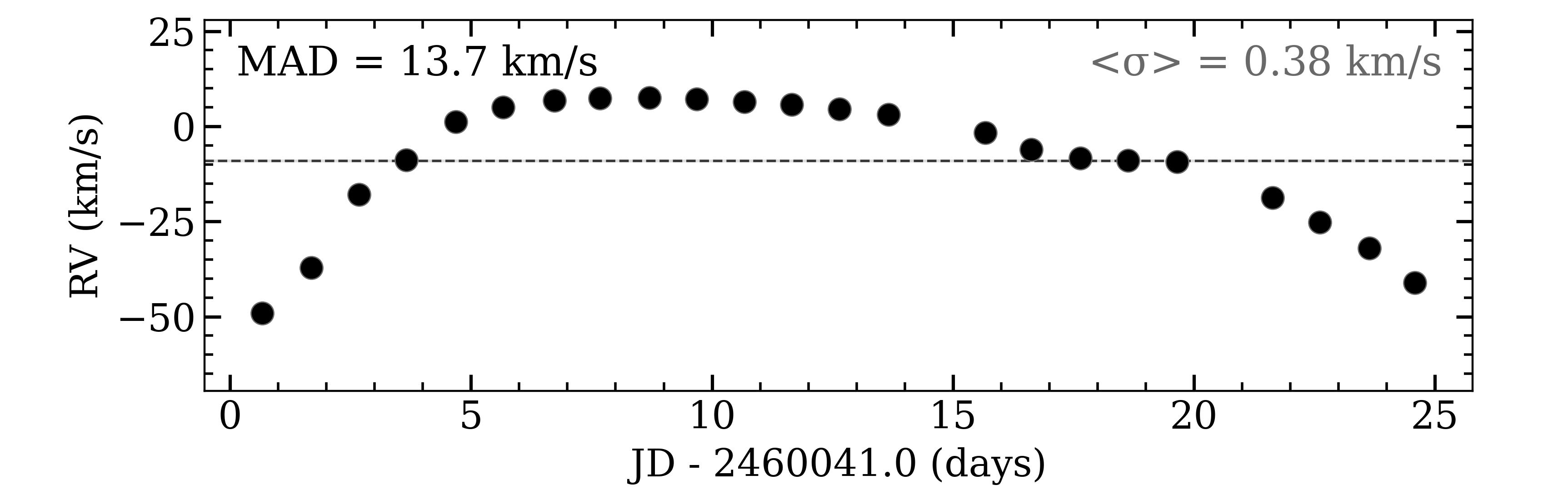}{0.49\textwidth}{(KX Com BC)}}
\figcaption{Radial velocity timeseries from CHIRON for four resolved targets. The weighted average RV value for each case is shown with a black dashed line and given in black in the top left (except KX~Com~BC listing the MAD instead), with the average single-point uncertainty spanning above and below in grey and given in the top right. Error bars are always shown, but appear smaller than the points for KX~Com~BC owing to the different scale. (Top) --- NLTT~44989~A and B show all 15 epochs across roughly 1.5 years, with neither star showing RV variations beyond the noise. (Bottom) --- KX~Com~A also appears non-varying within the noise for the eight available epochs. In contrast, KX~Com `B' shows the only case we find with RV variations, which we ascribe to orbital motion with an unresolved third component `C'. We only show 23 of the 30 total epochs available for KX~Com~BC so as to highlight the orbital arc captured from our sequential visits throughout April 2023. \label{fig:RV-Timeseries}}
\end{figure}

Of the eight components in four systems, only one shows RV variations beyond the measurement uncertainties, KX~Com~B. As examples, the flat radial velocity curves for NLTT~44989~A, NLTT~44989~B, and KX~Com~A are shown in Figure \ref{fig:RV-Timeseries}. The additional visits we acquired for KX~Com~B to further investigate its RV signal are shown in the bottom right of Figure \ref{fig:RV-Timeseries} and confirm the presence of a new orbiting companion we dub `C' --- this makes the overall KX~Com~A-BC system a hierarchical triple. Our data provide a lower limit of $\sim$24 days on the BC orbital period.

The $v\sin(i)$ measures for KX~Com~BC vary from 2.64 to 15.84~km/s and double-lined SB2 behavior is seen. Our RV analysis methodology did not explicitly take this into account for KX~Com~BC, but the H$\alpha$ spectra shown in Figure \ref{fig:Ha-spec} do not show marked RV misalignment, which might be expected if the RV measures were poor. Improved RVs from our data are possible for BC, but this is left as future work. The SB2 nature was further confirmed and the orbit analyzed using higher resolution spectra in recent work by P24, discussed further in \S\ref{subsec:KXCOMAB}.

\subsection{X-rays - \textit{Chandra} Observatory} \label{subsec:chandra-results}

Our \textit{Chandra} X-ray fluxes ($F_X$) in the 0.3--10~keV band and \textit{Gaia} DR3 parallaxes were used to calculate the X-ray luminosities ($L_X$) given in Table \ref{tab:ChandraTable}. The APEC-fit coronal normalization ($Norm$) values were converted to Volume Emission Measures\footnote{This is sometimes labeled $EM$ in other works instead of $VEM$. It parameterizes the coronal plasma's electron density and volume, effectively tracing the amount of emitting material.} ($VEM$) using $Norm = 10^{-14}$ $VEM/{4\pi d^2}$, per XSPEC/APEC documentation, and the VEM values are also given in Table \ref{tab:ChandraTable}.

There are individual distance measurements for each star in a given system, but components should be at functionally the same distance. We evaluated the implications of measured distance offsets by calculating $L_X$ and $VEM$ while fixing both components to the A then B star distances and found changes always $<$2\%, much smaller than the $L_X$ and $VEM$ uncertainties themselves, owing to the precise \textit{Gaia} parallaxes. Based on this, we chose to simply use the individual distance measurements for each star's calculations in our reported results. Our asymmetric uncertainties in flux ($\sigma_{F_X}$), $Norm$ ($\sigma_{Norm}$), and distance ($\sigma_d$) were used to calculate asymmetric $\sigma_{L_X}$ and $\sigma_{VEM}$ values using traditional error propagation methods on the upper and lower 1-$\sigma$ uncertainties in turn\footnote{This approach yields reasonable uncertainties despite its simplifying assumptions because the typically large $\sigma_{F_X}$ and $\sigma_{Norm}$ radically dominate over the \textit{Gaia} $\sigma_\pi$ errors that are $\sim$25--250$\times$ smaller fractionally, enabling nearly unchanged asymmetric fractional errors upon propagation.}.

\begin{figure}[!t]
\centering
\gridline{\fig{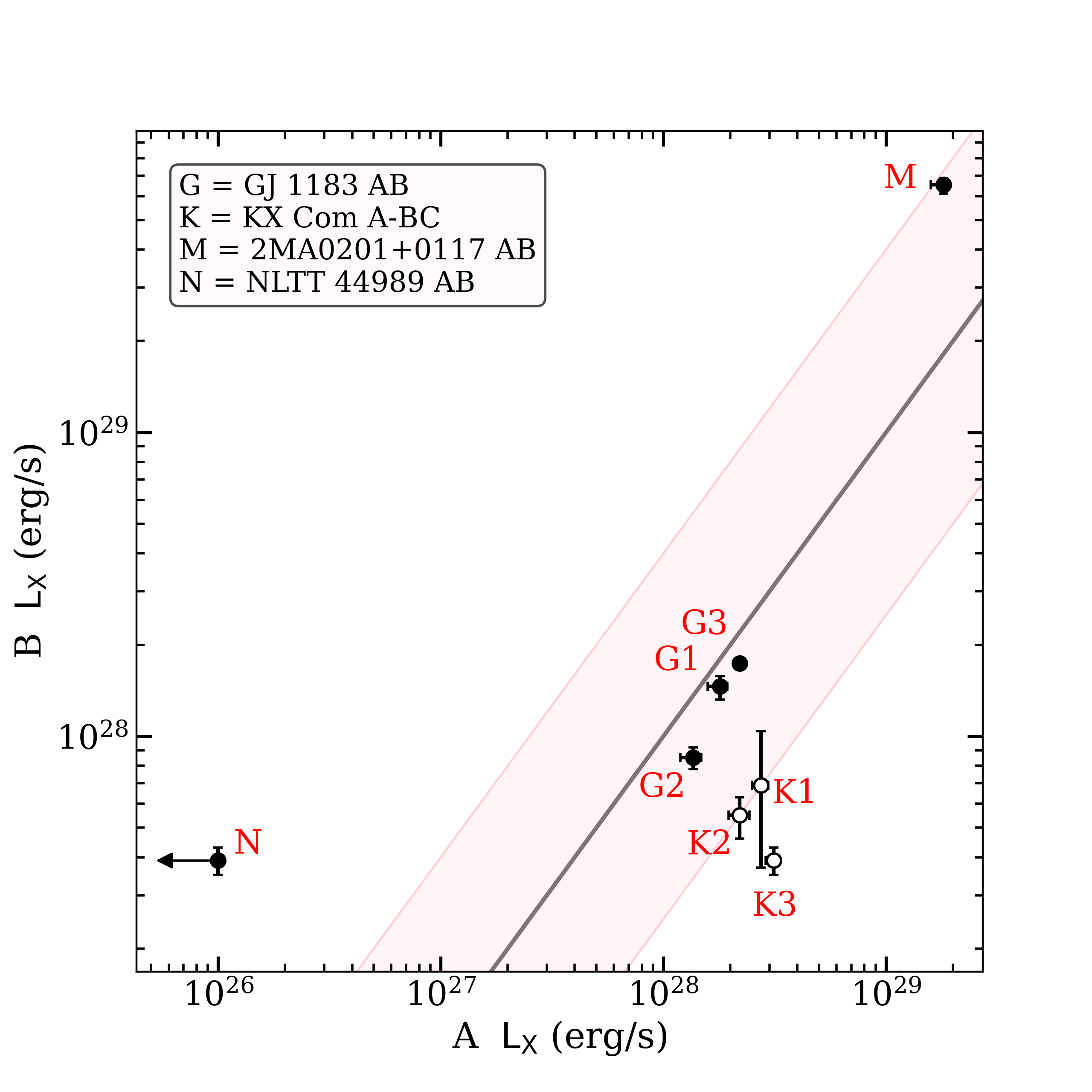}{0.49\textwidth}{(a)}
          \fig{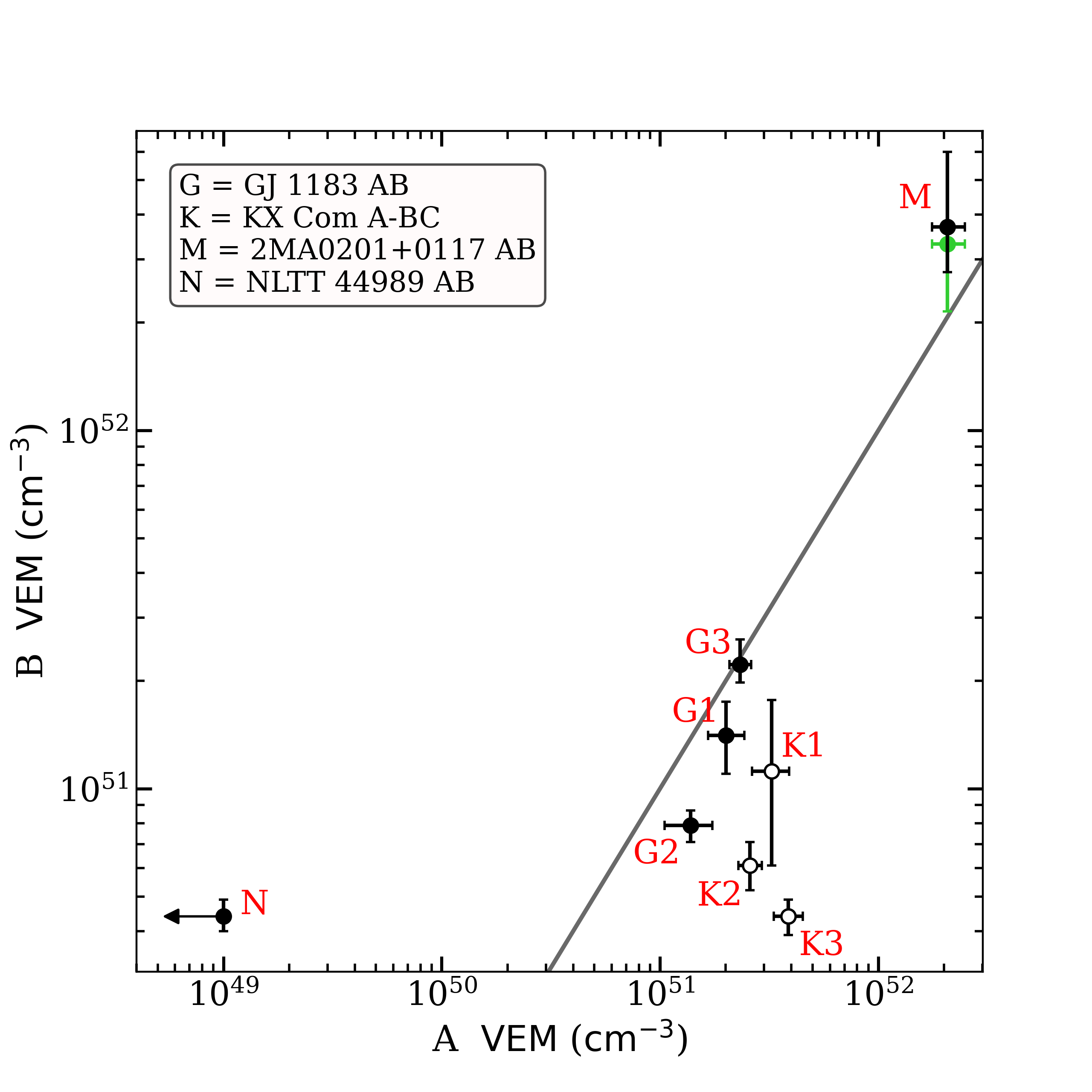}{0.49\textwidth}{(b)}}
\figcaption{One-to-one equivalency plots comparing (a) $L_X$ and (b) VEM between components in each of the four systems for each observation ID. The cases with three exposures from \textit{Chandra}, KX~Com~A-BC and GJ~1183~AB, have results from their individual epochs numbered after the associated system letter label following the designations in Table \ref{tab:ChandraTable}. Grey diagonal one-to-one lines are shown along with leftward arrows for the upper limits on NLTT~44989~A. Open symbols indicate the non-twin triple KX~Com~A-BC. In (a) the pink shaded region shows a factor of 4$\times$ both above and below the one-to-one line (see \S\ref{subsec:chandra-results} for discussion). In (b) a green point indicates the hotter component in the 2T model result for 2MA~0201+0117~B. \label{fig:LX-VEM-equal}}
\end{figure}

The left panel of Figure \ref{fig:LX-VEM-equal} shows a one-to-one plot comparing $L_X$ between components of the four systems at each observing epoch. The grey diagonal line traces equal strengths and the pink shaded region outlines a factor of 4$\times$ both above and below equality. This spread is based on the observed long-term $L_X$ behavior of M dwarfs reported in \cite{Magaudda_2022} that shows roughly a factor of 2$\times$ in variability scatter (see their Figure 13). Our twins could have one star sampled 2$\times$ lower and the other 2$\times$ higher, or the inverse, resulting in the 4$\times$ value used here. Theoretical work by \cite{Farrish_2021} also supports that a large portion of this scatter in $L_X$ could be legitimate astrophysical variation linked to stellar activity cycles in M dwarfs.

Each system is worthy of comment:

\textbf{NLTT~44989~AB} (N in the plot) is the one system in which the two stars show a complete mismatch in $L_X$. This pair breaks the identical twin paradigm because the two stars' $L_X$ is radically different, as was also the case for their H$\alpha$ and rotation behavior. The difference between A and B is well beyond the level ascribable to intrinsic X-ray variations, so this result is robustly capturing typical behavior for the two stars. The weak- or non-detection from A (see \S\ref{subsubsec:NLTT-A}) also means the X-ray activity mismatch may be even more pronounced than we measure here.

\textbf{GJ~1183~AB} (G) exhibits modest differences in $L_X$ that can possibly be explained by intrinsic X-ray variability for isolated stars due to activity cycles. The change in quiescent $L_X$ values between the first and second \textit{Chandra} observations of A and B is intriguing given the exposures occurred only a few hours apart. We speculate that this might be due to the rapid 0.86d and 0.68d rotation periods causing shifted views of the stellar coronae to be visible in each observation. The two exposure midpoints are separated by 10.68 hours, implying that the stars would have rotated by $\sim$186$^{\circ}$ and $\sim$236$^{\circ}$ respectively. However, the X-ray light curve for GJ~1183~B--1 shows a few minor peaks lasting several minutes each with slightly elevated counts that could be weak flares mixed in the noise (see Fig.~\ref{fig:all-xray-lc}). Excluding these points and re-analyzing GJ~1183~B--1, we find $F_X$=27.3$_{-8.2}^{+4.5} \times 10^{-14}$ erg s$^{-1}$cm$^{-2}$, now overlapping with GJ~1183~B--2 at $F_X$=23.0$_{-1.9}^{+2.0} \times 10^{-14}$ erg s$^{-1}$cm$^{-2}$, so the difference may instead stem from analysis uncertainty in what data points are truly ``quiescent".

\textbf{2MA~0201+0117~AB} (M) also exhibits a modest difference in $L_X$, explicable by intrinsic X-ray variability during each star's possible cycle, although it does lie near the upper edge of the 4$\times$ shaded region. This is a young pre-main sequence system, so its astrophysical X-ray behavior may be somewhat different than that of the dwarf stars used in \cite{Magaudda_2022}, and it might be expected to exhibit larger than average $L_X$ variations.

\textbf{KX~Com~A-BC} (K) is not a true twin given that one component has a close companion, so it is shown with open symbols in Figure \ref{fig:LX-VEM-equal} and subsequent X-ray Figures. Offsets from the one-to-one line may be due to interactions of the B and C stars and/or their different masses and rotation from the A component.

\begin{figure}[!t]
\centering
\includegraphics[scale=0.58]{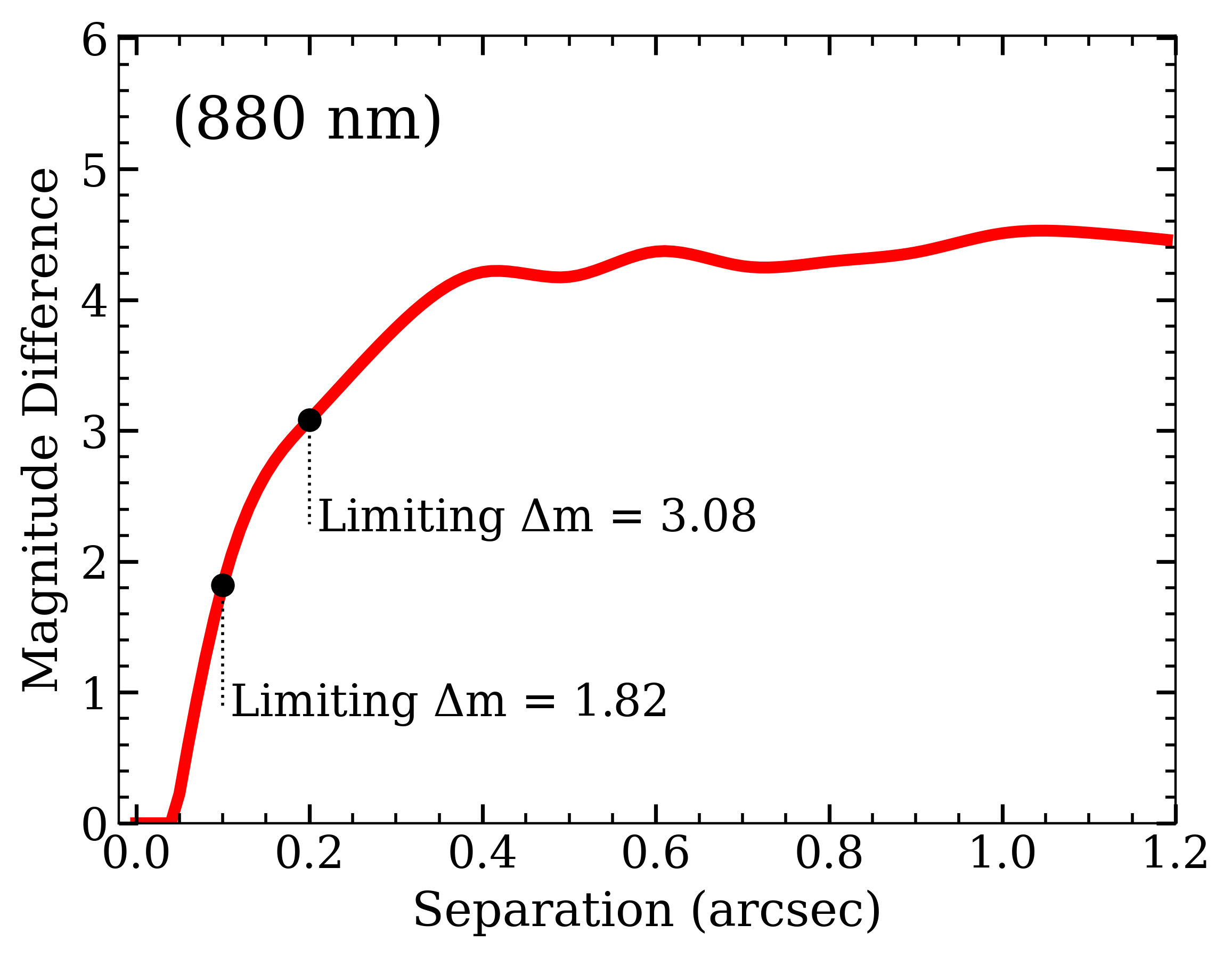}
\figcaption{An example speckle detection limit curve from the 880~nm channel of QWSSI on LDT for 2MA~0201+0117~A is shown. No new unresolved companions were found for this star, to limiting magnitude differences ($\Delta$m) of 1.82 mag at 0.1$\arcsec$, 3.08 mag at 0.2$\arcsec$, and 4 mag or more beyond $\sim$0.3$\arcsec$. A 0.1$\arcsec$ separation corresponds to 4.93~AU at the stellar distance of 49.3~pc. \label{fig:speckle}}
\end{figure}

The right panel of Figure \ref{fig:LX-VEM-equal} compares VEM between components, where the overall trends align well with $L_X$ --- this shows that where we find higher or lower X-ray luminosities we correspondingly measure higher or lower amounts of emitting coronal plasma. The coronal temperatures, not plotted here, show no definitive patterns beyond a possible slight indication of larger X-ray luminosities at hotter temperatures. A comparison of the coronal temperatures between the A and B components also yields no confident trends of note, with most temperatures being in the region of 8--13 MK.

\subsection{Speckle Imaging - SOAR \& LDT} \label{subsec:speckle-results}

For each component in the four systems, speckle observations found no additional companions down to subarcsecond separations. An example detection limit curve from QWSSI+LDT for 2MA~0201+0117~A is shown in Figure \ref{fig:speckle}. In the 880 nm band, the LDT results generally reached to roughly $\Delta$mag $\approx$ 1.9 at 0\farcs1, $\Delta$mag $\approx$ 3.4 at 0\farcs2, and $\Delta$mag $\approx$ 4 at $\gtrsim$0\farcs3. The RV companion to KX~Com~B was not detected by speckle at LDT due to the small separation of 0.13 AU, or 5 mas at 27 pc, expected for an orbital period of 25 days for these stars with masses of $\sim$0.2~$\textup{M}_\odot$ (see \S\ref{subsec:KXCOMAB}). In the $I$ band, the SOAR results typically reached $\Delta$mag $\approx$ 2.5 at 0\farcs15 and $\Delta$mag $\approx$ 3.3 at 1\farcs0. We did not detect the background contaminating source underneath NLTT~44989~B (discussed in \S\ref{subsec:contam}), presumably because it has separations from B of 0\farcs29 and 0\farcs23 at the epochs of the two SOAR visits and a $\Delta$$G$ of 3.75 mag, putting it beyond the detection limits. Overall, these non-detections in all four systems help preclude potentially unresolved companions that would break the twin natures of the pairs.

\begin{figure}[!t]
\centering
\includegraphics[scale=0.655,angle=0,origin=c]{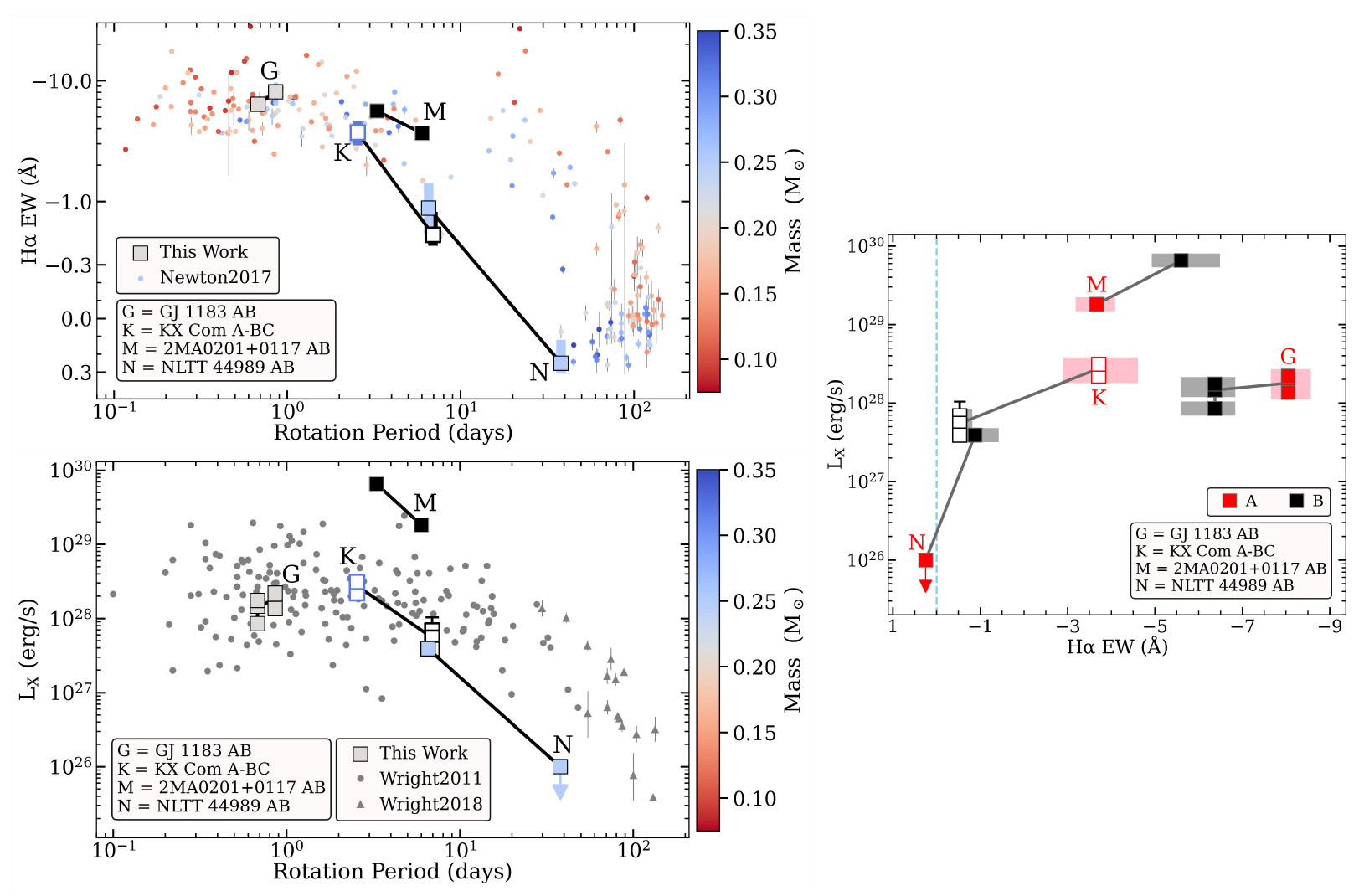}
\figcaption{In each panel, stars studied here are represented by squares and lines connect components in a pair.  Open squares indicate components A and BC in the non-twin triple KX~Com. (Top Left) --- $P_{rot}$ versus H$\alpha$ EW, with fully convective stars (M$<$0.35) from \cite{Newton_2017} underplotted. All points for bona fide twins are color coded by mass except our stars that have less reliable mass estimates, shown in black. Bars on the squares for the stars studied here show the range of observed H$\alpha$ values for each star, excluding flares or epochs without A and B observed back-to-back. System letter labels are placed next to the A components in each pair. (Bottom Left) --- The same as above, now with $L_X$ and underplotting fully convective stars from \cite{Wright2011} and \cite{2018MNRAS.479.2351W}. An arrow indicates the upper limit in $L_X$ for NLTT~44989~A (\S\ref{subsubsec:NLTT-A}). The multiple \textit{Chandra} epochs for GJ~1183 and KX~Com are shown with multiple connected points. (Right) --- H$\alpha$ EW versus $L_X$, again showing the observed H$\alpha$ ranges as shaded bars and multiple connected points for the multiple \textit{Chandra} visit cases. The vertical dashed blue line at EW=0 divides active emission and inactive absorption stars. An arrow again indicates the upper limit in $L_X$ for NLTT~44989~A. Overall, among twin pairs, strong H$\alpha$ emission is correlated with high $L_X$, and each of those observables is correlated with fast rotation. \label{fig:Prot-activity}}
\end{figure}

\section{Rotation-Activity Comparisons} \label{sec:rot-activity-results}

Many efforts have shown that the rotation-activity relationship for M dwarfs divides into rapidly-rotating stars with saturated activity and slower-rotating stars with unsaturated activity that follow a Skumanich-like trend \citep{1972ApJ...171..565S}. We plot our twins in this space in Figure \ref{fig:Prot-activity} for both chromospheric H$\alpha$ activity and coronal $L_X$ activity, underplotting results from \cite{Newton_2017}, \cite{Wright2011}, and \cite{2018MNRAS.479.2351W} to illustrate the saturated and unsaturated regimes. We re-derived $L_X$ values for the \cite{2018MNRAS.479.2351W} sources using \textit{Gaia} DR3 parallaxes and their reported \textit{Chandra} X-ray fluxes in an energy bandpass very similar to ours.

We plot $P_{rot}$, $L_X$, and H$\alpha$ EW directly, instead of the often used Rossby number ($R_o = P_{rot}/\tau_{conv}$), $L_X / L_{bol}$, and $L_{H\alpha} / L_{bol}$ parameters for a few reasons. Traditionally, the incorporation of $\tau_{conv}$ and $L_{bol}$ helps account for the luminosity and convective properties changing with mass when considering a collection of different mass stars; the minimum H$\alpha$ absorption depth is also a function of mass \citep{1986ApJS...61..531S,Newton_2017}. Luckily, our twin stars have nearly identical masses, meaning their $\tau_{conv}$ and $L_{bol}$ factors would be functionally the same --- a robust comparison between our twin components is therefore possible in the rotation-activity plane without applying these corrections. This has the added benefit of avoiding empirical relations often used to derive these parameters, which would otherwise add new assumptions and uncertainties at each step. For example, we previously showed in \cite{Jao2022_rossby} that the \cite{Wright2011} $\tau_{conv}$ relation is built upon a sample with poor mass estimates in the fully convective regime --- our underplotted FC sources from \cite{Wright2011} likely include many interloping PC M dwarfs as a consequence.

While this means care should be taken to only compare our twins within a pair and not across different types of M dwarfs without considering mass, the qualitative appearance of the saturated and unsaturated regimes is still evident enough for general comparison. The most striking result in Figure \ref{fig:Prot-activity} is again that of NLTT~44989~AB, where B appears on the lower envelope of the saturated sequence while A is well evolved towards the unsaturated, slowly-rotating inactive clump. The right side of Figure \ref{fig:Prot-activity} shows H$\alpha$ versus $L_X$, where a twin star appearing more active in H$\alpha$ is always the more active component in $L_X$ as well --- all lines trend diagonally upward. In the case of GJ~1183~AB, the three different \textit{Chandra} epochs do overlap in $L_X$ for A and B, but for each specific epoch A is always larger in $L_X$ than B (see Fig.~\ref{fig:LX-VEM-equal}).

\begin{figure}[t]
\centering
\gridline{\fig{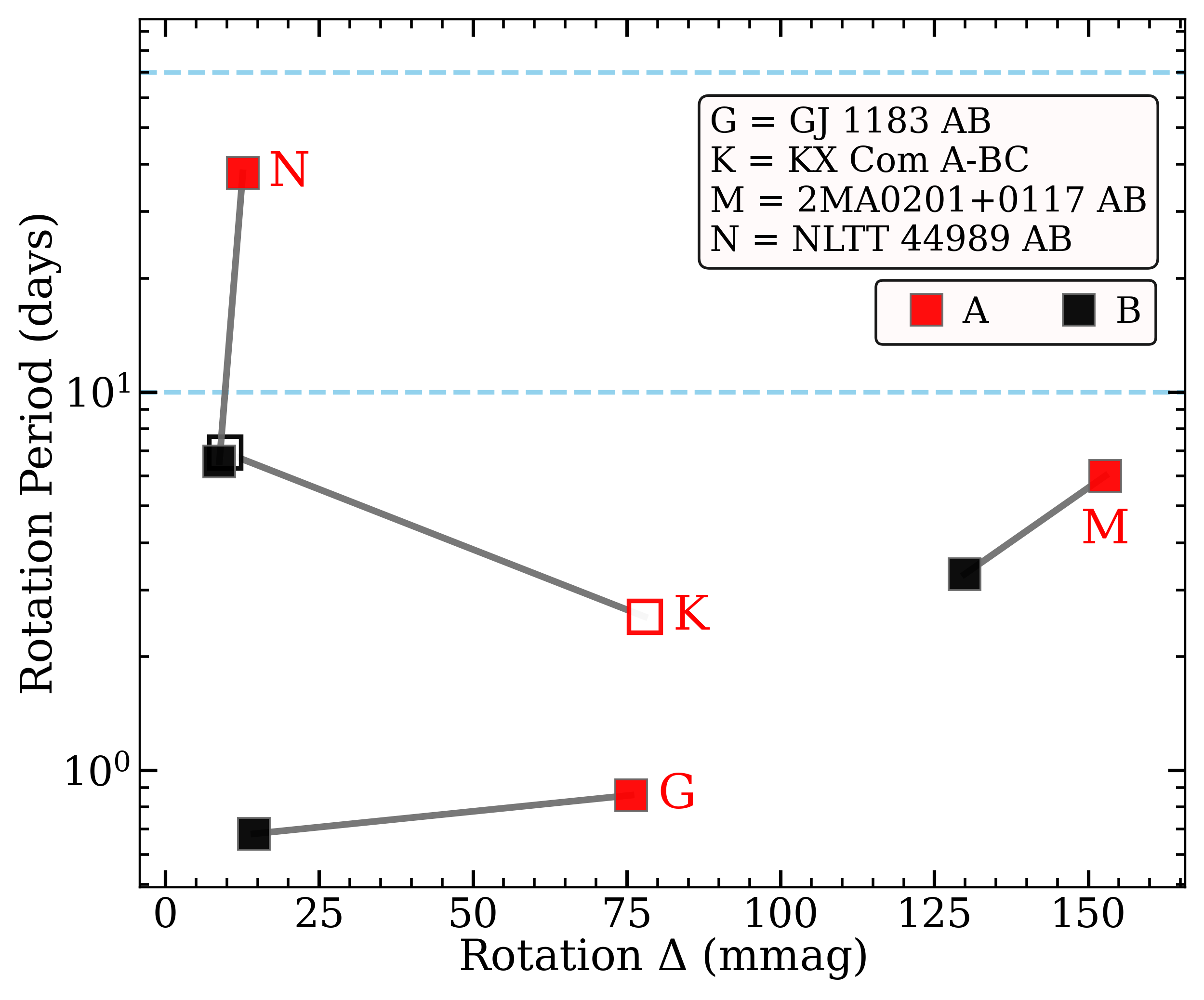}{0.33\textwidth}{(a)}
          \fig{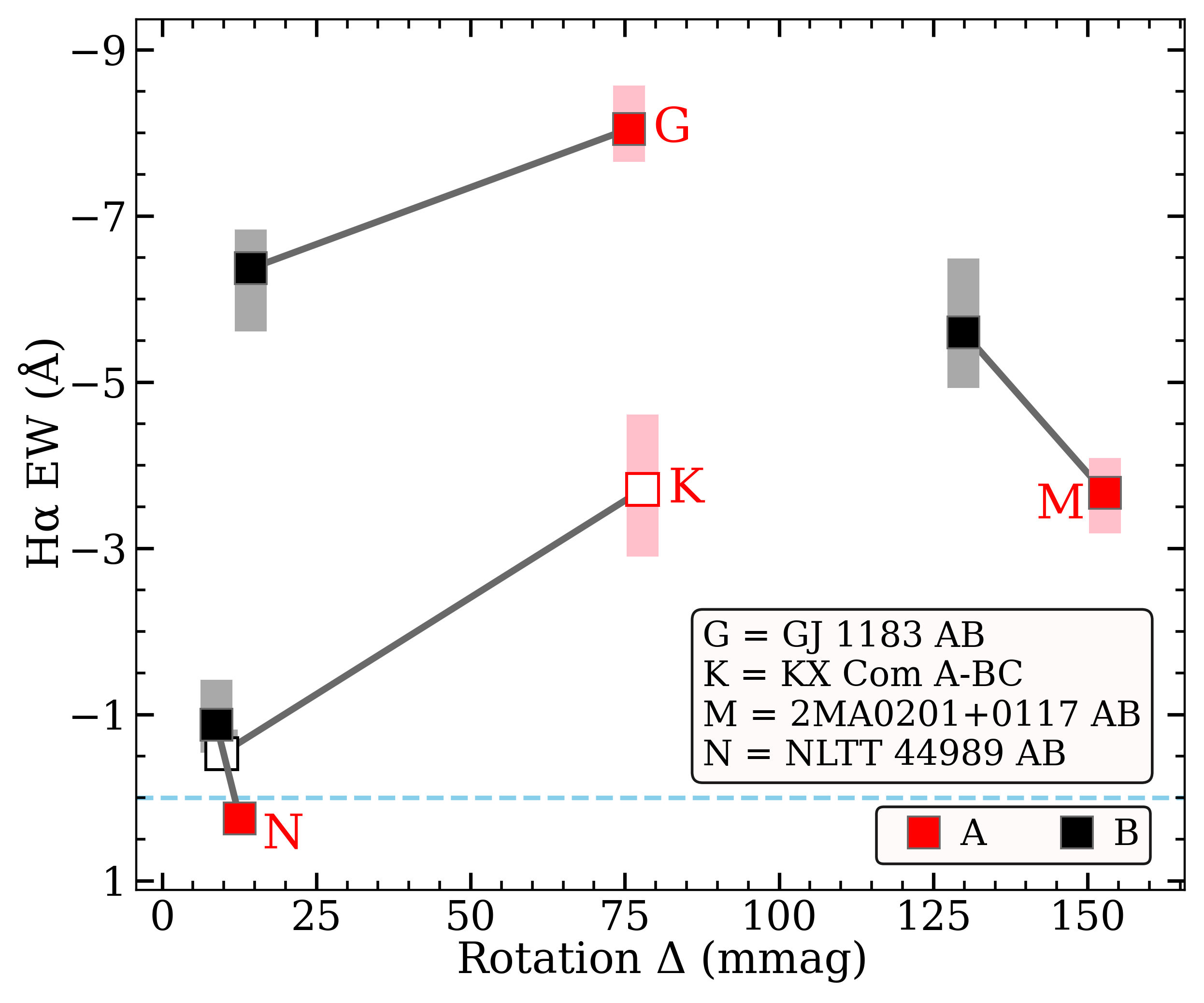}{0.33\textwidth}{(b)}
          \fig{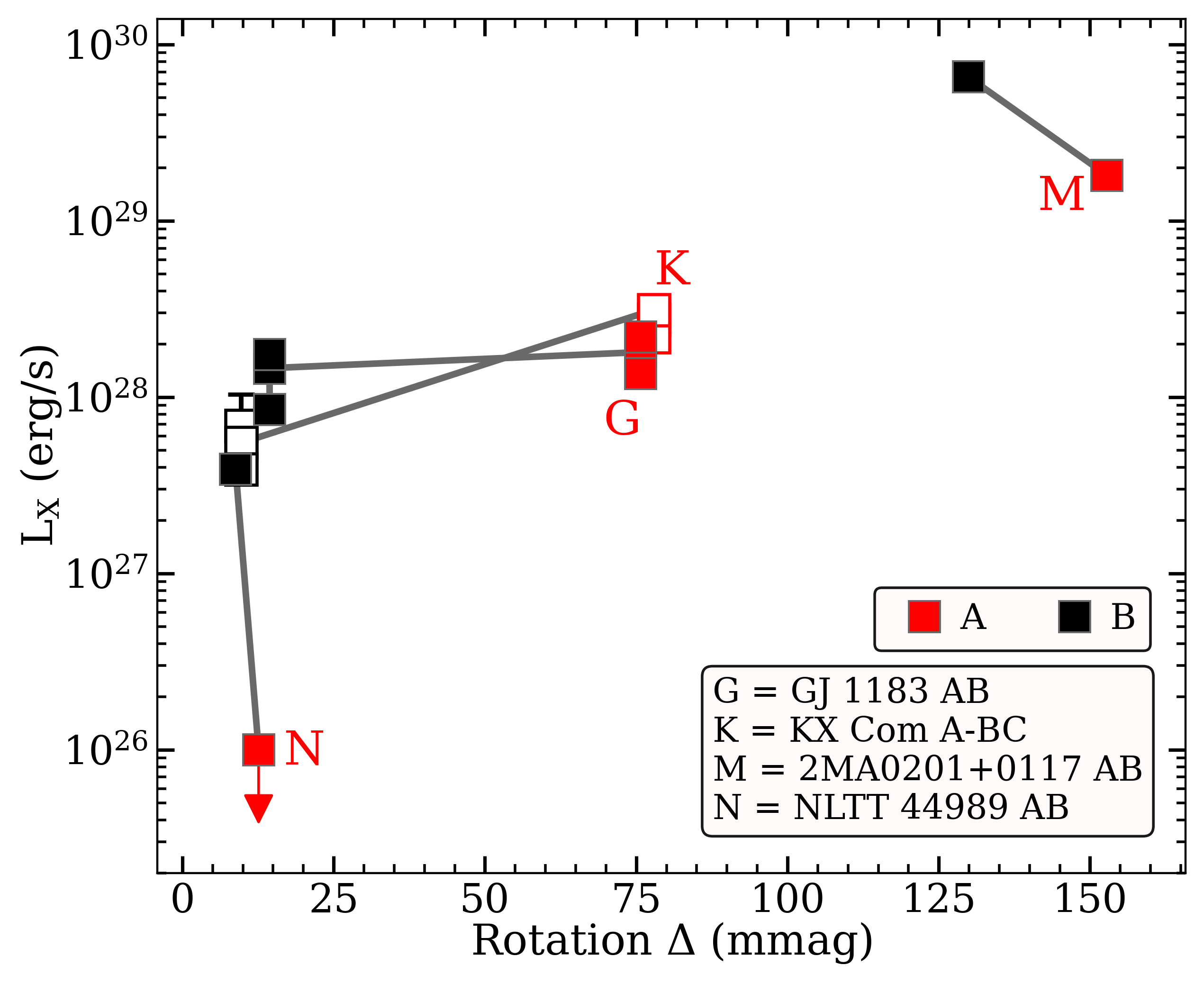}{0.33\textwidth}{(c)}}
\vspace*{0mm}
\figcaption{The 0.9m $V$-band rotation amplitudes (peak-to-peak $\Delta$) versus (a) the rotation period, (b) H$\alpha$ EW, and (c) $L_X$. Lines connect the A (red) and B (black) components in a pair, with open squares indicating the non-twin triple KX~Com~A-BC. Overall, no clear correlations are seen. Blue horizontal dashed lines in (a) at 10d and 70d bound the approximate region of intermediate rotation periods where few FC M dwarfs are found \citep{Newton_2018}. The blue horizontal dashed line in (b) at EW=0 marks the approximate transition between active H$\alpha$ emission and inactive H$\alpha$ absorption. Shaded red and grey bars in (b) are the ranges of observed H$\alpha$ EW values for each star, excluding flares or epochs without A and B observed back-to-back. Stars in (c) with three vertically stacked connected points show the three epochs of \textit{Chandra} $L_X$ measurements in those cases. An arrow in (c) indicates the upper limit in $L_X$ for NLTT~44989~A. \label{fig:Amp-activity}}
\end{figure}

We plot the remaining combinations of $L_X$, H$\alpha$ EW, $P_{rot}$, and rotation amplitude (peak-to-peak $\Delta$) against each other in Figure \ref{fig:Amp-activity}. Here it can be seen that unlike for H$\alpha$ and $L_X$, the amplitudes of magnitude changes during rotation do not always track with other activity parameters. The photometric rotation amplitudes are always larger in the A components than B components, but only weakly so for NLTT~44989~AB --- it remains unclear if this is a robust result or if the low number of three true twin systems randomly appeared this way given the changeable nature of spot modulation amplitudes. The trend is especially uncertain given the A and B component labels may not perfectly track which component is truly slightly more massive or not, as different magnitude measurements sometimes swap which star is brighter (see \S\ref{sec:Sample}, Table \ref{tab:SampleTable-phot}, and \S\ref{subsubsec:discussion-explanations-masses}). This primary component rotation amplitude trend will be explored with more twin systems in our forthcoming second paper (Couperus et al.~in prep).

\section{Additional System Notes} \label{sec:SysNotes}

Here we provide further system-specific details for each pair, including any additional insights into the activity, rotation, ages, or multiplicity.

\subsection{GJ 1183 AB} \label{subsec:GJ1183AB}

Both components are slightly elevated above the main sequence, as shown in Figure \ref{fig:HRD}, indicating that the stars may be young. However, our analysis with BANYAN $\Sigma$ found no reliable membership for the system in nearby young associations (see \S\ref{sec:Sample}). Our speckle search, high-resolution lucky imaging work by \cite{2017A&A...597A..47C}, and our RVs from the CHIRON spectra revealed no additional components.

H$\alpha$ indices (defined as $F_{H\alpha}/F_{cont.}$) of 12.42 for A and 6.22 for B were measured for the pair during 1993--1994 by \cite{1995AJ....110.1838R}. Stronger H$\alpha$ emission from A is consistent with our own mean EW results of A at $-$8.05~$\mathrm{\AA}$ and B at $-$6.37~$\mathrm{\AA}$, so the activity differences we find were evident nearly three decades ago. This could either be due to sustained differences over those three decades, or from varying activity strengths due to underlying activity cycles, as indicated in our long-term photometry of the system seen in Figure \ref{fig:longterm-lcs} --- flares in the \cite{1995AJ....110.1838R} measures are a possibility as well. This system was also observed for H$\alpha$ EWs by \cite{Pass_2024_ApJ} (P24), who found A=$-$9.067$\pm$0.057~$\mathrm{\AA}$ and B=$-$9.289$\pm$0.057~$\mathrm{\AA}$ from R=3000 FLWO-FAST spectra. Our CHIRON measurements with R=27000 from seven epochs spread over 2.1 years were taken earlier in time, yielding H$\alpha$ EWs of $-$7.65~$\mathrm{\AA}$ to $-$8.57~$\mathrm{\AA}$ for A and $-$5.61~$\mathrm{\AA}$ to $-$6.84~$\mathrm{\AA}$ for B, somewhat lower than the P24 results. These discrepancies may be due to systematic factors caused by different spectral resolutions and methods in determining EWs, or perhaps P24 captured stellar flaring events given both stars are quite active.

GALEX reports UV magnitudes for the stars that are consistent within the errors between components (A: FUV=20.53$\pm$0.25  NUV=19.20$\pm$0.10) (B: FUV=20.46$\pm$0.25  NUV=19.38$\pm$0.11) \citep{2017ApJS..230...24B}, which could indicate similar activity levels in the UV. However, the FWHMs reported for the FUV and NUV sources range from 5.2--12.4$\arcsec$, compared to the AB separation of 13.07$\arcsec$, so some blending is likely. Radio detections were also found for GJ~1183~A by \cite{2024MNRAS.529.1258P}, supporting the potential for future radio activity investigations of the pair.

GJ~1183~A was captured emitting two tremendous flares during our observations: the first in $V$ at the 0.9m where the star was seen about 2.6 magnitudes brighter than usual before dimming by 1.7~mag over 30 minutes (this flare is excluded from the 0.9m light curves shown here in Figures \ref{fig:longterm-lcs} and \ref{fig:09m-rot}), and the second in H$\alpha$ with an EW of $-$15.03~$\mathrm{\AA}$ compared to the average EW of $-$8.05~$\mathrm{\AA}$. The absence of comparably strong events in GJ~1183~B is not conclusive evidence for activity differences given the random nature of flares, but GJ~1183~A is clearly an extremely magnetically active star.

\subsection{KX Com A-BC} \label{subsec:KXCOMAB}

The KX~Com system is not a true twin pair but rather an A-BC triple, where we find C to be a close companion to B using RVs, as discussed in \S\ref{subsubsec:rv-results}. This system was also recently observed and discussed in detail by \cite{Pass_2024_ApJ} (P24), who use the alternate name LDS 942 from \cite{1969PMMin..21....1L}. Care must be taken as the components labeled A and B are swapped between their work and ours --- we find ``A-BC" that P24 report as ``AC-B", but we use our naming convention in the following text. The R=44000 TRES spectra from P24 overlap in time with our CHIRON spectra, and SB2 behavior for BC is evident in both sets of spectra. Their data enabled a robust orbital fit for BC and yielded $P_{orb}=25.274\pm0.016$ days; they estimated masses of 0.23~$\textup{M}_\odot$ and 0.20~$\textup{M}_\odot$ for the close binary components. This result is consistent with the $P_{orb}\gtrsim24$d lower limit we determined and indicates our own data very nearly wrapped the orbit. KX~Com~A did not display any close-in additional companions in Robo-AO high-resolution imaging by \citep{2015ApJ...798...41A}, in agreement with our speckle and RV results and the RV results of P24. The KX~Com~B component was observed with high-resolution lucky imaging by \cite{2012ApJ...754...44J}, who did not detect the very close C component and found no additional companions.

P24 report H$\alpha$ EWs of $-$4.15~$\mathrm{\AA}$ and $-$4.52~$\mathrm{\AA}$ from TRES and FAST spectra, respectively, for the more active component we call KX~Com~A, consistent with the range of H$\alpha$ variability of $-$2.90~$\mathrm{\AA}$ to $-$4.61~$\mathrm{\AA}$ we observed over 1.2 years. We see H$\alpha$ ranging from $-$0.43~$\mathrm{\AA}$ to $-$0.82~$\mathrm{\AA}$ in our blended KX~Com~BC results, in general agreement with P24's median measure of $-$0.83~$\mathrm{\AA}$ or $-$0.78~$\mathrm{\AA}$ depending on the spectra they considered. P24 were able to ascribe this emission to KX~Com~B with the C component appearing flat in H$\alpha$. Our X-ray results also align with H$\alpha$, finding KX~Com~A is much more active than BC. 

P24 suspected the 2.55d rotation period in \textit{TESS} belongs to KX~Com~A, which we have confirmed in our analysis here using higher resolution 0.9m data (\S\ref{subsec:rot-results}). We also find a period of 6.93d in our 0.9m photometry of KX~Com~BC, which we presume belongs to the H$\alpha$ active B star and not the H$\alpha$ inactive C star, with C having an unknown and likely even longer third period. While the B and C component mass estimates are similar at 0.23~$\textup{M}_\odot$ and 0.20~$\textup{M}_\odot$, they could still display a strong rotation mismatch akin to the one we observe in NLTT~44989~AB. Rotational spindown in FC M dwarfs in general --- and the various factors that could deviate spindown between otherwise twin stars --- are discussed later in \S\ref{subsec:discussion-spindown} and \S\ref{subsec:discussion-explanations}, but KX~Com~A-BC may also have dynamical triple-star interactions further disrupting the rotation and orbit angular momentum evolution of the entire system \citep{2023MNRAS.526.6168F}. Further observations to confirm our 6.93d rotation period in B and find the presumably longer rotation period in C could help elucidate the role these dynamical interactions play in hierarchical triple spindown.

Following the rotation periods, it is noteworthy that the slightly more massive single-star component, KX~Com~A, is more magnetically active in H$\alpha$ and $L_X$ than the lower mass, binary KX~Com~BC components; this phenomenon was noted by P24 as well. Traditional expectations are that close-in companions will tidally interact to sustain rapid rotation and high levels of activity beyond typical active lifetimes. However, the 25d orbital period for BC from P24 is much longer than the 6.93d brightness modulation pattern we see in the component's 0.9m light curve and the $\sim$7-day tidal circularization timescale of M dwarfs \citep{Vrijmoet_thesis_2023}, indicating B and C are not in tidal synchronization. This may be the result of the aforementioned dynamical interactions of hierarchical triples as discussed in \cite{2023MNRAS.526.6168F}. 

\subsection{2MA 0201+0117 AB} \label{subsec:2MA0201AB}

Both components in this system are elevated well above the main sequence, as shown in Figure \ref{fig:HRD}. This is the only twin system in this paper with an age estimate, in this case because it is a member of the 25-Myr old $\beta$ Pictoris association \citep{2015A&A...583A..85A, 2017A&A...607A...3M}. This means the $\sim$2$\times$ difference we find in component rotation periods is likely a result of their formative rotation periods and disk lifetimes, having more connection to rotation starting points than our other twins that have had more time to evolve.

\subsection{NLTT 44989 AB} \label{subsec:NLTT44989AB}

Of the several different known twin or near-twin systems with active/inactive mismatches found throughout the combination of this work, \cite{Pass_2024_ApJ}, and \cite{Gunning_2014}, NLTT~44989~AB is the only case with measured rotation periods for both components. The periods themselves also confirm that the enormous activity differences we see in H$\alpha$ and $L_X$ are aligned with strongly mismatched rotation in this case. To the best of our knowledge, we are the first to report rotation and activity measurements for each component in this fascinating twin system. Furthermore, no additional companions were uncovered for NLTT~44989~A or B from two speckle visits (\S\ref{subsec:speckle-methods}, \S\ref{subsec:speckle-results}), our 15 RV epochs over 1.5 years (Fig.~\ref{fig:RV-Timeseries}), or various other related checks (\S\ref{subsec:sample-companion-checks}), supporting their twin nature.

Both components reside toward the lower edge of the main sequence (see Fig. \ref{fig:HRD}), possibly implying an older age or lower metallicity for the pair compared to the other three systems --- this is in agreement with NLTT~44989~A hosting the longest rotation period here as well. We do not find any reported UV sources for the system in \cite{2017ApJS..230...24B}, but GALEX images of the field clearly show elevated counts near B and minimal counts near A, commensurate with our other activity signatures. Deeper studies of the UV activity are left for future work.

The system resides in a dense field, with particular care needed to assess potential contamination in any future observations --- this is especially important for long-term campaigns where proper motion may become relevant as well. The contaminating sources are much fainter in the optical (see \S\ref{subsec:contam} for a discussion of this in our own observations), but even these small deviations could be relevant for validating and interpreting results for the system, such as we discuss later in \S\ref{subsubsec:discussion-explanations-masses}.

\section{Discussion} \label{sec:discussion}

Here we focus on only the three true twin systems, GJ~1183~AB, 2MA~0201+0117~AB, and NLTT~44989~AB, and disregard KX~Com~A-BC because of its non-twin nature. See \S\ref{subsec:KXCOMAB} for a separate discussion of KX~Com~A-BC.

\subsection{Observed Differences and Possible Causes} \label{subsec:discussion-explanations}

We observe consistent activity differences in H$\alpha$ and $L_X$ beyond the uncertainties in each twin pair case. The most modest of these is GJ~1183~AB, where A is 58$\pm$9\% stronger in $L_X$ on average\footnote{Relative differences in $L_X$ and H$\alpha$ are calculated using simple ratios between components and standard uncertainty propagation techniques. $L_X$ used the larger of the asymmetric errors for propagation, the upper limit in $L_X$ with no error for NLTT~44989~A, and a weighted average of the three $L_X$ visits for GJ~1183~AB. H$\alpha$ used the mean and standard deviation of EWs for each star, only including non-flaring epochs with both stars successfully observed back-to-back.} over three visits and 26$\pm$9\% stronger in H$\alpha$ EW on average over seven visits. GJ~1183~A and B host functionally similar rotation periods of 0.86d and 0.68d --- these both fall in the saturated regime, and we indeed see both stars similarly located in the rotation-activity plane in Figure \ref{fig:Prot-activity}. 2MA~0201+0117~AB demonstrates the next largest set of differences, with B 3.6$\pm$0.5 times stronger in $L_X$ from one visit and 52$\pm$19\% stronger in H$\alpha$ on average from nine visits. This trend follows the rotation given that B is faster at 3.30d compared to A at 6.01d, a difference likely resulting from formation and disk evolution given the young $\sim$25~Myr age for this $\beta$ Pic association system. Finally, NLTT~44989~AB shows the strongest differences, with B $\geq$39$\pm$4 times stronger in $L_X$, $\sim$6 times faster in rotation period, and a complete A/B inactive/active mismatch in H$\alpha$ for 15 visits over 1.5 years. In every system's case, these differences are all despite component stars having the same mass, age, composition, and environment.

We next provide brief evaluations of the plethora of possible causes for the observed activity and rotation differences between components within these three twin systems. The potential causes are separated into characteristics of the stars themselves (\S\ref{subsubsec:discussion-explanations-rotation} to \S\ref{subsubsec:discussion-explanations-masses}), effects of companions past or present (\S\ref{subsubsec:discussion-explanations-evolution} to \S\ref{subsubsec:discussion-explanations-exoplanets}), or purely observational consequences (\S\ref{subsubsec:discussion-explanations-inclinations} to \S\ref{subsubsec:discussion-explanations-bkgsources}). We begin with the most straightforward explanation.

\subsubsection{Could different rotation periods be causing the different activity levels?} \label{subsubsec:discussion-explanations-rotation}

Different rotation speeds are the obvious explanation for our observed activity differences, and higher activity levels generally track with faster rotation periods in our results (Fig. \ref{fig:Prot-activity}). That said, this produces the obvious follow-up questions: \textit{why} are the rotation periods different? And, how well can we predict rotation periods and activity levels? We primarily address these questions later in \S\ref{subsec:discussion-spindown} and \S\ref{subsec:discussion-exoplanets}, though some of the enumerated factors considered next are relevant to rotation evolution as well.

\subsubsection{Could the observations be snapshotting stellar activity cycles?} \label{subsubsec:discussion-explanations-cycles}

Even if components have exactly the same rotation periods, out of phase stellar cycles could still manifest different levels of observed activity between otherwise twin stars at a snapshot in time. Long-term stellar activity cycles are known or strongly suspected to exist in a number of PC and FC M dwarfs \cite[e.g.,][Couperus et al.~in prep, and references therein]{1986A&A...154..171M, 2013ApJ...764....3R, 2016A&A...595A..12S, 2016ApJ...830L..27R, Wargelin_2017, Henry_2018, 2019MNRAS.483.1159I, 2019A&A...628L...1I, 2020A&A...644A...2I, 2023A&A...670A..71F, 2023ApJ...949...51I, 2023MNRAS.525.2015D, 2024MNRAS.527.4330L}. We even observe a candidate long-term photometric cycle in GJ~1183~A, in contrast to much lower amplitude changes in B over a decade baseline (Fig. \ref{fig:longterm-lcs}). That said, the activity differences between A and B may be the manifestation of even longer timescale stochastic variations in the dynamos as stellar cycles change in time --- for example, B could be in a Maunder Minimum--like low-spot-activity state in contrast to A \citep{1976Sci...192.1189E}.

Beyond cycles seen via optical photometry, X-ray activity cycles also exist in some stars, including a candidate X-ray cycle in the FC M dwarf Proxima Cen \citep{Wargelin_2017}. Theoretical work by \cite{Farrish_2021} suggests that such cycles in M dwarfs cause variability in $L_X/L_{bol}$ throughout spindown. Observationally, the long-term variability scatter in $L_X$ for M dwarfs is about a factor of 2 for most stars, according to the results of \cite{Magaudda_2022} (see their Fig. 13). \cite{dsouza_2023} found a comparable level of scatter in $L_X$ between similar components in M dwarf wide binaries. Similar behavior is also observed in the twin M dwarf system GJ 65 AB, where work by \cite{Wolk-UVCet-CoolStars21} reveals that X-ray flaring activity levels changed moderately for one component compared to observations taken nearly two decades earlier. X-ray cycles may therefore play a role in the $L_X$ differences we observe in GJ~1183~AB and 2MA~0201+0117~AB (see \S\ref{subsec:chandra-results} for additional details), whereas NLTT~44989~A and B differ well beyond the typical scatter seen for field stars and beyond the level likely attributable to cycles alone (see Fig. \ref{fig:Prot-activity}).

Long-term cycles in chromospheric H$\alpha$ activity have also been found for M dwarfs, with a few such example cases reported in \cite{2011A&A...534A..30G}, \cite{2013ApJ...764....3R}, and \cite{2023A&A...670A..71F}.  However, the expected amplitudes of the variations in H$\alpha$ EW for different FC M dwarfs are too poorly understood to offer robust constraints for informing our differences here. While there is H$\alpha$ EW scatter of up to several angstroms in low-mass field stars (see Fig. \ref{fig:Prot-activity}), it is unclear what proportion of this variability may be caused by cycles, and a significant portion is likely from very short-term variability \citep[see e.g.,][]{2012PASP..124...14B, Gunning_2014, Medina_2022_Ha}. It suffices to say that activity cycles may be partially responsible for some amount of the differences we see in H$\alpha$ EW between twin components, though again not in the extreme case of NLTT~44989~A/B where a total inactive/active mismatch is seen.

\subsubsection{Could a dynamo bistability exist in some M dwarfs?} \label{subsubsec:discussion-explanations-bistability}

Beyond observational snapshotting, the magnetic dynamos themselves may host underlying instabilities. Theoretical work by \cite{Gastine_2013} has suggested the possibility of a double-branched dynamo regime wherein late M dwarfs could fall into one of two dynamo states depending on the initial parameters. If a dynamo bistability exists, our twins may have had similar initial parameters but could have converged to two different dynamo states for some unknown duration, resulting in the mismatched activity and/or rotation we see today. In contrast, \cite{Kitchatinov_2014} again implicate oscillatory stellar cycles to explain the underlying observations. Overall, we generally favor the explanation of oscillatory cycles given the growing observational and theoretical evidence for such stellar activity cycles in fully convective M dwarfs \cite[e.g.,][Couperus et al.~in prep, and references therein]{2016A&A...595A..12S, 2016ApJ...830L..27R, 2016ApJ...833L..28Y, Henry_2018, Brown_2020, 2020A&A...644A...2I, 2023A&A...670A..71F, 2023ApJ...949...51I, 2023MNRAS.525.2015D, 2024MNRAS.527.4330L}. This is also supported in our own data here given the candidate photometric cycle in GJ~1183~A (Fig. \ref{fig:longterm-lcs}). However, separate observational X-ray activity results from \cite{2014ApJ...785...10C} and \cite{2024A&A...687A..95M} support the presence of a bimodal dynamo in very-low-mass very-rapidly-rotating stars, so this factor ultimately remains a possibility in some parameter regimes. Multi-epoch Zeeman–Doppler imaging observations of our twins to reconstruct their magnetic field characteristics over cycle timescales would provide very strong evidence to investigate this further.

\subsubsection{Could metallicity be changing the activity and rotation?} \label{subsubsec:discussion-explanations-metallicity}

Stellar activity can change with composition (i.e., metallicity), as shown in the results of \cite{2021ApJ...912..127S}. Different metallicities could result in discordant rotation periods via different amounts of activity and subsequent magnetic braking over evolutionary timescales. However, our twins should have functionally identical compositions as members of the same wide binaries, as supported by the work of \cite{2020MNRAS.492.1164H}, who found that wide binary components typically have [Fe/H] matched to within 0.02 dex. Future work yielding detailed abundance measurements from our spectra could validate this assumption, but for now we note that no significant differences have been seen when inspecting overlapping spectra for components in the twin systems, see, e.g., Figure \ref{fig:Ha-spec}.

\subsubsection{Could slightly different component masses be responsible?} \label{subsubsec:discussion-explanations-masses}

Our selection for pairs being equal-mass is exclusively based on requiring $BP$, $RP$, \textit{J}, \textit{H}, and $K_s$ to all match within $<$0.10~mag between components. Differences of 0.1 mag would correspond to slightly different masses and subsequently mildly deviated spindown timescales, possibly explaining any activity or rotation mismatches in our pairs, so we attempt to quantify this. The three true twin pairs here have an average difference of 0.04 mag between components across all five required filters, with estimated masses for A and B always differing by $<$0.005\(\textup{M}_\odot\) if we consider precisions higher than reported in Table \ref{tab:SampleTable-phot}. The \cite{Benedict_2016} $M_V$ MLR we use has an rms scatter of 0.19 mag or 0.023\(\textup{M}_\odot\), so we are functionally at or within the MLR precision limits. \cite{Pass_2024_ApJ} calculated a 0.02\(\textup{M}_\odot\) difference would yield a 3.86\% chance of observing a roughly twin binary pair during an active/inactive mismatch in their results due to just mass-dependent spindown. We observed 27 of our twin systems for H$\alpha$ activity, 25 if we remove the two known or suspected higher-order multiples, and found only NLTT~44989~AB had an active/inactive mismatch (Couperus et al.~in prep). This represents 1/25=4\% of our systems, very similar to the \cite{Pass_2024_ApJ} estimate, so we cannot rule out very slight mass differences as a possible explanation for NLTT~44989~AB. Furthermore, the NLTT~44989~B $BP$ measurement we use for our mass estimate has up to $\sim$6.3\% extra contaminating flux from background sources (\S\ref{subsubsec:contam-mags} and Fig.~\ref{fig:NLTT-field}), which if removed shifts B about 0.066~mag farther from A in terms of brightness (but still within the MLR scatter level), possibly favoring the slight mass difference explanation for this system specifically.

Despite this possibility, the present mass estimates we do have for NLTT~44989~A and B differ by only 0.0017\(\textup{M}_\odot\), giving a much lower 0.33\% chance to be observed with an active/inactive mismatch by chance from just mass-dependent spindown based on Eq.~(3) in \cite{Pass_2024_ApJ}, so we disregard this factor for the remainder of the discussion and treat them as true twins. Higher precision MLRs or mass measurements for our twins, or a larger sample of twins, would allow us to investigate this further.

\subsubsection{Could non-standard evolutionary scenarios explain NLTT 44989 AB?} \label{subsubsec:discussion-explanations-evolution}

It is worth considering if the extreme case of NLTT~44989~AB is a complex outlier. While we assume our binary components are the same age and formed together, we cannot strictly rule out the possibility that two stars with similar masses and compositions in a binary have different ages because of dynamical many-star interactions in the distant past \citep[e.g.,][and references therein]{1991ARA&A..29....9V, 2011ASPC..447...47K}. Another extreme possibility is past mass transfer from more-massive companions followed by the smaller low-mass star(s) subsequently being ejected via dynamical interactions --- past stellar mergers could play a similar role \citep[e.g.,][]{2024Sci...384..214F}. These scenarios are somewhat analogous to blue stragglers \citep[e.g.,][]{1989AJ.....98..217L, 2006ApJ...647L..53F, 2006MNRAS.373..361M, 2011AJ....141..142D} but now hypothetically appearing in leftover low-mass stars. In these scenarios we might expect different compositions between the components, which could be tested with a detailed abundance analysis of our spectra. We leave this for future work, and again simply note the extremely congruent overlapping continuum features in Figure \ref{fig:Ha-spec} as evidence favoring similar compositions.

The strong activity level mismatch in NLTT~44989~AB has similar counterparts in H$\alpha$ activity found by \cite{Pass_2024_ApJ} and \cite{Gunning_2014}, as well as the results for BL+UV Ceti outlined in \S\ref{sec:intro}. While these binaries may all have such complex evolutionary histories that invalidate their twin natures, we consider this scenario highly unlikely in light of the more probable alternate explanations available involving spindown properties of FC M dwarfs, as discussed in \S\ref{subsec:discussion-spindown} below.

\subsubsection{Could there be dynamical binary interactions at play?} \label{subsubsec:discussion-explanations-dynamics}

Even assuming our stars formed together and are the same age, they may still be dynamically interacting in ways distinct from isolated stars. Our systems are presently in wide $>$80~AU configurations, so this should generally not be the case. However, it is important to consider the possibility of the duplicity interfering with disks during formation, which could result in shorter disk lifetimes or otherwise impact star-disk rotational coupling and subsequently produce different stellar rotation periods. The rotation differences could then propagate and, depending on the age of the stars, remain today, resulting in both rotation and activity differences. Various efforts have found that these disk disruptions can occur in wide binaries out to component separations of $\sim$80--100~AU \cite[e.g.,][]{1996ApJ...458..312J, Meibom2007, 2009ApJ...696L..84C, 2012ApJ...751..115H, 2017A&A...607A...3M, 2019A&A...627A..97M, 2023ASPC..534..275O}, even in low-mass M stars. GJ~1183~AB and 2MA~0201+0117~AB have projected separations well beyond this limit, at 229~AU and 515~AU respectively, but NLTT~44989~A and B are closer at 87~AU. This means that disk disruption effects may be at play in the NLTT system, though the true separation is likely larger than the projected value. The two stars are functionally the same in mass as well, so could hypothetically have somewhat equal and opposite impacts on each other's disks. However, the extent to which twin M stars form with twin disks is not well informed, so future studies examining disk architectures of twin PMS M stars would prove insightful.

Secondarily, \cite{2019MNRAS.489.5822E} and \cite{2022ApJ...933L..32H} propose that an excess population of twin wide binaries may form close together via circumbinary disk accretion with subsequent dynamical widening to their present-day wider separations --- we conjecture that this could hypothetically alter the initial rotation periods or early rotational evolution of our twin binary stars compared to isolated single stars. This twin excess fraction appears stronger at masses $<$0.6\(\textup{M}_\odot\) and extends out to binary separations of $\sim$10,000~AU \citep{2019MNRAS.489.5822E}, covering a very large portion of our broader sample. Finding only our single case with strongly deviated behaviors may therefore disfavor this hypothesis, but a robust statistical comparison of our sample against the results of \cite{2019MNRAS.489.5822E} is needed to investigate this hypothesis further. We leave this as future work given the significant uncertainty in how much this formation pathway would or would not disrupt the long-term rotational evolution of a given pair.

\subsubsection{Could there be hidden unresolved companions?} \label{subsubsec:discussion-explanations-companions}

Higher-order multiplicity beyond simple duplicity is also a concern. Any relatively massive unresolved companions, be they black holes, neutron stars, white dwarfs, red dwarfs, or brown dwarfs could break our twin comparisons and possibly explain any observed differences in either rotation or activity. We have tried to rule out this scenario as much as possible using speckle imaging (\S\ref{subsec:speckle-results}), timeseries RVs (\S\ref{subsubsec:rv-results}), various \textit{Gaia} parameters (\S\ref{subsec:sample-companion-checks}), and literature checks (\S\ref{subsec:sample-companion-checks} \& \S\ref{sec:SysNotes}), all of which uncovered no additional companions to our three true twin systems. We also do not see clear evidence for intermixed photometric patterns in our resolved 0.9m data (Fig.~\ref{fig:longterm-lcs} \& Fig.~\ref{fig:09m-rot}) that could suggest multiple blended stars, if they existed. That said, there are regions of parameter space not ruled out, such as a close-in companion orbiting at less than $\sim$0.7~AU (the SOAR and LDT $\sim$40~mas speckle limits at the closest distance case of GJ 1183 B) in a face-on configuration that would result in no detectable RV signature. Unresolved sources could also have combined brightnesses that might not appear obviously elevated on the main sequence, or be similar brightness and thus less identifiable via the {\it Gaia} RUWE value. Altogether, while we cannot entirely disregard these various possibilities, we consider unseen massive companions relatively unlikely given our multifaceted investigation.

\subsubsection{Could exoplanets orbiting the stars change the stellar behaviors?} \label{subsubsec:discussion-explanations-exoplanets}

Beyond hidden massive companions, planets could also drive mismatches, either via disk disruptions during formation or from star-planet interactions. In the former, planets could hypothetically impact the lifetime of a circumstellar disk and the subsequent star-disk locking duration, thereby possibly changing the rotation period at which a star forms and the consequent period we see today. For the latter, be they tidal or magnetic star-planet interactions, there is evidence to suggest impacts on the stellar rotation and/or activity are possible and observable \citep[e.g.,][]{2016A&A...591A..45P, 2022MNRAS.513.4380I, 2023NatAs...7..569P, 2023arXiv230500809T, 2024MNRAS.527.3395I}. Finally, in an extreme case, past planetary mass transfer or complete engulfment into the host star could also disrupt the stellar rotation \citep{2011ApJ...732...74G, 2012MNRAS.425.2778M}. While low-mass M dwarfs host very few massive planets \citep{2023AJ....165...17G, 2023AJ....166...11P, 2023MNRAS.521.3663B}, a rare case might be the reason we only found a single case with significant rotation mismatches among our 13 twin systems with rotation results (Couperus et al.~in prep). Furthermore, stars with stellar companions (such as our binary targets) are actually one of the possible explanations responsible for dynamically placing gas giants very close in around M dwarfs \citep[see e.g.,][and references therein]{2023AJ....166...30C}. Overall, the extent and timescale of these various planetary effects on the host star are active areas of investigation and depend heavily on the system configuration \citep[e.g.,][]{2003ApJ...589..605W, 2011ApJ...732...74G, 2012MNRAS.425.2778M, 2015ApJ...799...27P}.

This all implies a significant result --- if planet interactions with the disk/star cause detectable disruptions in stellar spindown, this could enable the discovery of planets based on spin comparisons in the future. Future work obtaining higher-precision RV observations of NLTT~44989~A and B to determine if one has a massive close-in planet while the other does not would be valuable to investigate these hypotheses.

\subsubsection{Could the stars have different rotational inclinations?} \label{subsubsec:discussion-explanations-inclinations}

Observational viewing angles are relevant because inclinations can significantly alter spot modulation amplitudes. This could explain the mismatched long-term photometric activity levels in GJ~1183~AB (Fig. \ref{fig:longterm-lcs}), but in this case, $P_{rot}$ and $v\sin(i)$ are similar for A and B, and with similar twin radii we therefore do not expect their inclinations to differ markedly. Mismatched inclinations also cannot explain the cases we find with different rotation rates, except for specific spot configurations as discussed next in \S\ref{subsubsec:discussion-explanations-spots}. $L_X$ and H$\alpha$ emission from active stars are largely the result of distributed magnetic heating in the chromosphere and corona and they are therefore generally treated as largely independent of inclination. The role of inclination may be significantly more important in the context of theoretical results by \cite{Brown_2020} that demonstrated dynamo action strongly manifesting in a single hemisphere when fully convective M dwarfs are modeled; these results remain to be verified observationally, but if correct, would have remarkable implications for studies of FC M dwarf activity.

\subsubsection{Could specific spot configurations be changing our measured rotation periods from the true periods?} \label{subsubsec:discussion-explanations-spots}

Spots rotating in and out of view can be an imperfect technique for measuring periods in certain cases, creating yet another observational effect related to viewing angle. Our measured rotation periods could be impacted by similar spot configurations on opposite sides of the stars manifesting similar-amplitude modulations; these would masquerade as a periodic pattern twice as fast as the true rotation period (see recent examples of this for \textit{TESS} data in \cite{2024A&A...687A.180R}). This effect can only produce periods appearing falsely faster, primarily twice as fast, but not slower, assuming spot configurations do not change rapidly relative to the rotation timescales. So, if this case were occurring, the true rotation periods of our stars would be about double the duration of our measured periods.

GJ~1183~A and B have measured periods of 0.86d and 0.68d, where either doubling would still result in periods broadly similar and in the very active regime. In 2MA~0201+0117~A and B we measured 6.01d and 3.30d periods respectively, where doubling B's 3.30d period would repair their factor of $\sim$2 difference. For NLTT~44989~A and B with adopted periods of 38d and 6.55d respectively, doubling the shorter 6.55d period to 13.10d would still present a marked mismatch in rotation and the two stars would still appear in different active/inactive regions of the rotation-activity diagrams in Fig.~\ref{fig:Prot-activity}. The impact on our overall results would therefore be relatively minor; the interpretation of GJ~1183~AB would not markedly change, 2MA~0201+0117~AB might no longer suggest different rotation during ongoing formation, and our most important result of NLTT~44989~AB would have the same general interpretation. The H$\alpha$ and $L_X$ activity differences in each pair would also be unaffected.

We cannot entirely rule out these spot configuration possibilities, but we consider them quite unlikely given our multiple data sources from distinct and sometimes multi-year spans in time that often recover very similar periods for many of our stars (see \S\ref{subsec:rot-results}). Longer baseline observations that allow time for spot configurations to change could investigate this further.

\subsubsection{Could distant background sources of impactful brightness be lurking directly behind our stars?} \label{subsubsec:discussion-explanations-bkgsources}

There could also be contamination from astrophysically unassociated sources distinct from unseen orbital companions. Background sources more distant than our stars but of considerable brightness --- such as evolved luminous stars, high surface brightness galaxies, or active galactic nuclei --- would corrupt our results if aligned on the sky with our stars during our observations. Nearby contaminants assessed with \textit{Gaia} are discussed in \S\ref{subsec:contam}, and our speckle observations probe even closer to $\lesssim$0\farcs1, but that still leaves area directly behind the stars unexplored. However, this possibility can be assessed in all cases because the pairs have large proper motions of 70--367~mas/yr that substantially change the sky positions over human timescales. Our visual assessment used archival images within Aladin from DSS2-Red \citep[Ep.~1984--1998;][]{1996ASPC..101...88L, 2004AJ....128.3082G}, SkyMapper R-band \citep[Ep.~2014--2015;][]{2018PASA...35...10W}, and ZTF DR7 r-band \citep[Ep.~2018--2021;][]{ZTF_data}, finding that at the epochs of our new observations the pairs are not directly overlapping with any bright background sources to the extent the proper motions allow us to check (except the cases already discussed for NLTT~44989~AB in \S\ref{subsec:contam}). In addition, the general astrophysical behaviors of the stars in our observations are consistent with that of pre- or main sequence M stars, instead of, for example, active galactic nuclei.

\subsection{Implications for Fully Convective M Dwarf Spindown} \label{subsec:discussion-spindown}

Considerable progress has been made in recent years towards understanding the spindown of fully convective M dwarfs \citep[e.g.,][]{2014ApJ...789..101B, Newton_2016, Newton_2017, Garraffo2018, Newton_2018, Medina_2022_spindown, Pass2022, Pass_2023, Jao_2023, 2023ApJ...954L..50E, 2023MNRAS.526..870S, 2024NatAs...8..223L, Pass_2024_ApJ}. To summarize: the stars typically begin and stay relatively rapidly rotating at $P_{rot} <$ 10d for roughly 1--3 Gyrs \citep{Medina_2022_spindown, Pass2022}, around 2.4$\pm$0.3 Gyrs undergo very rapid spindown during a phase of strong rotational braking \citep{Medina_2022_spindown}, are settled into slow rotation at $P_{rot} >$ 90d by 12.9$\pm$3.5 Gyrs \citep{Medina_2022_spindown}, and can ultimately reach periods at least as long as $\sim$180 day \citep{Medina_2022_spindown}. The intriguing `fast braking phase' is supported by the observed dearth of field FC M dwarfs with intermediate 10--70d rotation periods \citep{Newton_2016, Newton_2017, Newton_2018}, which is visible in the top-left panel of Fig.~\ref{fig:Prot-activity} as a clustering into two groups with $P_{rot}$ $\lesssim$10d and $\gtrsim$70d. The starting age of the fast braking phase is primarily set by stellar mass, with lower mass stars exhibiting a greater span between their fast and slow rotation distributions. However, there is clear variability to this overall process because some stars have spun down considerably by $<$1 Gyr \citep{Pass2022}. This may be caused in part by different initial rotation periods, potentially the result of different birth environments \citep[][]{Pass_2024_ApJ}. The fast braking phase is also possibly linked to elevated flaring and H$\alpha$ emission \citep{2019ApJ...870...10M, Pass_2023}, but little is known about magnetic activity in this transitional time period due to the paucity of targets within it. The transition between different rotational evolutionary stages may ultimately be driven by changes in stellar magnetic morphology and dynamo state, as described in \cite{Garraffo2018}, though alternative discussions are given in \cite{2019ApJ...886..120S} and \cite{2023MNRAS.526..870S}.

The key result we find here is the case of NLTT~44989~A and B, where twin FC M dwarfs with the same age/mass/composition/environment present rotation periods of 38d and 6.55d respectively, with correspondingly strong mismatches in $L_X$ and H$\alpha$. Our favored explanation is that NLTT~44989~A has already begun and progressed a moderate amount through its fast braking phase, while NLTT~44989~B has either not yet begun or only minimally progressed into the phase. This explanation is supported by the positions of each star in the top-left panel of Figure \ref{fig:Prot-activity}, where we see B on the lower envelope of the saturated regime and A already far along its transition to the slowly-rotating inactive clump. The position of B being noticeably below the activity level of other similar-mass and similar-rotation field stars suggests it may indeed have already begun its transition into the fast braking phase but not yet markedly slowed its rotation. However, the aforementioned possibility of elevated activity during the fast braking phase could contradict this --- the evolving strength of different activity tracers throughout the entirety of the transition phase clearly needs more study. The lack of field stars in the region directly between A and B tracks with the form of the saturated regime and implies that B will maintain a similar (or potentially greater) H$\alpha$ emission strength while it spins down to $\sim$30--50d, after which its H$\alpha$ activity will decrease. Component A already lacks H$\alpha$ activity, comparable to the similar-mass slowly-rotating field stars, suggesting its H$\alpha$ activity will remain largely the same going forward as it spins down from 38d towards $\sim$100d and beyond. The $L_X$ activity for both stars will likely follow similar pathways.

Our case of NLTT~44989~AB is akin to the two recently-reported FC M dwarf wide binary systems in \cite{Pass_2024_ApJ} whose components display mismatched active/inactive H$\alpha$ despite very similar masses\footnote{One of these two systems is also in our full twins sample and will be included in our future paper.}. These three (near) twin systems all exhibit strong differences in H$\alpha$, as well as in $L_X$ and rotation in NLTT~44989~AB. Each case controls for the age, composition, and mass, implying that the spindown process is not a function of any combination of these factors exclusively; this is in agreement with the observed dispersion in spindown epoch at similar mass and age as discussed in \cite{Pass2022} and \cite{Pass_2024_ApJ}. Scatter in rotation periods before and after the star-disk locking phase \citep[star-disk locking ---][]{1991ApJ...370L..39K, 1993AJ....106..372E} could play a significant role in setting stars' future spindown evolution as well. High-energy environments at birth may impact disk lifetimes \citep{2017AJ....153..240A}, and consequently, rotation rates \citep{2021MNRAS.508.3710R}, but should not be a differentiating factor in these three binary systems. The observed activity and rotation differences must therefore be manifesting at least in part from some combination of (1) initial rotation periods changing by factors other than birth energy environment, e.g., disk sizes and/or masses, (2) a stochasticity in the onset of the fast braking phase, or (3) other unknown parameters relevant to overall spindown evolution.

There are no present-day observables that can reach back in time to pinpoint initial rotation periods for stars, but our twin cases provide some context. We find very similar periods of 0.86d and 0.68d in GJ~1183~AB, and in our forthcoming paper we will report results for several more twin systems having components with similar rotation periods (Couperus et al.~in prep). On the other hand, in 2MA~0201+0117~AB we see periods of 6.01d and 3.30d at $\sim$25~Myr, so rotation periods can differ at a very young age --- this is in agreement with the scatter in rotation M stars show in young clusters as well \citep[e.g.,][]{2021ApJ...916...77P}. For example, some stellar components could have had more massive disks with longer disk lifetimes and disk-locking durations, leading to subsequently slower stellar rotation periods after contraction finishes. The spindown models of \cite{2023MNRAS.526..870S} can reproduce the very different rotation periods of NLTT~44989~AB at a range of ages $\gtrsim$2.5~Gyr if different configurations of initial period and disk lifetime are assumed for each 0.25~$\textup{M}_\odot$ component (see their Fig. 5). However, these models still generally struggle to match the very long periods seen in low-mass field stars, as well as the \cite{Pass2022} results. Altogether, differences in initial rotation rates remains a plausible hypothesis to explain present-day differences.

We summarize that possible culprits for generating these spin mismatches in NLTT~44989~AB and 2MA~0201+0117~AB could be early formation factors as discussed above, dynamical binary disk interactions ( \S\ref{subsubsec:discussion-explanations-dynamics}), planetary impacts on the disk and/or host star (\S\ref{subsubsec:discussion-explanations-exoplanets}), or complex dynamo behaviors (\S\ref{subsubsec:discussion-explanations-bistability}). These are joined by several other less likely explanations considered throughout \S\ref{subsec:discussion-explanations}. We recommend that the key follow-up investigations to continue disentangling these various possibilities are higher-precision RV exoplanet searches of NLTT~44989~AB, Zeeman–Doppler imaging magnetic reconstructions of NLTT~44989~AB and other twin FC M dwarf binaries with activity mismatches, and studies of disks in twin PMS M stars. Future work should also try to obtain a more precise measurement of the NLTT~44989~A rotation signal\footnote{\textit{TESS} is planned to observe the NLTT~44989~AB system again in 2025 during sectors 91 and 92, aiding future studies.}, as it potentially has a period even longer than we measured here giving an even stronger mismatch (see \S\ref{subsec:rot-results}).

\subsection{Implications for Exoplanet Host Activity Predictions} \label{subsec:discussion-exoplanets}

It is desirable to reconstruct the complete stellar activity history of a given exoplanet host star to help model the impacted planetary atmospheric evolution and evaluate habitability factors. This is especially true for M dwarfs, given that they host key candidates for exoplanet atmospheric characterization with existing and upcoming observatories, while also exhibiting significant stellar activity. Alas, our results highlight the ongoing challenges in modeling M dwarfs' activity evolution. For example, one could know the precise age, mass, and composition of the components in NLTT~44989~AB but would still be unable to predict their present-day rotation periods to within a factor of $\sim$6, or their X-ray luminosities to within a factor of $\sim$40 or more. The rotation difference in particular has significant implications for any attempts to apply gyrochronology in FC M dwarfs as well. Even if rotation periods are known and match, GJ~1183~AB indicates intrinsic differences of at least 58$\pm$9\% in $L_X$ and 26$\pm$9\% in H$\alpha$ can exist. Further muddying our interpretations of activity, these intrinsic scatters are derived from multi-epoch observations of just the activity \textit{now}, not a fully reconstructed average history. {\it Our results alone don't determine if most FC M dwarfs typically deviate by these amounts, but do demonstrate that presumed twin stars can differ by at least this much.}

There is also likely a phase around the rapid spindown epoch during which the scatter in initial rotation periods drives scattered spindown epochs for even equal mass FC M dwarfs, thus resulting in a degraded ability to predict any small star's rotation and activity within the fast braking time period. This claim is supported by \cite{Pass_2024_ApJ}, who found a 1$\sigma$ dispersion upper limit of 0.5~Gyr in otherwise mass-directed FC M dwarf spindown. It is therefore important to further investigate NLTT~44989~AB and similar mismatched systems to determine if they represent a phase all FC M dwarfs go through --- perhaps briefly making it rare to see --- or if they are complex outliers.

Efforts continue to gradually improve our understanding and ability to estimate activity over M dwarf stellar lifetimes, but for now it is clear that considerable caution should be employed for any exoplanet studies relying on activity reconstructions for specific FC M dwarf hosts, such as Proxima Cen or TRAPPIST-1.

\section{Conclusions} \label{sec:conclusions}

We have presented newly acquired long-term light curves (\S\ref{subsec:longterm-methods}, \S\ref{subsec:longterm-results}), rotation periods (\S\ref{subsec:rot-methods}, \S\ref{subsec:rot-results}), H$\alpha$ EWs (\S\ref{subsec:chiron-methods}, \S\ref{subsubsec:ha-results}), radial velocities (\S\ref{subsec:chiron-methods}, \S\ref{subsubsec:rv-results}), X-ray luminosities (\S\ref{subsec:chandra-methods}, \S\ref{subsec:chandra-results}), coronal parameters (\S\ref{subsec:chandra-methods}, \S\ref{subsec:chandra-results}), and speckle imaging observations (\S\ref{subsec:speckle-methods}, \S\ref{subsec:speckle-results}) for four fully convective M star twin wide binaries. We found one system, KX~Com~A-BC, to be a hierarchical triple, in agreement with the recent results of \cite{Pass_2024_ApJ} --- the other three systems present as true twin binaries with the same age/mass/composition/environment. The main takeaways from this work are the following:

\begin{itemize}
    \item We uncover consistent activity differences in $L_X$ and H$\alpha$ for all three true twin pairs. NLTT~44989~A/B shows a remarkable inactive/active disagreement between components, while long-term stellar activity cycles may be influencing the relative strength of the observed mismatches in other cases.
    \item In each twin pair, the component more active in $L_X$ is also the more active star in H$\alpha$, while photometric rotation amplitudes do not always follow this trend.
    \item NLTT~44989~AB has a strong rotation rate mismatch of 38d versus 6.55d, 2MA~0201+0117~AB has a moderate mismatch at 6.01d and 3.30d, and GJ~1183~AB hosts similar rotation periods of 0.86d and 0.68d.
    \item The discrepant rotation periods and activity levels in NLTT~44989~AB likely stem from one component having begun and progressed moderately through its fast braking phase before the other, despite both components having the same age, mass, composition, and environment. We hypothesize that material interactions may be responsible, either through disk interference at formation producing very different initial stellar rotation periods or from longer term star-planet interactions as the system has evolved.
    \item GJ~1183~AB shows mismatched spot activity levels throughout a decade of photometry and demonstrates that twin FC M dwarfs with very similar rotation periods can still deviate in photometric variability properties, as well as in $L_X$ and H$\alpha$ that differ by 58$\pm$9\% and 26$\pm$9\% respectively.
    \item 2MA~0201+0117~AB is a pre-main-sequence twin pair in the $\beta$ Pictoris association that provides a valuable system for future theoretical comparisons. In particular, this system indicates that differences may be present in twin stars at a very young age of only $\sim$25 Myr.
    \textbf{\item Overall, the differences found here for twin stars indicate that it is {\it very} difficult to anticipate the integrated historical environments provided by fully convective M dwarfs for any orbiting planets.}
\end{itemize}

This work has considered four systems from our broader sample of 36 total systems that span the entire partially and fully convective M dwarf sequence. Rotation and activity results for the remaining 32 twin pairs will be presented in our forthcoming paper.

\vskip20pt

RECONS CTIO/SMARTS 0.9m light curve data shown in this work are available as Data behind the Figure (DbF) products in the online journal. X-ray data are available from the \textit{Chandra} Data Archive. CHIRON spectra can be obtained from the NOIRLab Data Archive.


\begin{acknowledgments}

A. Couperus thanks the following individuals for conversations that enhanced this work: Leonardo Paredes, Emily Pass, and Russel White. We also thank Andrei Tokovinin for his assistance in collecting and reducing the SOAR speckle results. This work has been supported by the NSF through grants AST-141206, AST-1715551, and AST-2108373, as well as via NASA/\textit{Chandra} grant GO1-22013B. We have used data from the SMARTS 0.9m telescope, which is operated as part of the SMARTS Consortium by RECONS (www.recons.org) members, and with the assistance of staff at Cerro Tololo Inter-American Observatory. This work has made use of data from the European Space Agency (ESA) mission \textit{Gaia}, processed by the \textit{Gaia} Data Processing and Analysis Consortium (DPAC). Funding for the DPAC has been provided by national institutions, in particular the institutions participating in the \textit{Gaia} Multilateral Agreement. This publication makes use of data products from the Two Micron All Sky Survey, which is a joint project of the University of Massachusetts and the Infrared Processing and Analysis Center/California Institute of Technology, funded by the National Aeronautics and Space Administration and the National Science Foundation. This paper includes data collected by the \textit{TESS} mission, which are publicly available from the Mikulski Archive for Space Telescopes (MAST) at \dataset[doi:10.17909/0cp4-2j79]{http://dx.doi.org/10.17909/0cp4-2j79}. This research has made use of NASA’s Astrophysics Data System (ADS), as well as the SIMBAD database \citep{SIMBAD} and VizieR catalog access tool \citep{VizieR} operated at CDS, Strasbourg, France. This research has also made use of software provided by the \textit{Chandra} X-ray Center (CXC) in the application packages CIAO and Sherpa.

\end{acknowledgments}

%

\facilities{CTIO:0.9m, CTIO:1.5m(CHIRON), CTIO:2MASS, FLWO:2MASS, \textit{Gaia}, \textit{TESS}, CXO, PO:1.2m, SOAR, LDT}


\software{IRAF \citep{10.1117/12.968154,1993ASPC...52..173T}, SExtractor \citep{1996A&AS..117..393B}, CIAO \citep{Ciao2006}, Sherpa \citep{Sherpa2001}, unpopular \citep{unpopular}, Astropy \citep{2013A&A...558A..33A,Astropy2018}, Matplotlib \citep{Matplotlib2007}, NumPy \citep{NumPy2020}, and Aladin \citep{2000A&AS..143...33B,2014ASPC..485..277B}.}




\appendix
\section{\textit{Chandra} X-ray Figures} \label{sec:appendix}
\begin{figure}[!ht]
\centering
\gridline{\fig{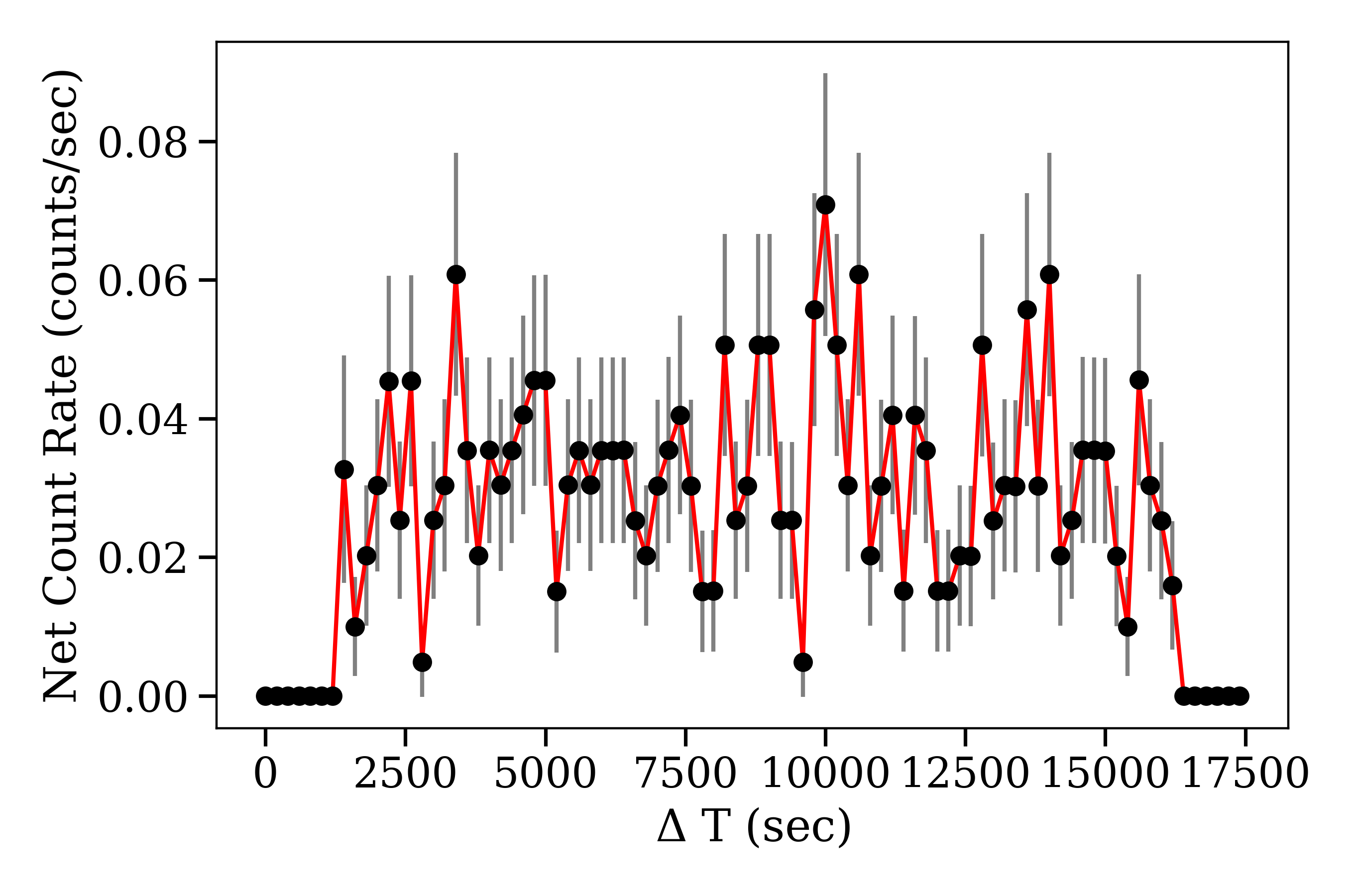}{0.33\textwidth}{(GJ 1183 A--1)}
          \fig{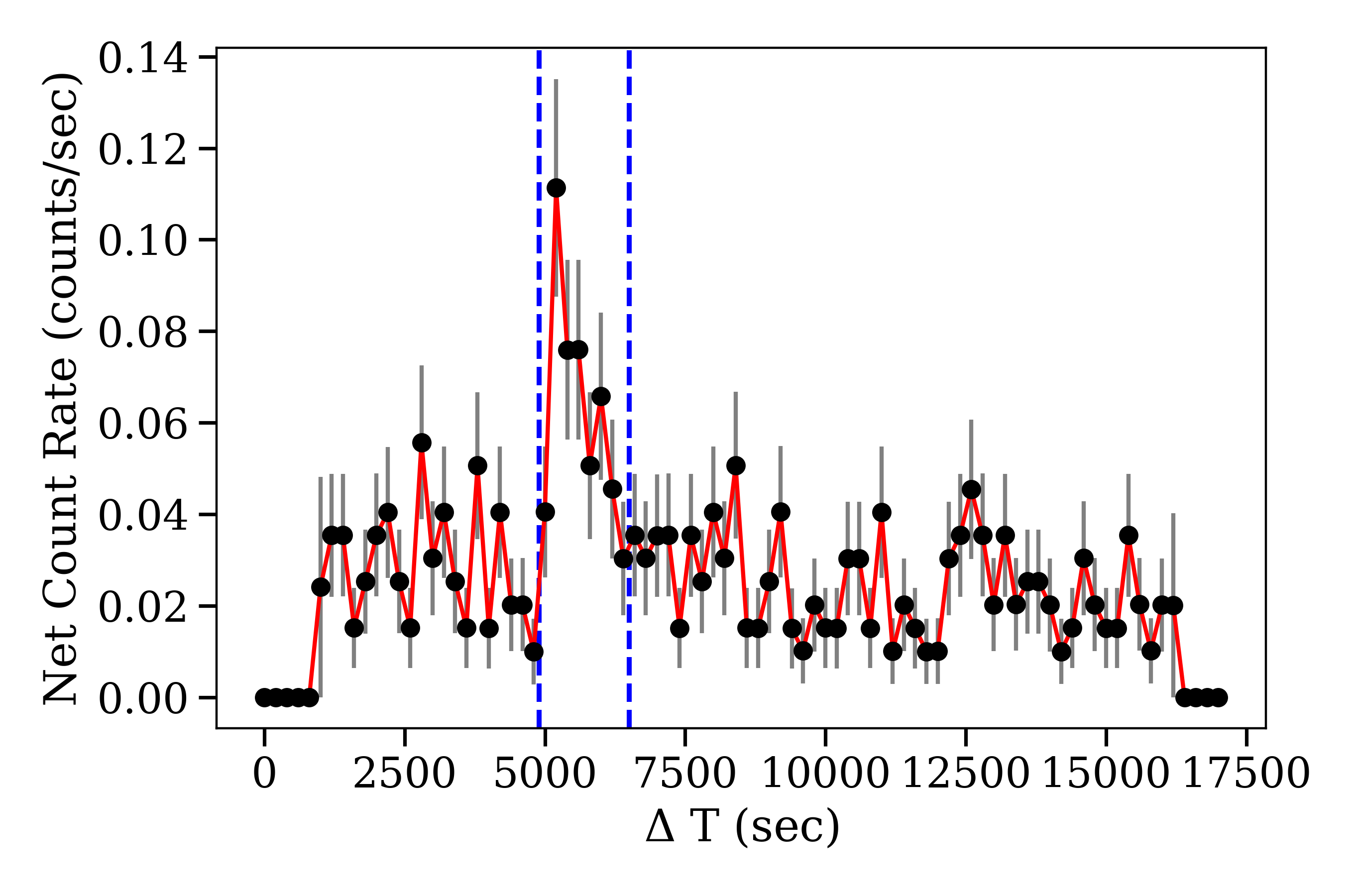}{0.33\textwidth}{(GJ 1183 A--2)}
          \fig{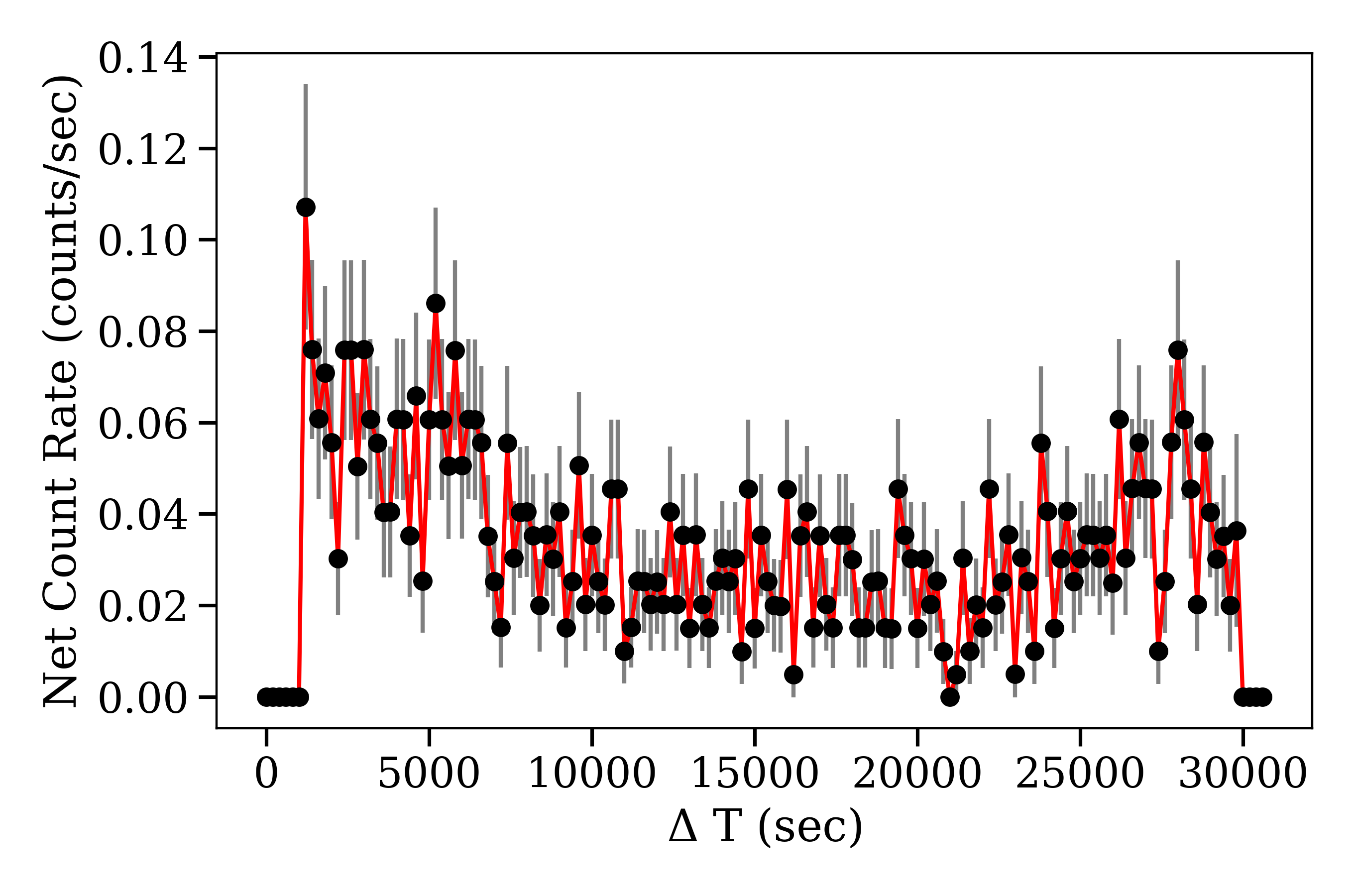}{0.33\textwidth}{(GJ 1183 A--3)}}
\vspace*{-4mm}
\gridline{\fig{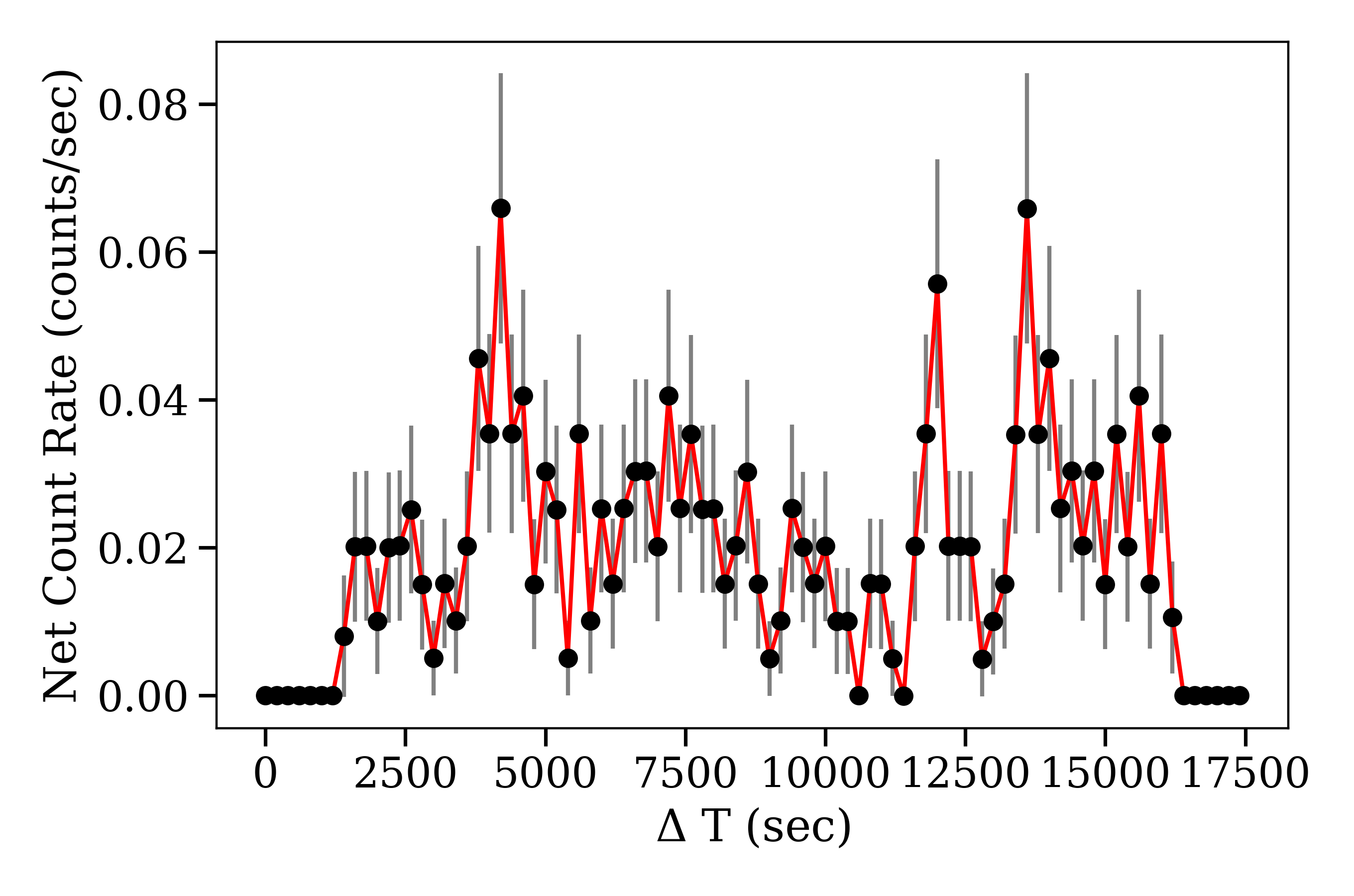}{0.33\textwidth}{(GJ 1183 B--1)}
          \fig{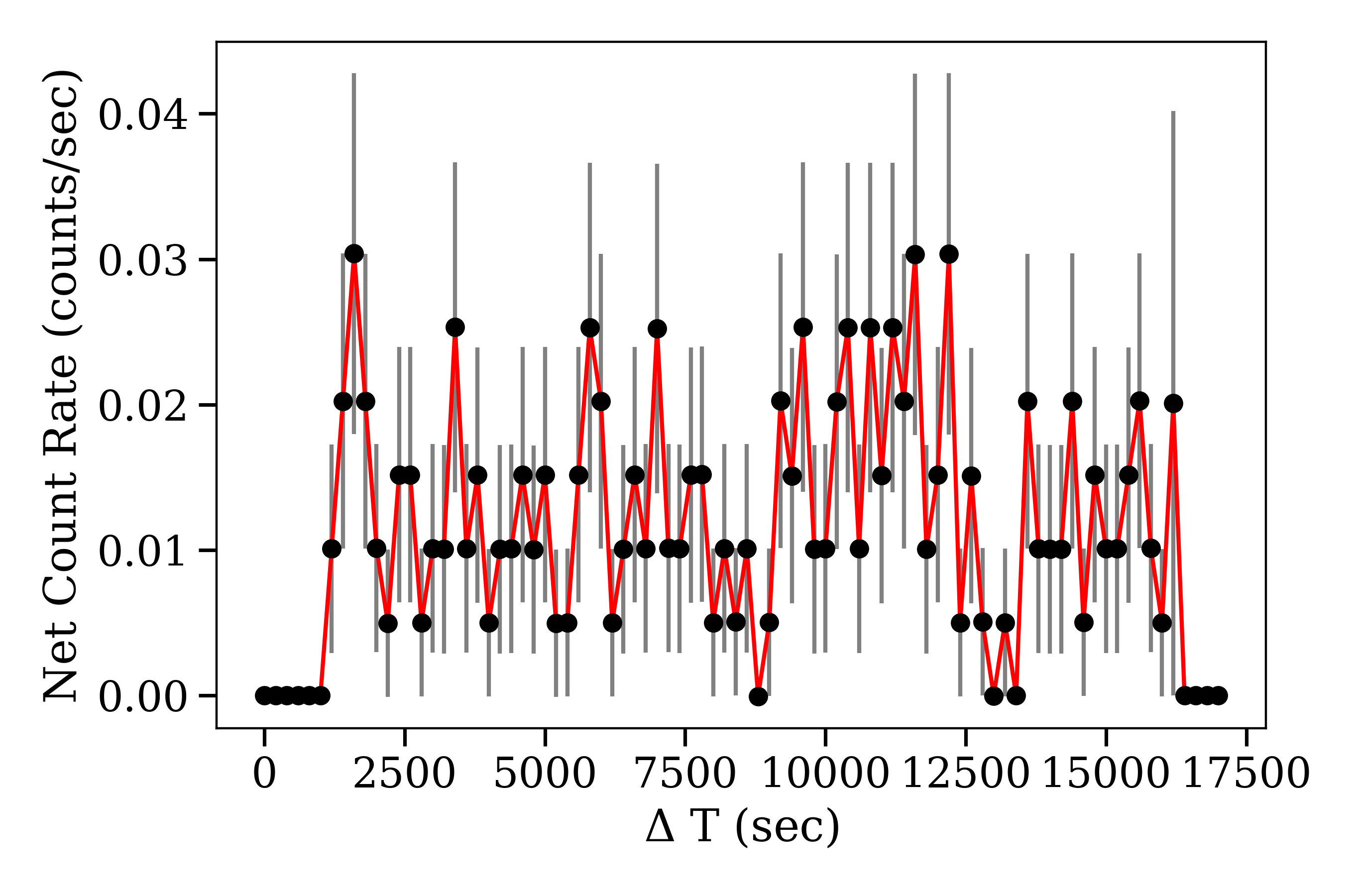}{0.33\textwidth}{(GJ 1183 B--2)}
          \fig{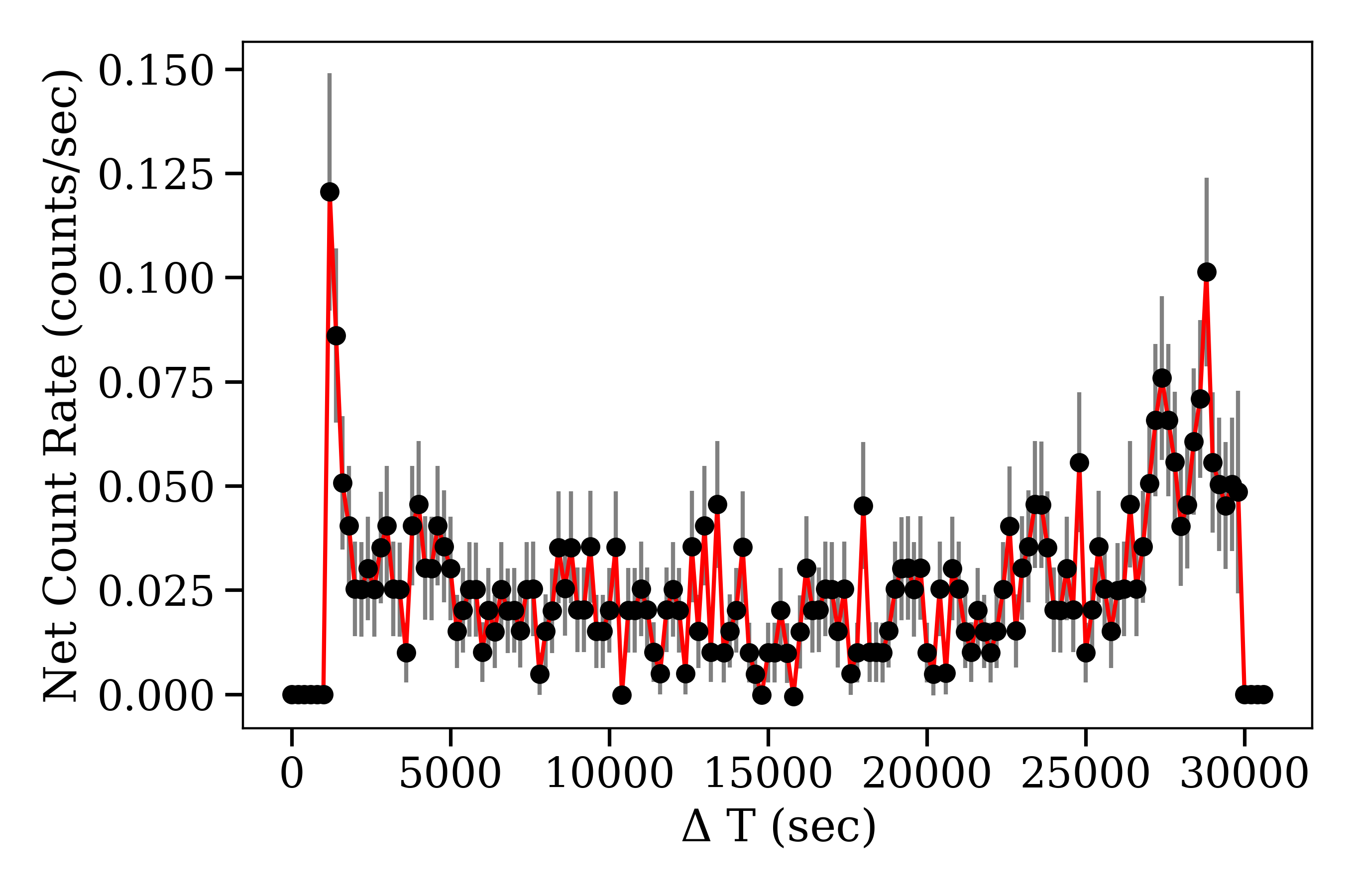}{0.33\textwidth}{(GJ 1183 B--3)}}
\vspace*{1mm}
\gridline{\fig{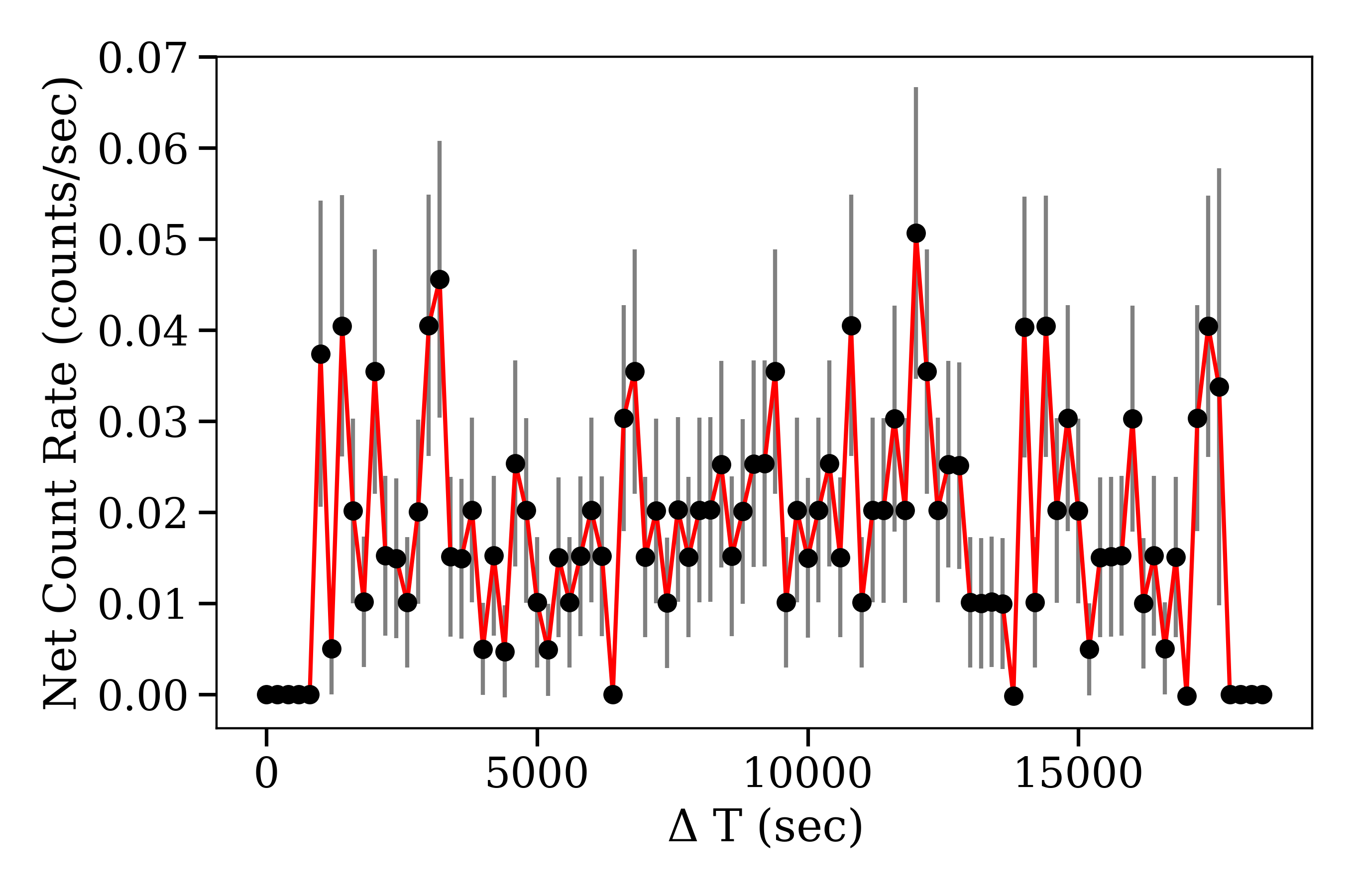}{0.33\textwidth}{(KX Com A--1)}
          \fig{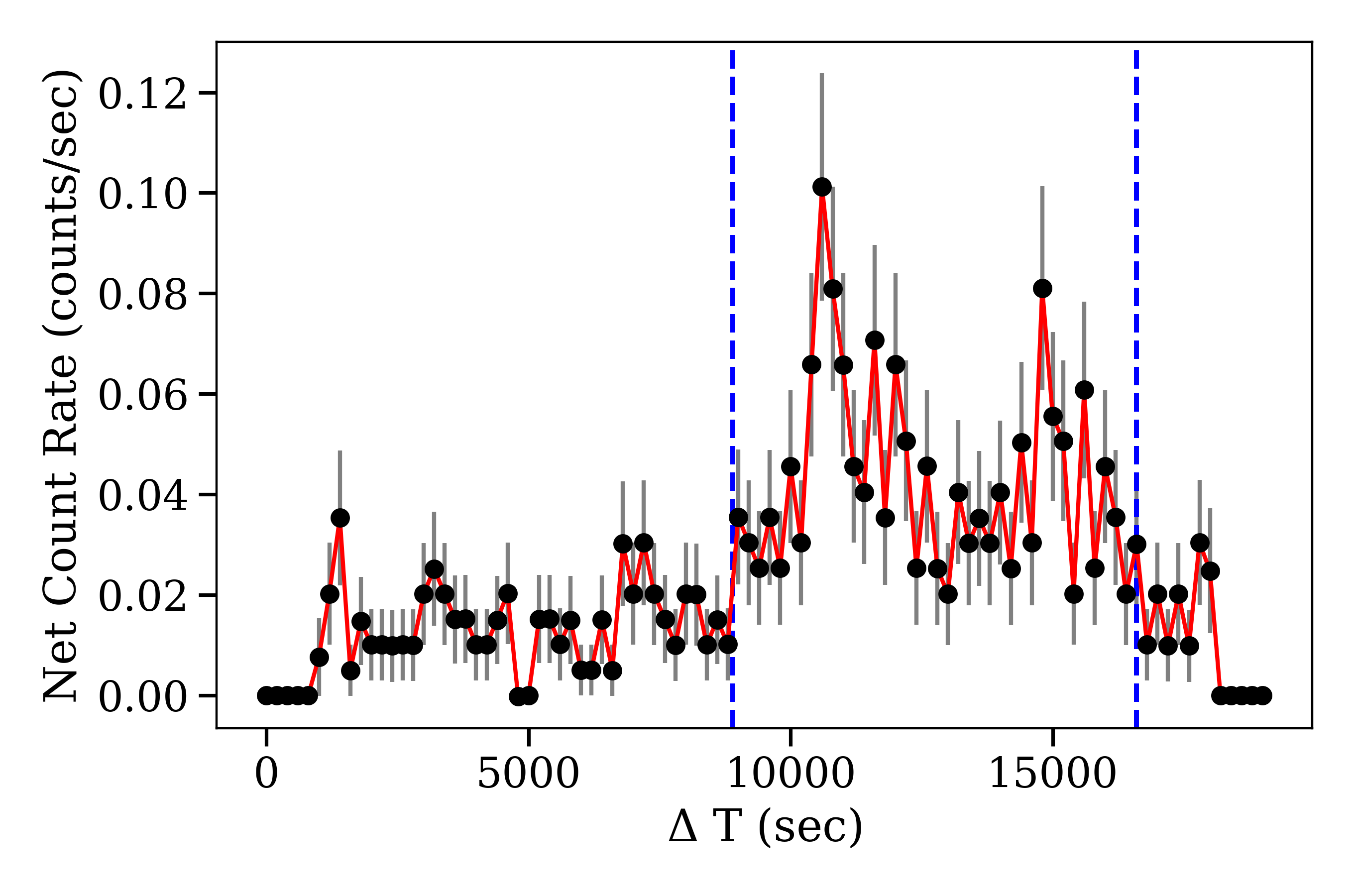}{0.33\textwidth}{(KX Com A--2)}
          \fig{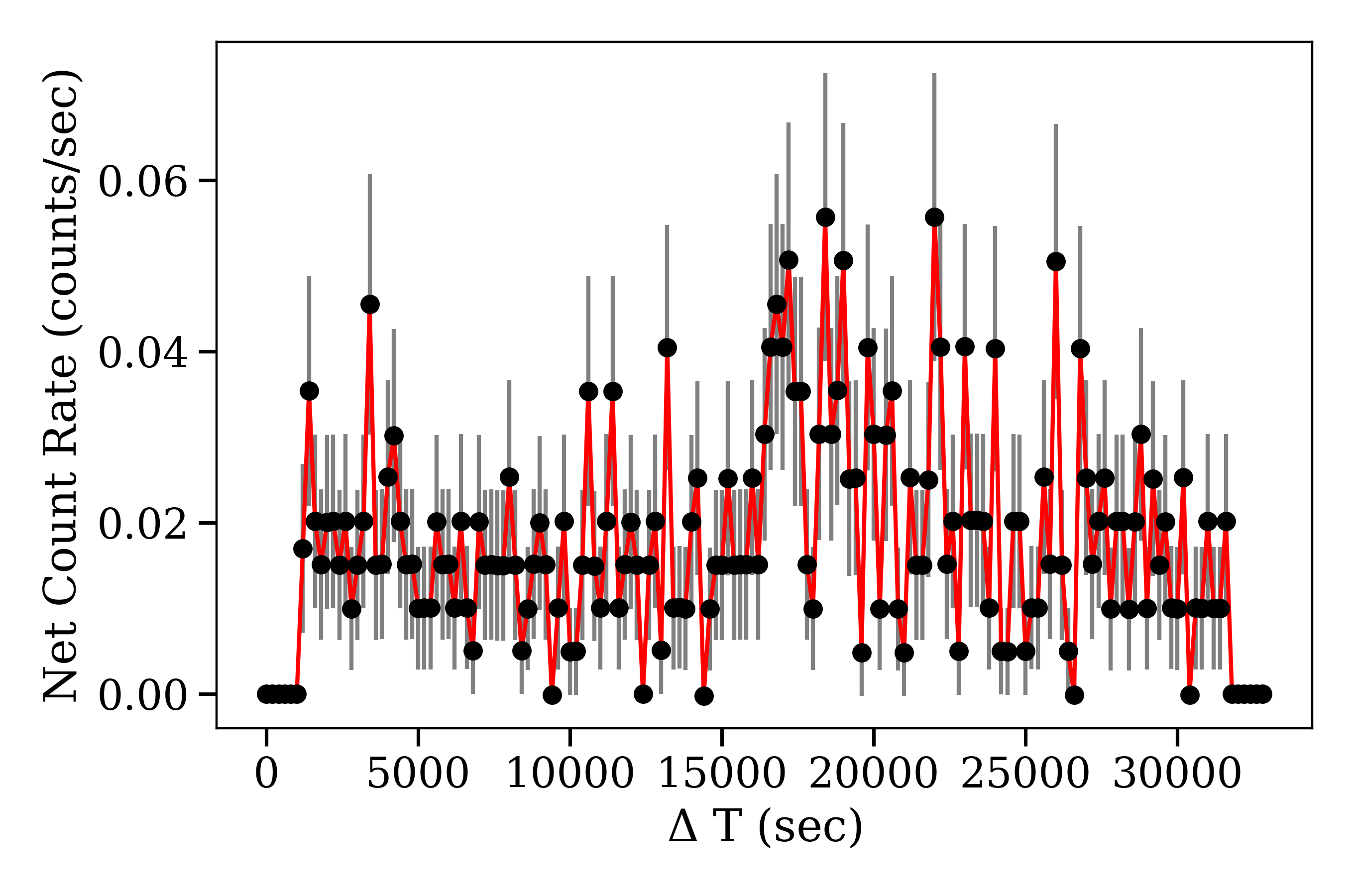}{0.33\textwidth}{(KX Com A--3)}}
\vspace*{-4mm}
\gridline{\fig{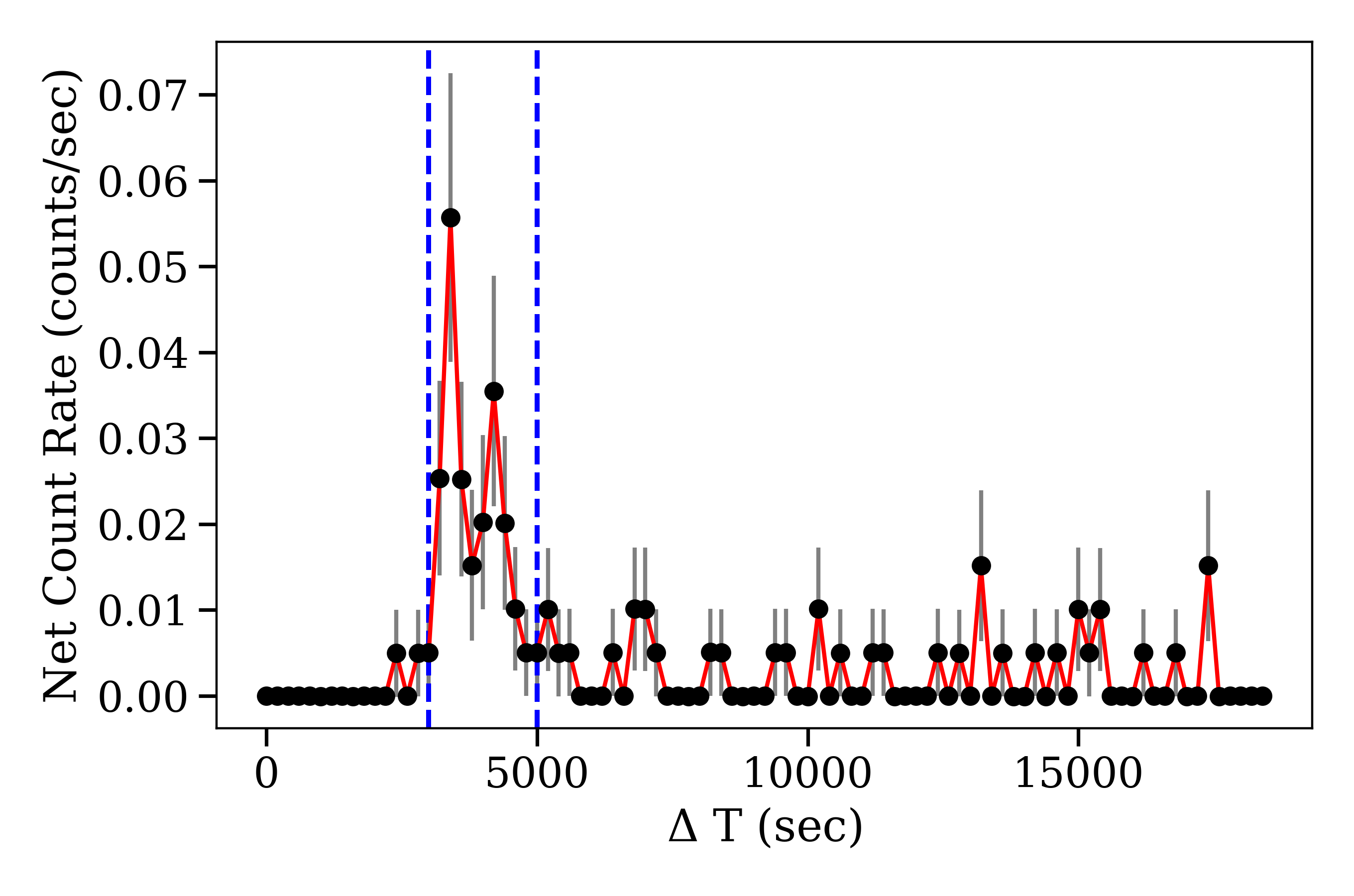}{0.33\textwidth}{(KX Com BC--1)}
          \fig{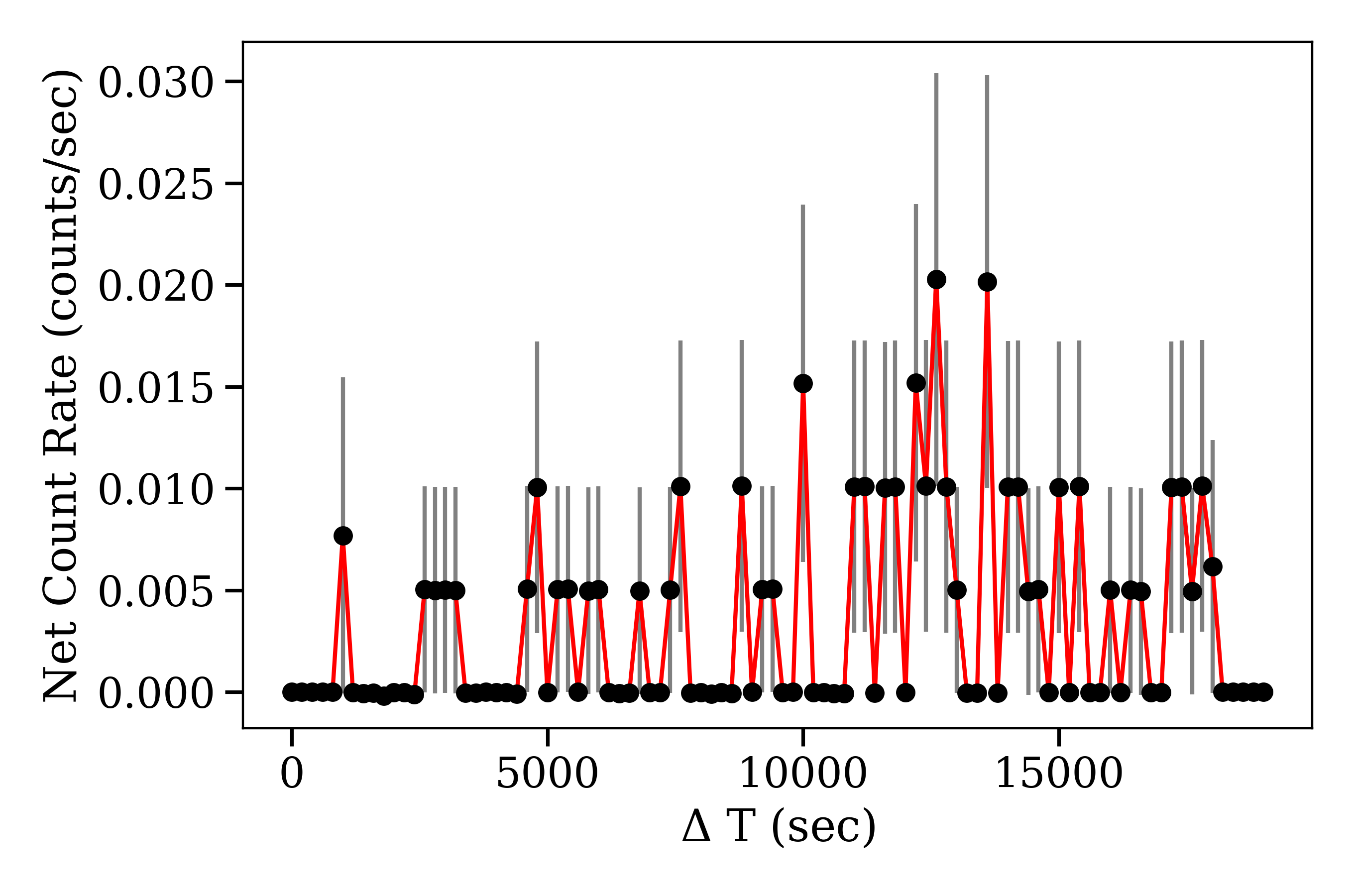}{0.33\textwidth}{(KX Com BC--2)}
          \fig{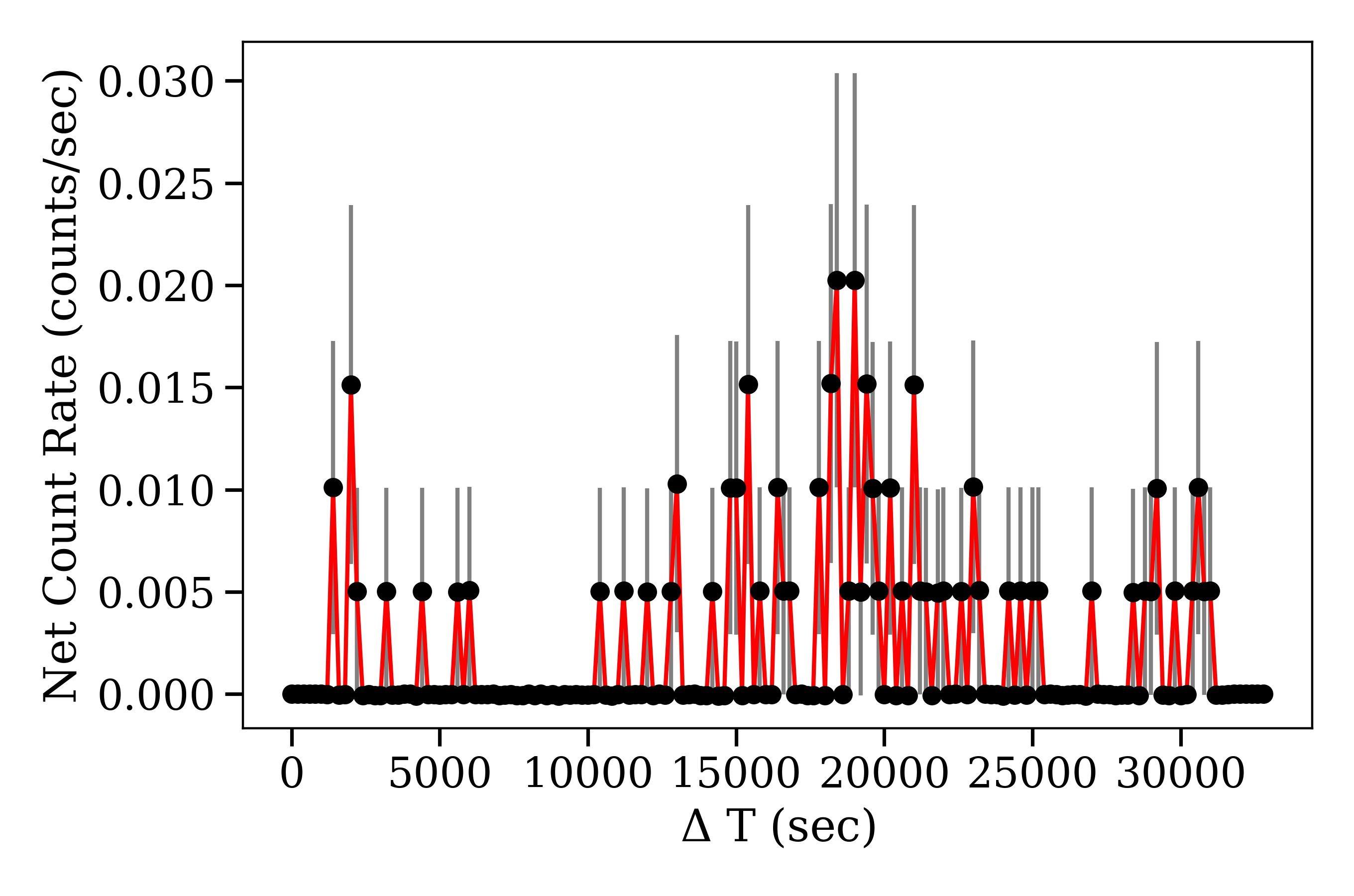}{0.33\textwidth}{(KX Com BC--3)}}
\vspace*{1mm}
\gridline{\fig{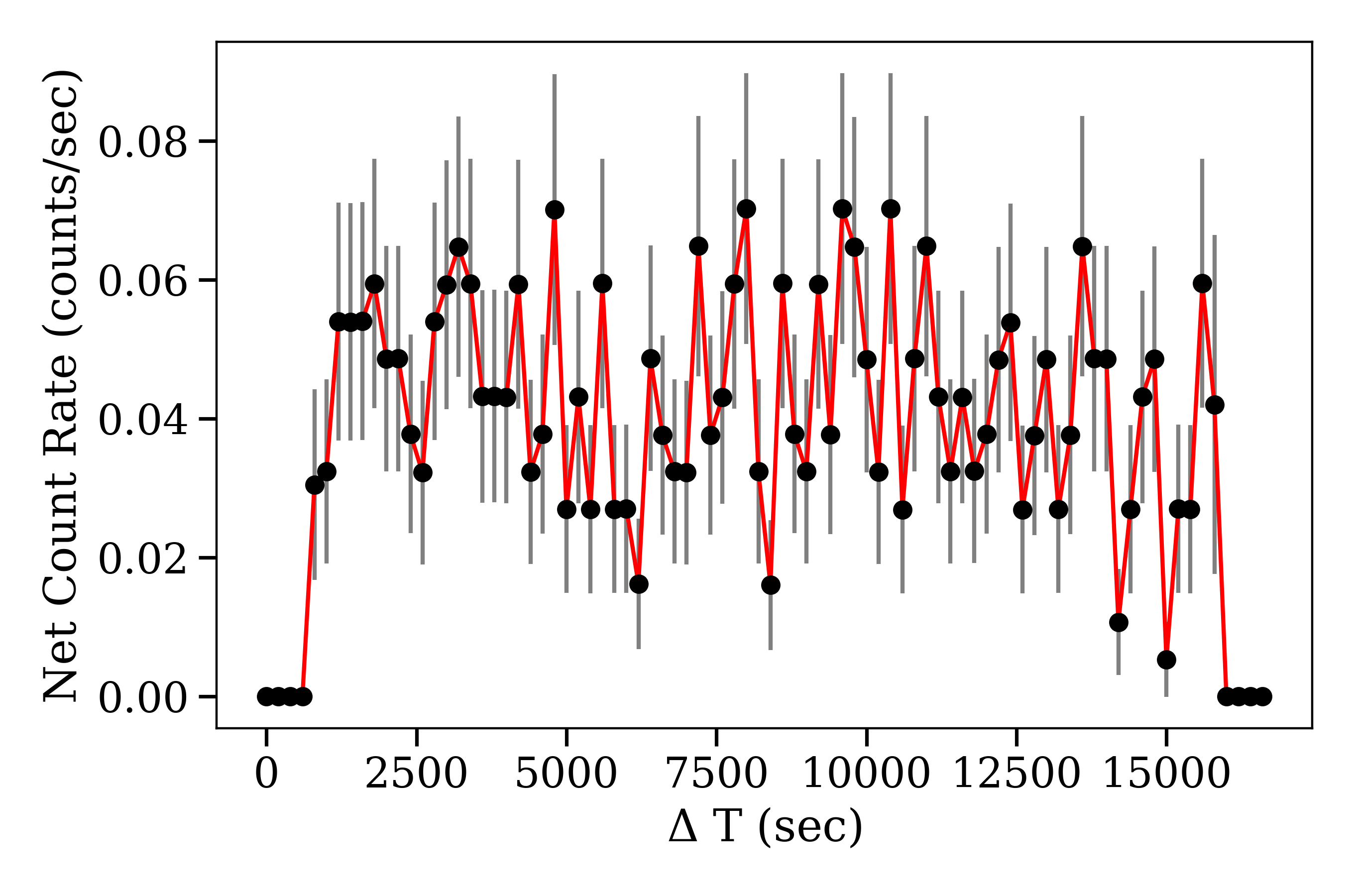}{0.33\textwidth}{(2MA 0201+0117 A)}
          \fig{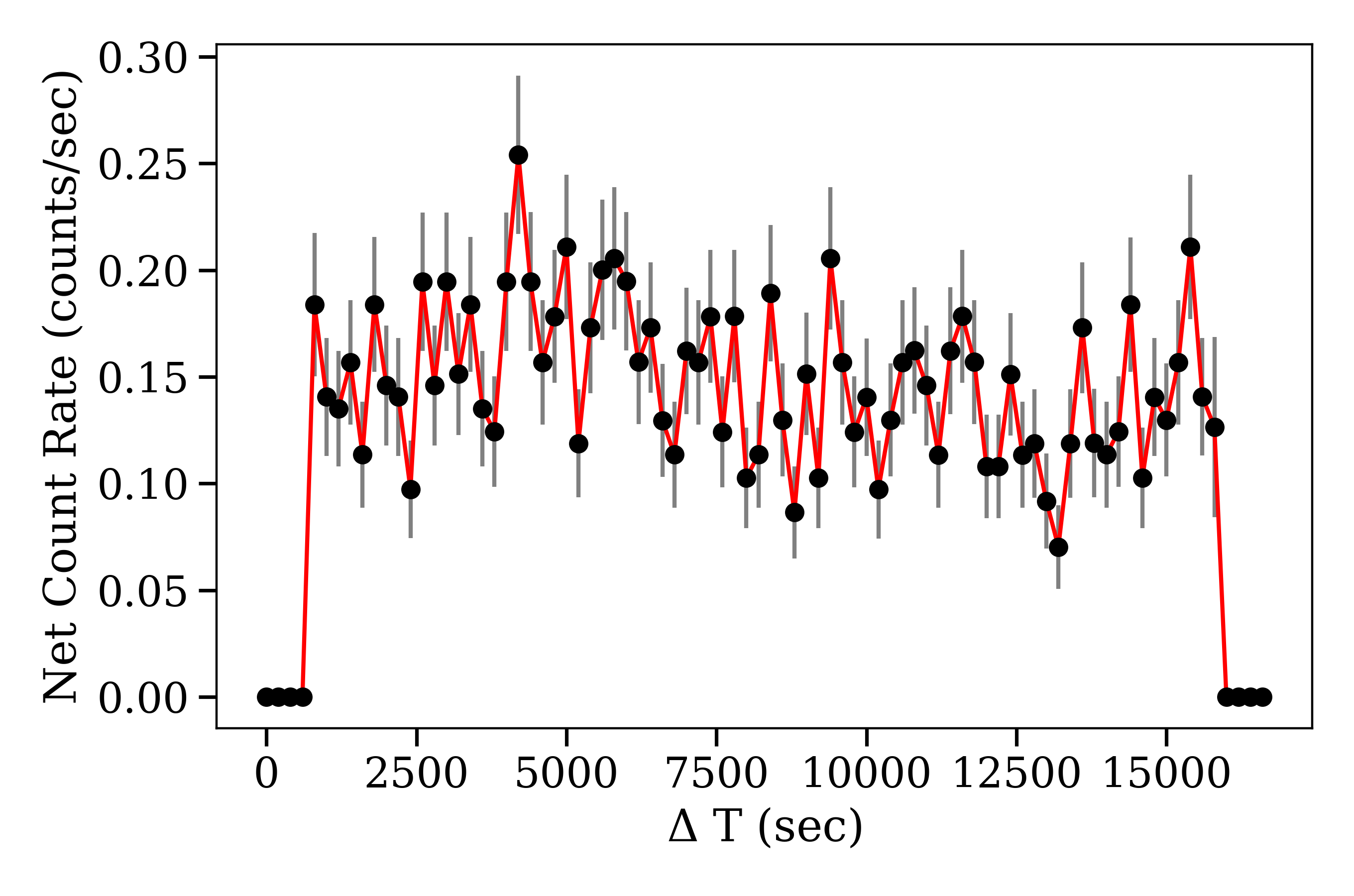}{0.33\textwidth}{(2MA 0201+0117 B)}}
\vspace*{-2mm}
\figcaption{The same as Figure \ref{fig:NLTT-B-LC}, now for all of the remaining X-ray datasets. NLTT~44989~A is not shown given its lack of a confident detection (see \S\ref{subsubsec:NLTT-A}). \label{fig:all-xray-lc}}
\end{figure}

\begin{figure}[!ht]
\centering
\gridline{\fig{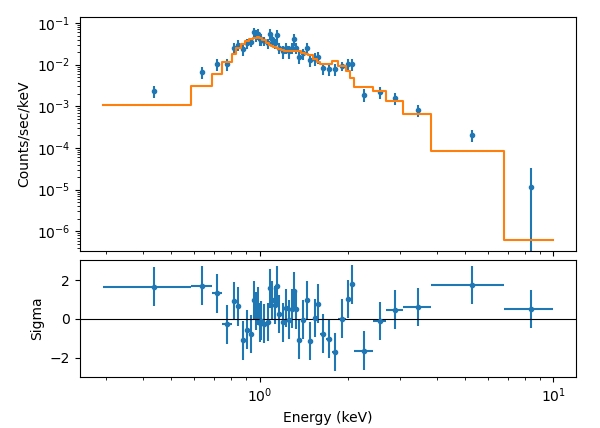}{0.31\textwidth}{(GJ 1183 A--1)}
          \fig{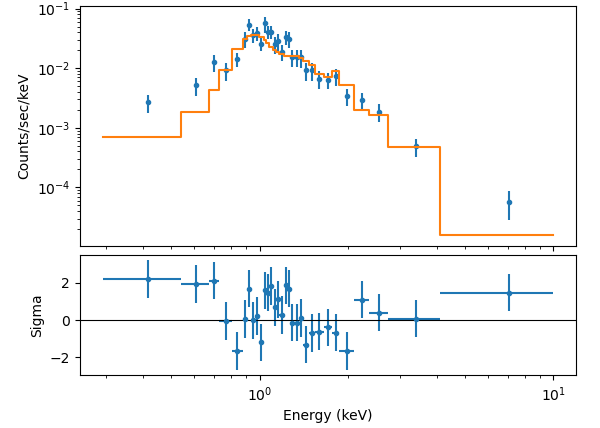}{0.31\textwidth}{(GJ 1183 A--2)}
          \fig{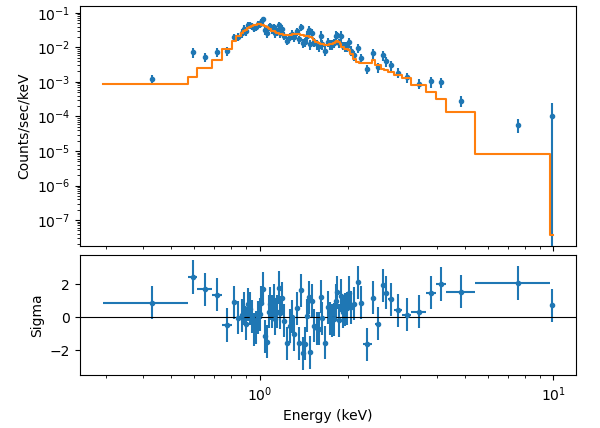}{0.31\textwidth}{(GJ 1183 A--3)}}
\vspace*{-3mm}
\gridline{\fig{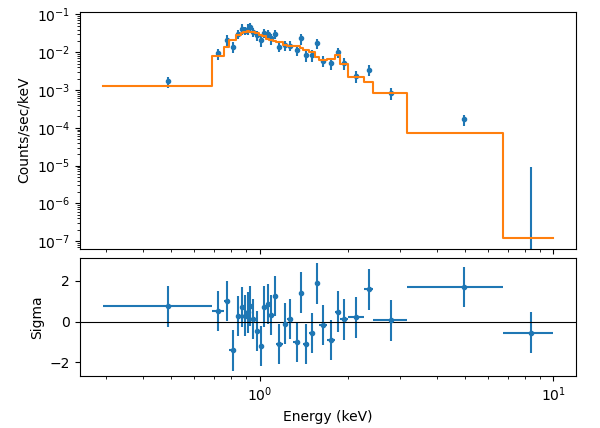}{0.31\textwidth}{(GJ 1183 B--1)}
          \fig{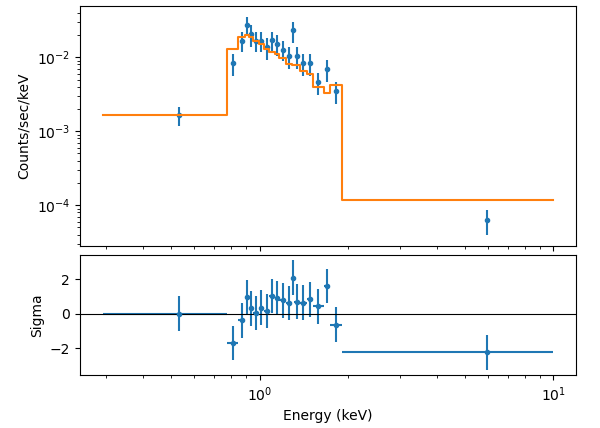}{0.31\textwidth}{(GJ 1183 B--2)}
          \fig{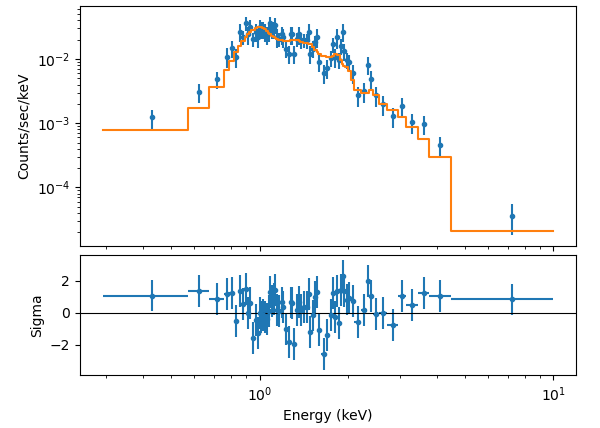}{0.31\textwidth}{(GJ 1183 B--3)}}
\vspace*{1mm}
\gridline{\fig{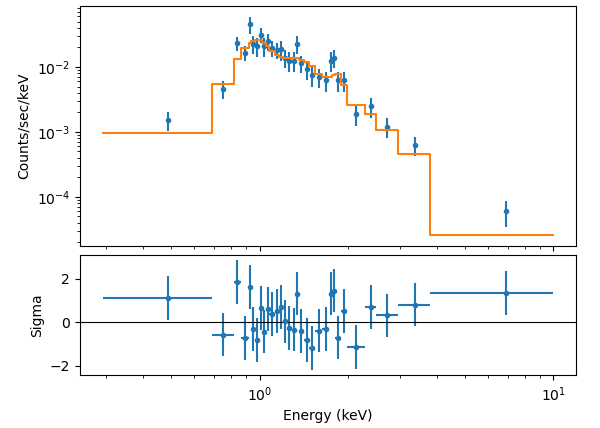}{0.31\textwidth}{(KX Com A--1)}
          \fig{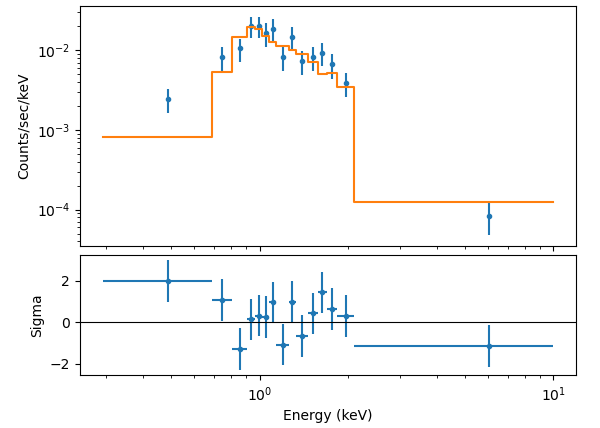}{0.31\textwidth}{(KX Com A--2)}
          \fig{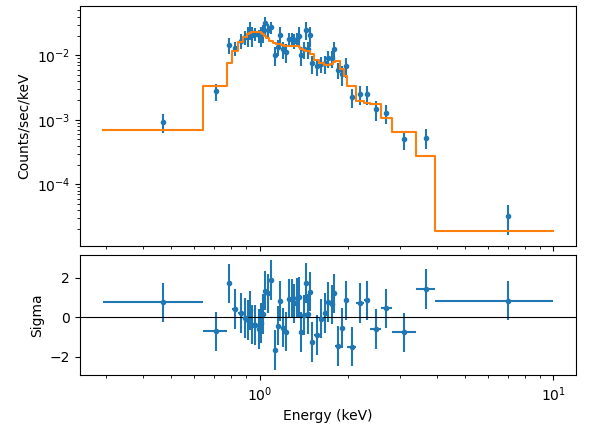}{0.31\textwidth}{(KX Com A--3)}}
\vspace*{-3mm}
\gridline{\fig{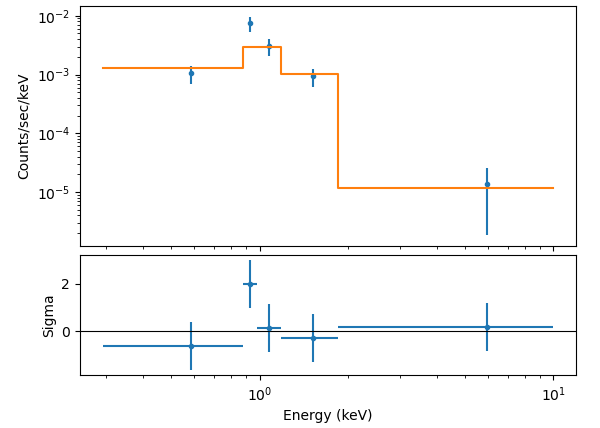}{0.31\textwidth}{(KX Com BC--1)}
          \fig{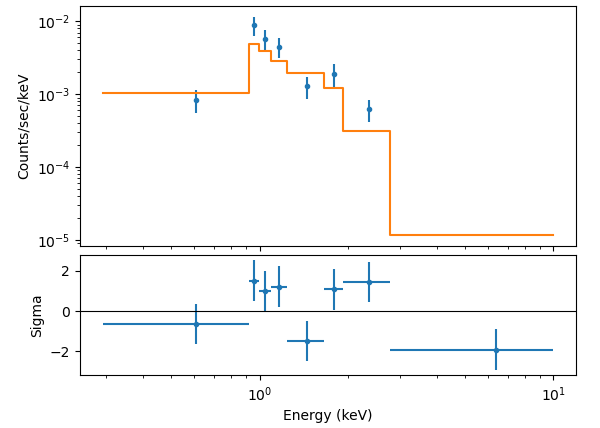}{0.31\textwidth}{(KX Com BC--2)}
          \fig{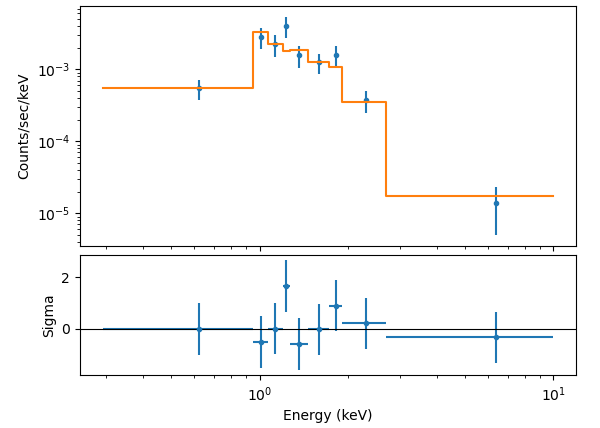}{0.31\textwidth}{(KX Com BC--3)}}
\vspace*{1mm}
\gridline{\fig{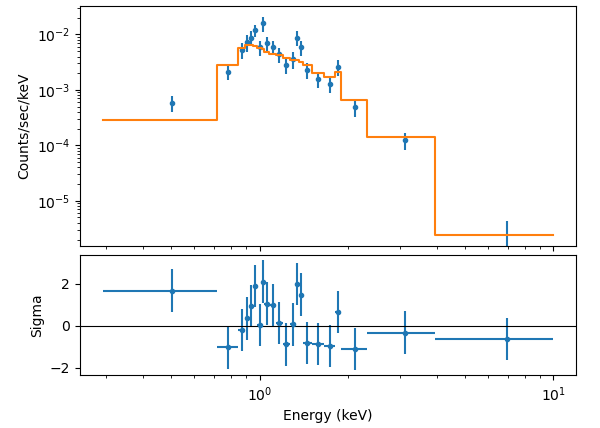}{0.31\textwidth}{(NLTT 44989 B)}}
\vspace*{-2mm}
\figcaption{The same as Figure \ref{fig:2MA0201-SpectraFit}, now for all of the remaining quiescent X-ray datasets. NLTT~44989~A is not shown given its lack of a confident detection (see \S\ref{subsubsec:NLTT-A}). \label{fig:all-xray-spectra}}
\end{figure}

This appendix provides all of the X-ray light curves (Fig.~\ref{fig:all-xray-lc}) and quiescent X-ray spectral fits (Fig.~\ref{fig:all-xray-spectra}) from our \textit{Chandra} observations, except those already shown earlier in Figure \ref{fig:NLTT-B-LC} and Figure \ref{fig:2MA0201-SpectraFit}.

\newpage
\bibliography{References}{}
\bibliographystyle{aasjournal}



\end{document}